\documentclass[preprint,showpacs,titlepage,aps,prd,tightenlines,amsmath,byrevtex,nofootinbib]{revtex4}

\usepackage{graphicx}

\newcommand{\mn}{\Delta m_{21}^2}
\newcommand{\mt}{\Delta m_{31}^2}
\newcommand{\si}{s_{12}}
\newcommand{\sn}{s_{23}}
\newcommand{\st}{s_{13}}
\newcommand{\ci}{c_{12}}
\newcommand{\cn}{c_{23}}

\def\lsim{\raise0.3ex\hbox{$\;<$\kern-0.75em\raise-1.1ex
\hbox{$\sim\;$}}}
\def\gsim{\raise0.3ex\hbox{$\;>$\kern-0.75em\raise-1.1ex
\hbox{$\sim\;$}}}

\begin{document}
\title{
Resolving CP Violation by Standard and Nonstandard Interactions and 
Parameter Degeneracy in Neutrino Oscillations
}
\author{A.~M.~Gago$^{1}$}
\email{agago@fisica.pucp.edu.pe}
\author{H.~Minakata$^{2}$}
\email{minakata@tmu.ac.jp}
\author{H.~Nunokawa$^{3}$}
\email{nunokawa@fis.puc-rio.br} 
\author{S.~Uchinami$^{2}$}
\email{uchinami@phys.metro-u.ac.jp}
\author{R.~Zukanovich Funchal$^{4}$}
\email{zukanov@if.usp.br}
\affiliation{
$^1$Secci\'on F\'{\i}sica, Departamento de Ciencias, Pontificia
Universidad Cat\'{o}lica del Per\'{u}, Apartado 1761, Lima, Per\'{u} \\
$^2$Department of Physics, Tokyo Metropolitan University, Hachioji, 
Tokyo 192-0397, Japan \\
$^3$Departamento de F\'{\i}sica, Pontif{\'\i}cia Universidade Cat{\'o}lica 
do Rio de Janeiro, C. P. 38071, 22452-970, Rio de Janeiro, Brazil  \\ 
$^4$Instituto de F\'{\i}sica, Universidade de S\~ao Paulo, 
 C.\ P.\ 66.318, 05315-970 S\~ao Paulo, Brazil
}

\date{December 23, 2009}

\vglue 1.4cm

\begin{abstract} 

  In neutrino oscillation with non-standard interactions (NSI) the
  system is enriched with CP violation caused by phases due to NSI in
  addition to the standard lepton Kobayashi-Maskawa phase $\delta$. In
  this paper we show that it is possible to disentangle the two CP
  violating effects by measurement of muon neutrino appearance by a
  near-far two detector setting in neutrino factory experiments.
  Prior to the quantitative analysis we investigate in detail the
  various features of the neutrino oscillations with NSI, but under
  the assumption that only one of the NSI elements, $\varepsilon_{e
    \mu}$ or $\varepsilon_{e \tau}$, is present.  They include synergy
  between the near and the far detectors, the characteristic
  differences between the $\varepsilon_{e \mu}$ and $\varepsilon_{e
    \tau}$ systems, and in particular, the parameter degeneracy.
  Finally, we use a concrete setting with the muon energy of 50 GeV and
  magnetized iron detectors at two baselines, one at $L=3000$ km and
  the other at $L=7000$ km, each having a fiducial mass of 50 kton to
  study the discovery potential of NSI and its CP violation effects.
  We demonstrate, by assuming $4 \times 10^{21}$ useful muon decays
  for both polarities, that one can identify non-standard CP violation
  down to $\vert \varepsilon_{e \mu} \vert \simeq \text{a few} \times
  10^{-3}$, and $\vert \varepsilon_{e \tau} \vert \simeq 10^{-2}$ at
  3$\sigma$ CL for $\theta_{13}$ down to $\sin^2 2\theta_{13} =
  10^{-4}$ in most of the region of $\delta$. The impact of the
  existence of NSI on the measurement of $\delta$ and the mass
  hierarchy is also worked out.

\end{abstract} 

\maketitle

\section{Introduction}
\label{sec:intro}

Discovery and establishment of neutrino mass and the lepton flavor
mixing \cite{MNS} by the atmospheric, the solar, and the reactor
experiments \cite{review} triggers a tantalizing question of whether
neutrinos have interactions outside the Standard Model of particle
physics.
In fact, the possibility has been extensively discussed starting from
the early works in \cite{wolfenstein,valle,guzzo,roulet} and in
conjunction with expectation of new physics at TeV energy scale
\cite{grossmann,berezhiani}.

To our current understanding, one of the most significant features of
fundamental matter is the quark-lepton parallelism. Then, the crucial
question is to what extent it prevails and in what respect it fails
\cite{moha-smi}.  For example, quark and lepton flavor mixing is
described by the CKM \cite{cabibbo,KM} and the MNS \cite{MNS}
matrices, respectively, similar in structure but with very different
values of the mixing angles \cite{PDG}.
It is well known that in the quark sector the Kobayashi-Maskawa (KM)
theory \cite{KM} for CP violation works perfectly to the accuracy of
the experiments achieved to date \cite{CPV-review}.  We still do not
know if a natural extension of the KM theory to the lepton sector
provides a correct description of nature~\cite{CPV-review-neutrino}.

In the presence of non-standard interactions (NSI) of neutrinos,
however, the task of exploring lepton CP violation is inevitably
confused or enriched by the coexistence of possible another source of
CP violation induced by NSI \cite{concha1}.
In this paper we discuss the problem of how to resolve the confusion
caused by the two sources of CP violation in the context of
measurement of the lepton mixing parameters in neutrino factory
experiments~\cite{nufact}.
Of course, discovery of NSI will open an entirely new window to
physics beyond the Standard Model of particle physics.  Therefore, in
this paper, we aim at achieving the two goals: Illuminating how to
resolve the confusion between the standard interaction (SI) and the
NSI parameters, and exploring the discovery potential of NSI.  In this
paper, we mean by SI parameters $\theta_{13}$ and $\delta$ assuming
that the other standard lepton mixing parameters are determined to
reasonable accuracies.

The problem of confusion between the SI and the NSI parameters has
been raised in the form of $\theta_{13}-$NSI confusion in
\cite{confusion1,confusion2}.  In a previous paper \cite{NSI-nufact}
we have shown that it can be resolved by a two-detector setting, one
at baseline $L=3000$ km and the other at $L=7000$ km, in neutrino
factory experiments, enjoying an intense neutrino beam from a muon
storage ring.
In a natural continuation of the work, we discuss the problem of
two-phase confusion in this paper by extending our treatment to allow
the NSI elements $\varepsilon_{\alpha \beta} $ (for definition, see
Sec.~\ref{sec:nsifeatures}) to have phases.
For a related work on the same subject see
\cite{winter-nonstandardCP}.
The similar question of distinguishing two kind of CP violation in the
context of ``unitarity violation'' approach \cite{unitarity-violation}
has also been investigated \cite{altarelli-meloni,antusch-etal}.

One of the most important questions on NSI is how large their effects
are.  Given the extreme success of the Standard Model at low energies
the natural framework to address the question is via higher
dimensional operators \cite{weinberg}.  Assuming the new physics scale
of $M_{NP} \simeq$1 TeV, it may be $\vert \varepsilon_{\alpha \beta} \vert \sim
\left( M_{W} / M_{NP} \right)^2 \simeq 10^{-2}$ and $\sim \left( M_{W}
  / M_{NP} \right)^4 \simeq 10^{-4}$ for dimension six and eight
operators, respectively.
Possible forms of theses operators are further narrowed down by the severe
constraints on four charged lepton processes~\cite{kuno-okada}, the
$SU(2)$ partners of the neutrino interactions on leptons, the problem
raised in \cite{berezhiani}.
In view of the fact that the operators which are free from the four
lepton counterpart are very limited \cite{gavela-etal}, it may be the
right attitude to anticipate and cover a wide range of magnitudes
$10^{-4} \lsim \vert \varepsilon_{\alpha \beta} \vert \lsim 10^{-2}$ (or less if
possible) as our target for hunting the NSI effects.  For this reason,
we try to cover the region down to $\varepsilon_{\alpha \beta} \sim
10^{-4}$ in our analysis in this paper.
The constraints on magnitude of NSI on neutrinos has been investigated
\cite{davidson,enrique1}, and the authors of \cite{enrique2} made
major advances in understanding  the loop constraints.

As is well known, the effects of NSI exist not only in propagation in
matter but also in production and detection processes of neutrinos
\cite{grossmann,confusion1,confusion2,ota1}.  In this paper, however,
we only deal with the effect of NSI in neutrino propagation in matter.
The reason for this limitation is mainly technical; The features of
neutrino flavor transformation are sufficiently complicated so that it
deserves separate treatment.  For the same reason, we restrict
ourselves to systems with a single NSI element, either $\varepsilon_{e
  \mu}$ or $\varepsilon_{e \tau}$, at one time.
Thanks to these limitations, we can have a transparent view of the
parameter degeneracy~\cite{intrinsic,MNjhep01,octant} in systems with
NSI~\cite{NSI-perturbation} in its limited aspect that can be seen
more clearly under the simplified setting.
In this paper, we will try to give a complete understanding of the
characteristic features of the system, not just showing the
sensitivity contours.  In particular, we will clarify the reasons why
the sensitivities are so different between the systems with
$\varepsilon_{e \mu}$ and $\varepsilon_{e \tau}$.

Taking $\vert \varepsilon_{\alpha \beta} \vert \sim 10^{-4} - 10^{-2}$
as the target region, it is unlikely that reactor
\cite{reactor-proposal} or superbeam \cite{superbeam} experiments are
powerful enough to have significant discovery potential for NSI.
Thus, we are left with either the neutrino factory \cite{nufact} or
the beta beam \cite{beta} as the possible technologies within our
current knowledge.  In fact, there exist a number of articles which
are devoted to discuss the capabilities of discovering or constraining
NSI in a neutrino factory experiment, for instance, see
Ref.~\cite{confusion1,confusion2,ota1,NSI-nufact,NSI-accelerator-nufact}.
Because of the clean event reconstruction capability, we focus on the 
appearance channels $\nu_{e} \rightarrow \nu_{\mu}$ and
$\bar{\nu}_{e} \rightarrow \bar{\nu}_{\mu}$, 
the so called golden channels for the neutrino factory \cite{golden}. 
We assume, following \cite{NSI-nufact}, a high energy muon beam 
of $E=50$ GeV and two magnetized iron detectors at $L=3000$ and 7000 km.
Notice that in the $\nu_{e}$-induced appearance channels, this or
$\nu_{e} \rightarrow \nu_{\tau}$, the only relevant NSI elements are
either $\varepsilon_{e \mu}$ or $\varepsilon_{e \tau}$ as shown in a
perturbative framework in \cite{NSI-perturbation}, which is nothing
but the natural extension of the one developed in~\cite{golden}.

Of course, it is important to place constraints on NSI from all
possible available means. In fact, the bounds (to be) placed on NSI
have been discussed in the context of accelerator
neutrinos~\cite{NSI-accelerator-not-nufact} (excluding the neutrino
factory), atmospheric neutrinos~\cite{NSI-atmospheric}, reactor or
spallation beam neutrinos \cite{NSI-reactor}, solar neutrinos
\cite{NSI-solar}, as well as of astrophysical neutrinos
\cite{NSI-astrophysical}.

\section{Features of Neutrino Oscillation with NSI }
\label{sec:nsifeatures}

Neutrino oscillation in the presence of NSI is a highly nontrivial
problem.  We try to give an introductory discussions on it in the light
of some recent progresses.

\subsection{Neutrino propagation with NSI; General framework }
\label{subsec:frame}

We consider NSI involving neutrinos of the type 
\begin{eqnarray}
{\cal L}_{\text{eff}}^{\text{NSI}} =
-2\sqrt{2}\, \varepsilon_{\alpha\beta}^{fP} G_F
(\overline{\nu}_\alpha \gamma_\mu P_L \nu_\beta)\,
(\overline{f} \gamma^\mu P f),
\label{LNSI}
\end{eqnarray}
where $G_F$ is the Fermi constant, $f$ stands for the index running
over fermion species in the earth, $f = e, u, d$, and $P$ stands for a
projection operator which is either $P_L\equiv \frac{1}{2}
(1-\gamma_5)$ or $P_R\equiv \frac{1}{2} (1+\gamma_5)$, and $\alpha,
\beta = e, \mu$ and $\tau$.

To summarize its effects on neutrino propagation it is customary to
introduce the effective $\varepsilon$ parameters, which are defined as
$\varepsilon_{\alpha\beta} \equiv \sum_{f,P} \frac{n_f}{n_e}
\varepsilon_{\alpha\beta}^{fP}$, where $n_f$ ($n_e$) denotes the
$f$-type fermion (electron) number density along the neutrino
trajectory in the earth.
Using such $\varepsilon$ parameters the evolution equation of neutrinos
in the flavor basis is given by
\begin{eqnarray} 
i {d\over dt} \left( \begin{array}{c} 
                   \nu_e \\ \nu_\mu \\ \nu_\tau 
                   \end{array}  \right)
 = \frac{1}{2E} \left[ U \left( \begin{array}{ccc}
                   0   & 0          & 0   \\
                   0   & \Delta m^2_{21}  & 0  \\
                   0   & 0           &  \Delta m^2_{31}  
                   \end{array} \right) U^{\dagger} +  
                  a \left( \begin{array}{ccc}
            1 + \varepsilon_{ee}     & \varepsilon_{e\mu} & \varepsilon_{e\tau} \\
            \varepsilon_{e \mu }^*  & \varepsilon_{\mu\mu}  & \varepsilon_{\mu\tau} \\
            \varepsilon_{e \tau}^* & \varepsilon_{\mu \tau }^* & \varepsilon_{\tau\tau} 
                   \end{array} 
                   \right) \right] ~
\left( \begin{array}{c} 
                   \nu_e \\ \nu_\mu \\ \nu_\tau 
                   \end{array}  \right)\, ,
\label{general-evolution}
\end{eqnarray}
where $U$ is the MNS matrix \cite{MNS} (for which we use the standard
notation \cite{PDG}), $a\equiv 2 \sqrt 2 G_F n_e E$ \cite{wolfenstein},
$E$ is the neutrino energy and $\Delta m^2_{i j} \equiv m^2_{i}
- m^2_{j}$ with $m_{i}$ ($i=1-3$) the neutrino mass.

Here are some cautionary remarks: 
We note that since our discussion in this paper ignores effects of NSI
in production and detection processes of neutrinos, the results here
must be interpreted with care.  Yet, the system is complicated enough
so that the analysis of NSI effects in neutrino propagation itself
seems to be a meaningful step.
We also note that since we concentrate on NSI effects in neutrino
propagation in matter only the vector type interactions can be probed.
Moreover, the types of NSI that can be studied may be limited to a sub
class of generic NSI \cite{kopp2}.

\subsection{Neutrino oscillation probability with NSI }
\label{subsec:nsiosc}

We focus on $\nu_{e} \rightarrow \nu_{\mu}$ and $\bar{\nu}_{e}
\rightarrow \bar{\nu}_{\mu}$ appearance channels to analyze effects of
NSI, anticipating intense $\nu_{e}$ and $\bar{\nu}_{e}$ beam from
either a neutrino factory or beta beam.  All the quantitative analyses
will be done assuming muon storage ring and magnetized iron detectors
with muon charge identification capability.
For simplicity, we do not consider disappearance channels in this paper
as we believe that the inclusion of them would not modify substantially 
our results. 
For qualitative understanding of various features of the analysis we
use the oscillation probability derived in $\epsilon$ perturbation
theory \cite{NSI-perturbation}.  We assume that $\theta_{13}$ is small
as it is natural for the neutrino factory setting, though the current bound
is still rather mild \cite{CHOOZ}.  In this framework, $\epsilon
\equiv \frac{ \Delta m^2_{21} }{ \Delta m^2_{31} } \sim \sin
\theta_{13} \sim \vert \varepsilon_{e \alpha} \vert$ ($\alpha = \mu,
\tau$) are regarded as small expansion parameters of the same order,
while $\frac{a}{\Delta m_{31}^2}$ is regarded as of order unity.

It is shown that to second order in $\epsilon$ the relevant NSI
elements are only $\varepsilon_{e\mu}$ and $\varepsilon_{e\tau}$ in
the $\nu_{e} \rightarrow \nu_{\alpha}$ channel ($\alpha = \mu,
\tau$).  Then, the oscillation probability in the $\nu_{e} \rightarrow
\nu_{\mu}$ channel can be written in the form of an absolute square as
\cite{NSI-perturbation}
\begin{eqnarray}
&& \hspace{-6mm} P(\nu_e \to \nu_{\mu}; \varepsilon_{e \mu}, \varepsilon_{e \tau}) 
\nonumber \\
&=&
4 \Biggl | 
\ci \si  \cn \frac{\mn}{a} \sin \left( \frac{aL}{4E} \right) e^{-i \Delta_{31}} + 
\st  \sn e^{-i\delta }\frac{\mt}{a} \biggl (\frac{a}{\mt -a}\biggr )\sin \left( \frac{\mt -a}{4E}L \right) 
\nonumber \\
&& \hspace{20mm} 
+ \varepsilon_{e\mu } 
\left[ \cn^2 \sin \left( \frac{aL}{4E} \right) e^{-i \Delta_{31}} 
+ \sn^2 \biggl (\frac{a}{\mt -a}\biggr )\sin \left( \frac{\mt -a}{4E}L \right) \right] 
\nonumber \\
&& \hspace{20mm} 
- \cn \sn \varepsilon_{e\tau } 
\left[ \sin \left( \frac{aL}{4E} \right) e^{-i \Delta_{31}} - 
\biggl (\frac{a}{\mt -a}\biggr )\sin \left( \frac{\mt -a}{4E}L \right) \right]
\Biggr |^2, 
\label{Pemu}
\end{eqnarray}
where $c_{ij} \equiv \cos \theta_{ij}$, 
$s_{ij} \equiv \sin \theta_{ij}$, and 
$\Delta_{31} \equiv \frac{\Delta m^2_{31} L}{4E}$.

As we will learn in the next subsection, the system with
$\varepsilon_{e\mu}$ and $\varepsilon_{e\tau}$ is too complicated to
do a full analysis, having six unknown parameters including three CP
violating phases.
In this paper, therefore, we work on systems with only a single type
of NSI, $\varepsilon_{e\mu}$ or $\varepsilon_{e\tau}$, as a step to
understand the features of neutrino oscillations with NSI.

The expression of the oscillation probability in (\ref{Pemu})
illuminates a notable difference between the systems with
$\varepsilon_{e\mu}$ and $\varepsilon_{e\tau}$.  Namely, the
sensitivity to $\varepsilon_{e\mu}$ is generally better than that to
$\varepsilon_{e\tau}$ because the two kinematic factors in the square
bracket multiplied by the latter (former) tend to cancel (add up).

\subsection{Bi-probability plot in neutrino oscillation with NSI} 
\label{subsec:biprob}

Unfortunately, the reduced system with a single type of $\varepsilon$
is still quite complicated.  Therefore, we try to understand the
characteristic features of neutrino oscillations with
$\varepsilon_{e\alpha}$ ($\alpha = \mu, \tau$) in this subsection.
What makes this system so complicated is the presence of two CP
violating phases, the lepton KM phase \cite{KM} in the MNS matrix
\cite{MNS}, and the phase $\phi_{e\alpha}$ of the NSI element
$\varepsilon_{e\alpha} = \vert \varepsilon_{e\alpha} \vert e^{ i
  \phi_{e\alpha} } $ ($\alpha = \mu, \tau$).
An appropriate tool to illuminate the properties of neutrino
oscillation with emphasis on the role played by phases is the
bi-probability plot in $P( \nu_{e} \rightarrow \nu_{\mu} ) - P(
\bar{\nu}_{e} \rightarrow \bar{\nu}_{\mu})$ (for short $P - \bar{P}$)
space proposed in \cite{MNjhep01}.

The appearance oscillation probability in the neutrino channel
can be generically written as,
\begin{eqnarray}
P &\equiv& P( \nu_{e} \rightarrow \nu_{\mu} )  =  
\mathcal{A} + 
( \mathcal{R} + \mathcal{B} \cos \delta + \mathcal{C}  \sin \delta ) \cos \phi + 
( \mathcal{I} - \mathcal{B}  \sin \delta + \mathcal{C} \cos \delta ) \sin \phi, 
\label{Pemu-biP-phi}
\end{eqnarray}
where the expressions of the coefficients in (\ref{Pemu-biP-phi}) can be
easily derived from (\ref{Pemu}), and are given in
Appendix~\ref{Pemu-biPform}. The corresponding expression for anti-neutrino 
channel can be obtained by doing the transformations 
$a \to -a$, $\delta \to - \delta$, and $\phi \to - \phi$. 
In principle, there are two ways to draw the bi-probability plot,
varying $\phi$ while holding $\delta$ fixed, or varying $\delta$ while
holding $\phi$ fixed, the ``traditional'' way.  (See
(\ref{Pemu-compact}) in Appendix~\ref{Pemu-biPform}.)  The expression
in (\ref{Pemu-biP-phi}) anticipate the former way.
We use both ways for convenience to help to illuminate the physical
properties of the system.

\begin{figure}[bhtp]
\begin{center}
\vglue -0.5cm
\includegraphics[width=0.85\textwidth]{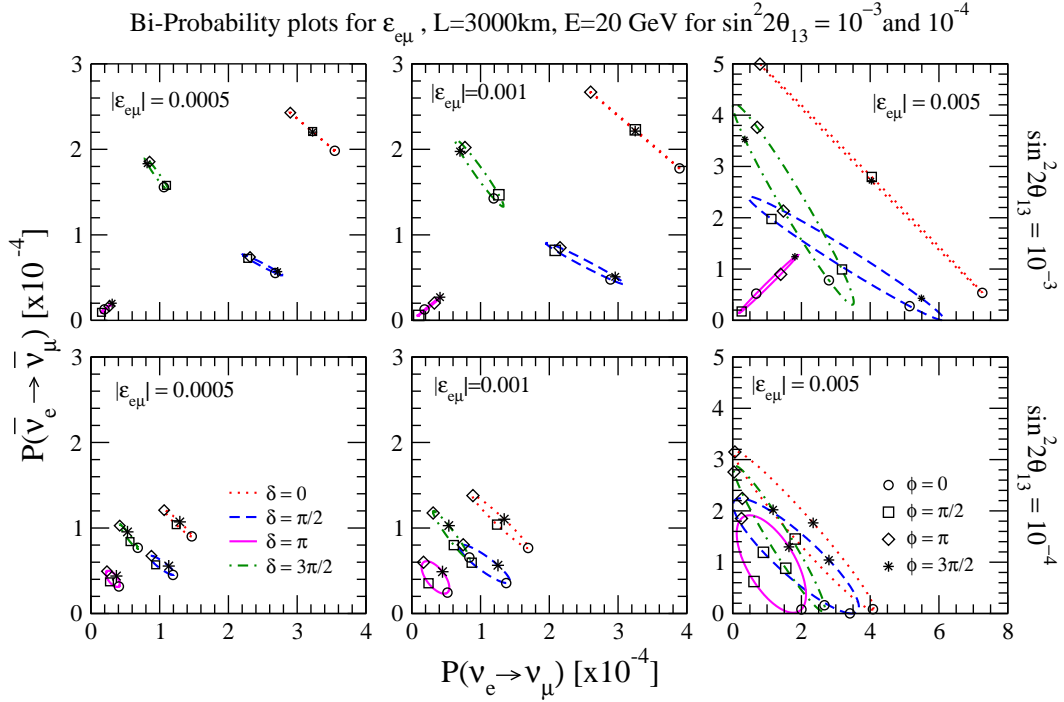}
\end{center}
\vglue -0.75cm
\caption{
Bi-probability plots for the systems with $\varepsilon_{e\mu}$ drawn 
by continuously varying $\phi_{e\mu}$, for four different values 
of $\delta = 0, \pi/2, \pi$ and $3\pi/2$. 
The panels in the upper and lower rows correspond to $\sin^2
2\theta_{13} = 10^{-3}$ and $10^{-4}$, respectively, whereas the ones
in the left, middle and right columns correspond to
$|\varepsilon_{e\mu}| = 5\times 10^{-4}, 10^{-3}$ and $5\times
10^{-3}$, respectively.  The normal hierarchy was assumed and the
values of the other mixing parameters used are
$\sin^2\theta_{12}=0.31$, $\Delta m^2_{21}= 8 \times 10^{-5}$ eV$^2$,
$\sin^2\theta_{23}=0.5$ and $|\Delta m^2_{31}|= 2.5 \times 10^{-3}$
eV$^2$.  We use these four values for the rest of the paper.  }
\label{biP-3000km-em-20GeV}
\end{figure}
%
\begin{figure}[bhtp]
\begin{center}
\vglue -0.4cm
\includegraphics[width=0.85\textwidth]{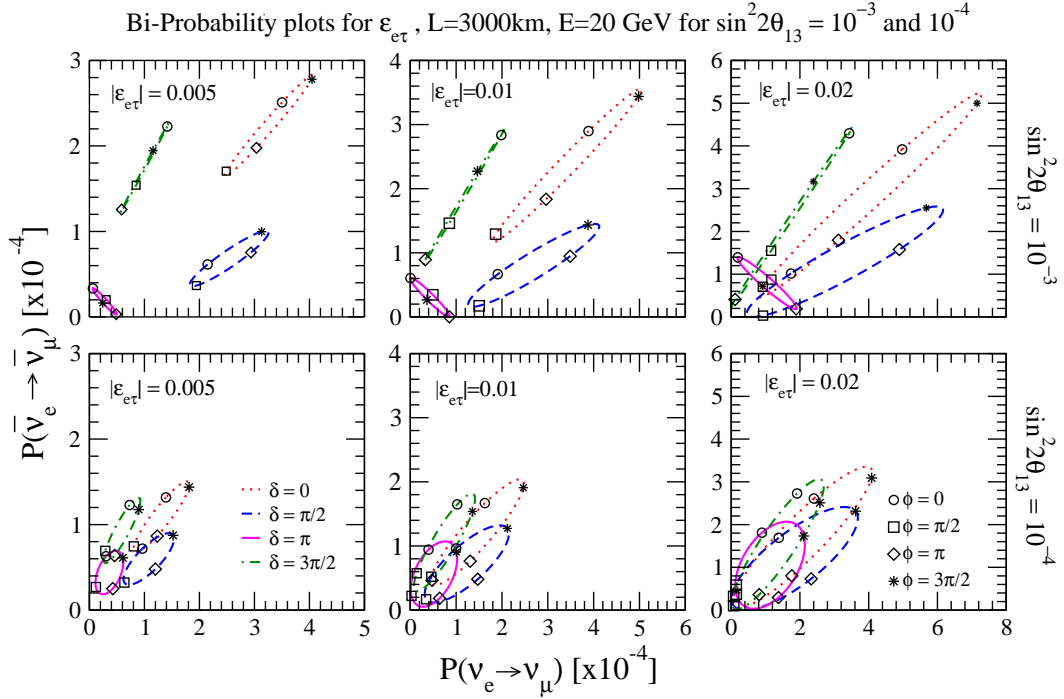}
\end{center}
\vglue -0.75cm
\caption{
Similar plots as shown in Fig.~\ref{biP-3000km-em-20GeV}
but for the systems with $\varepsilon_{e\tau}$. 
The panels on the left, middle and right columns correspond to
$|\varepsilon_{e\tau}| = 5\times 10^{-3}, 10^{-2}$ and $2\times
10^{-2}$, respectively.  }
\label{biP-3000km-et-20GeV}
\vglue -0.2cm
\end{figure}
%

Let us start to proceed step by step to understand the system with
NSI.
In Figs.~\ref{biP-3000km-em-20GeV} and~\ref{biP-3000km-et-20GeV}, 
we present the bi-probability plots for the systems with 
$\varepsilon_{e\mu}$ or $\varepsilon_{e\tau}$, respectively. 
We use the former way, varying
$\phi$ while holding $\delta$ fixed, to draw these figures.
The neutrino energy is arbitrarily chosen as $E = 20$ GeV, but we have 
checked that the main features of the bi-probability plots are similar 
for other values of the energy in the region 
$10~\text{GeV} \leq E \leq 40~\text{GeV}$, 
which are relevant in our setting.  
The values of NSI parameters in both systems are chosen so that 
the appearance probabilities take values comparable with each other.

There are some distinctive features in the bi-probability plots which
are notable in Figs.~\ref{biP-3000km-em-20GeV} and
\ref{biP-3000km-et-20GeV} including difference between the systems
with $\varepsilon_{e\mu}$ and $\varepsilon_{e\tau}$.
\begin{itemize}

\item The slope of the major axis is mostly positive and negative in
  the $\varepsilon_{e\tau}$ and $\varepsilon_{e\mu}$ systems,
  respectively, apart from the region around $\delta=\pi$.
  Furthermore, ellipses rotate when $\delta$ is varied.
  It is the very origin of the dynamical behavior of the system with
  two CP violating phases which spans a wide range in the
  bi-probability space.
  The reason for this behavior is explained in
  Appendix~\ref{rotating}, and this property will be important to
  understand the difference between the $\varepsilon_{e\tau}$ and
  $\varepsilon_{e\mu}$ systems in sensitivities to the parameters. In
  particular, when the two phases are varied, the ellipses span over
  almost the whole triangular region in $P - \bar{P}$ space in the
  $\varepsilon_{e\mu}$ system. See the left panel of
  Fig.~\ref{Prob_vary_delta_theta13} in Sec.~\ref{sec:overview}.

\item A glance over Figs.~\ref{biP-3000km-em-20GeV} and
  \ref{biP-3000km-et-20GeV} indicates a global feature, the size of
  the ellipses strongly (mildly) depends on $\vert \varepsilon \vert$
  ($\sin^2 2\theta_{13}$).  Also, the relative distance between
  ellipses is greater for smaller $\vert \varepsilon \vert$ and larger
  $\sin^2 2\theta_{13}$.  In regions where overlapping between
  ellipses is significant we must expect stronger two-phase confusion
  and parameter degeneracy.  Whereas in regions where ellipses are
  sparse, we expect less degeneracy and better sensitivities for the
  determination of parameters.
\end{itemize}

We will revisit these features of the bi-probability plot in our
overview of the sensitivities in Sec.~\ref{sec:overview}, thereby
meriting understanding some characteristic features.

\subsection{Near-far two-detector setting in a neutrino factory} 
\label{subsec:2detector}

In this subsection, we briefly review the two-detector setting in a
neutrino factory for determination of the NSI and SI parameters
proposed in \cite{NSI-nufact}.  We denote, hereafter, the detectors at
baselines $L\sim 3000$ km and $L\sim 7000$ km the near and the far
detectors, respectively.
The basic idea behind it is that the far detector, which is located
close to the so called ``magic baseline'',\footnote{
  The naming is due to \cite{huber-winter}.  The distance has been
  known as the matter refraction length \cite{wolfenstein}, as can be
  seen, for example in \cite{smirnov}.  The authors of \cite{BMW02}
  pointed out that $\delta$ (CP phase) dependence goes away at the
  magic baseline.  One of our motivations for placing the second
  detector at the distance in \cite{NSI-nufact} was that it is the
  best place, roughly speaking, to detect the effect of matter density
  to which the NSI elements are proportional \cite{mina-uchi}.  }
%
plays a role complementary to the near detector;\footnote{
%
  Though proposed in a variety of contexts, the basic idea for the
  two-detector setting here is quite different from the one in reactor
  $\theta_{13}$ experiment \cite {reactor-proposal}, in which the
  cancellation of systematic errors between the detectors is the key
  element. Ours here is to seek a complementary role, and hence it is
  closer to the one for measurement of CP violation \cite{MNplb97}.
  The idea has also been applied to the Kamioka-Korea identical
  two-detector complex to determine the mass hierarchy as well as
  discovering CP violation (a possible option for upgrading the T2K
  experiment \cite{T2K}) in which these two aspects are unified
  \cite{T2KK1st,T2KK2nd}.
}
%
The synergy between the two detectors greatly strengthens the
potential of parameter determination and help solving the confusion
between the NSI and the SI parameters.
This is very similar to and concordant with the idea of using the far
detector as degeneracy solver in the measurement of the standard
oscillation parameters without NSI \cite{intrinsic,huber-winter}.  The
setting is now considered to be a ``standard one'' in an international
effort for designing a neutrino factory \cite{ISS-nufact}.

\begin{figure}[bhtp]
\begin{center}
\vglue -0.8cm
\includegraphics[width=0.65\textwidth]{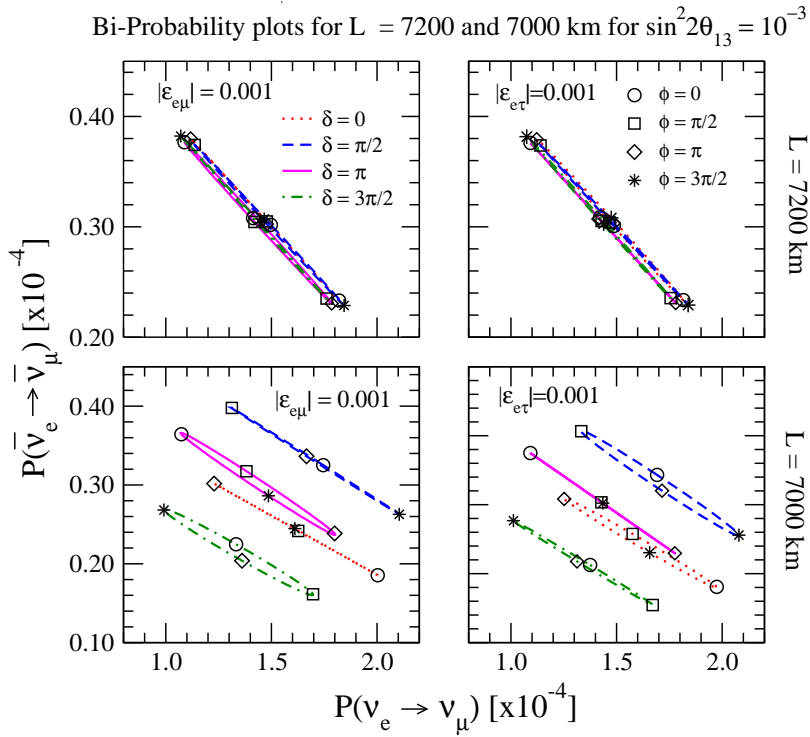}
\end{center}
\vglue -0.75cm
\caption{Similar bi-probability plots as shown in 
Figs.~\ref{biP-3000km-em-20GeV}
and ~\ref{biP-3000km-et-20GeV} (where $\phi$ is varied continuously) 
but at the magic baseline $L_{\text{magic}}
  = 7200$ km (upper panels) and at the actual baseline used in 
this paper $L = 7000$ km (lower panels), for $\sin^2 2\theta_{13} =  10^{-3}$.  
The left (right) panels are for the case where
$|\varepsilon_{e\mu}| = 0.001$ $(|\varepsilon_{e\tau}| = 0.001)$.
Comparison between the upper and the lower panels indicates the
difference of the behavior at $L_{\text{magic}}=7200$ km and $L=7000$
km.  }
\label{biP-magic}
\end{figure}

In our setting of turning on a single NSI element,
$\varepsilon_{e\mu}$ or $\varepsilon_{e\tau}$, the oscillation
probability has special features at the magic baseline.  It takes the
form in the system with $\varepsilon_{e\mu}$ of
\begin{eqnarray}
&& P(\nu_e \to \nu_{\mu}; \varepsilon_{e \mu}) =  
4 s^2_{23} s^2_{13} 
\left( \frac{ \Delta m^2_{31} }{ a - \Delta m^2_{31} }  \right)^2 
\sin^2 \left( \frac{ \Delta m^2_{31} }{ 4E } L \right) 
\nonumber \\
&+& \frac{ 4 a s_{23}^3}{(a - \Delta m_{31}^2)^2}
 \Bigl[  2 \Delta m_{31}^2 s_{13} 
 \vert \varepsilon_{e\mu} \vert 
  \cos ( \delta + \phi_{e \mu} )
  + s_{23} a |\varepsilon_{e\mu}|^2 \Bigr] 
\sin^2 \left( \frac{ \Delta m^2_{31} }{ 4E } L \right). 
\label{Pemu-emu-magic}
\end{eqnarray}
$P(\nu_e \to \nu_{\mu}; \varepsilon_{e \tau}) $ can be obtained by the
transformation $s_{23} \varepsilon_{e\mu} \rightarrow c_{23}
\varepsilon_{e\tau}$ at the magic baseline, in accord with the prescription 
given in Appendix~\ref{Pemu-biPform}.
Notice that the two phases come together in the form $\delta + \phi_{e
  \mu}$ in (\ref{Pemu-emu-magic}); The physical phase must be unique at
the magic baseline because the solar $\Delta m^2_{21}$ effect is
absent and the system is effectively two generations.\footnote{
%
The reasoning was spelled out in \cite{NSI-nufact}, and it has
subsequently obtained an analytic proof as a general theorem of phase
reduction in generic systems with all NSI elements
\cite{NSI-perturbation}.
}
%
Clearly, the bi-probability plot shrinks into a line with only a
cosine dependence on the CP phase, similar to the case at the
effective oscillation maximum \cite{KMN02}, and the length of the
shrunk ellipse is proportional to $\vert \varepsilon_{e \alpha}
\vert$.
This behavior is clearly seen in Fig.~\ref{biP-magic}, and its
consequence in some of the allowed contours in
Sec.~\ref{sec:overview}, for example, in
Fig.~\ref{mixed-sign-degeneracy-et-1} and
Fig.~\ref{modest-synergy-et-2}.

Here, let us add a comment; The baseline $L = 7000$ km chosen is not
exactly the magic baseline which is more like $L = 7200$ km. We have
chosen this distance anticipating that it is not always possible to
have a site which coincides to the exact magic baseline.  The
difference between the two baselines is illustrated in
Fig.~\ref{biP-magic}.

\section{Analysis Method}
\label{sec:method}

We describe the method which will be used in our analyses in the
following sections.
In this paper, we make the same assumptions as we made in our previous
paper~\cite{NSI-nufact} for the neutrino factory set up.
We assume an intense muon storage ring which can deliver $10^{21}$
useful decaying muons per year with the muon energy taken to be $50$
GeV, two magnetized iron detectors with a fiducial mass of 50 kton
each, one at baseline $L=3000$ km and the other at $L=7000$ km.
In this work, we assume that each detector can receive neutrino flux
corresponding to $10^{21}$ useful decaying muons per year.
We also assume data will be taken 4 years in neutrino mode and
another 4 years in anti-neutrino one.  We consider only the golden
channels, $\nu_e \to \nu_{\mu}$ and $\bar{\nu}_e \to \bar{\nu}_{\mu}$
in this paper.  For simplicity, we use the constant matter density
approximation, and take the earth matter density along the neutrino
trajectory as $\rho = 3.6 \text{ g/cm}^3$ and $\rho= 4.5 \text{
  g/cm}^3$ for baselines $L=3000$ km and $L=7000$ km, respectively.
The electron fraction $Y_e$ is taken to be 0.5.  We believe that using
more realistic earth matter density profile will not change much our
results.

We always use the normal mass hierarchy as an input for the results
shown in this paper, unless stated otherwise.  We, however, vary 
over the two different mass hierarchies during the fit of the data.
We will consider the setting in which only one element of NSI, either
$\varepsilon=\varepsilon_{e \mu}$ or $\varepsilon=\varepsilon_{e
  \tau}$, where $\varepsilon$ stands for $\vert \varepsilon \vert$ and
its complex phase $\phi$, is turned on.
For simplicity, we fix the values of oscillation parameters relevant
for solar and atmospheric neutrinos (apart from the mass hierarchy
which is assumed to be unknown) to their current best fit values as
$\sin^2\theta_{12}=0.31$, $\Delta m^2_{21}= 8 \times 10^{-5}$ eV$^2$,
$\sin^2\theta_{23}=0.5$ and $|\Delta m^2_{31}|= 2.5 \times 10^{-3}$
eV$^2$.
The last approximation may affect the estimated sensitivities quantitatively, 
but not in a significant way. 

Our $\chi ^2$ function, which is similar to the one used 
in \cite{NSI-nufact} apart from the part which include 
systematic uncertainties and background, is given by, 

\begin{equation}
\chi ^2 \equiv 
\min_{\theta_{13},\delta,\varepsilon,\rm sign(\Delta m^2_{31})}
\sum_{i=1}^3
\sum_{j=1}^2
\sum_{k=1}^2
\frac{
\left[ N^{\text{obs}}_{i,j,k} -
N^{\text{theo}}_{i,j,k}(\theta_{13},\delta,\varepsilon,\rm sign(\Delta m^2_{31})) 
\right]^2 }
{N^{\text{obs}}_{i,j,k}
+ 
(\sigma_{\text{sys}}N^{\text{obs}}_{i,j,k})^2 +
(\sigma_{\text{BG}}N^{\text{\text{BG}}}_{i,j,k})^2 
}\ ,
\label{eq:chi2}
\end{equation}
where $N^{\text{obs}}_{i,j,k}$ is the number of observed (simulated)
events computed by using the given input parameters and
$N^{\text{theo}}_{i,j,k}$ is the theoretically expected number of
events to be varied in the fit by freely varying the standard mixing
and NSI parameters.
The summations over the indices $i,j$ and $k$ imply summing over 3
energy bins, 2 baselines (3000 km or 7000 km), and 2 modes (neutrino
or anti-neutrino), respectively.  The intervals of 3 energy bins
considered are 4-8 GeV, 8-20 GeV, and 20-50 GeV for neutrinos and 4-15
GeV, 15-25 GeV, and 25-50 GeV for anti-neutrinos.
We refer the readers to Ref.~\cite{NSI-nufact} for details on 
how $N^{\text{obs}}_{i,j,k}$ and $N^{\text{theo}}_{i,j,k}$ are
computed.

Using the $\chi^2$ function defined in Eq.~(\ref{eq:chi2}), we define
the allowed regions in terms of 2 parameter space (shown in Sec.
\ref{sec:overview}) by the commonly used condition, $\Delta \chi^2
\equiv \chi^2-\chi^2_{\text{min}}$ = 2.3, 6.18 and 11.83 for 1, 2 and
3 $\sigma$ confidence level (CL), respectively, for 2 degrees of
freedom (DOF).  For the sensitivity plots which will be shown in the
next section, we used the condition $\Delta \chi^2 \equiv
\chi^2-\chi^2_{\text{min}}$ = 1, 4 and 9 for 1, 2 and 3 $\sigma$ CL,
respectively, for 1 DOF.

To illuminate the physical properties of the systems with NSI and to
indicate discovery potential of the two-detector setting in neutrino
factory we think it is revealing to utilize the following two choices 
for the experiment related assumptions in the $\chi^2$ function in
(\ref{eq:chi2}):

\vspace{2mm}
\noindent {\bf Choice R}: We take, following \cite{ISS-nufact},
$\sigma_{\text{sys}} = 2.5$ \% for the signal and $\sigma_{\text{BG}}
= 20$ \% for the background systematic uncertainties.  
We assume an energy independent detection efficiency of 70 \% for both
neutrino/antineutrino modes.  This is a rough approximation to the
efficiency curve obtained in \cite{Abe:2007bi}, which is presented in
the left panel of Fig. 20 of this reference.
The background fractions (efficiencies) are also taken 
from the same reference, presented in the right two 
panels of Fig. 20 of \cite{Abe:2007bi}. 
%

\vspace{2mm}
\noindent {\bf Choice O}: As a second choice we assume no systematic
uncertainties, no background, 100\% detection efficiency.  Namely,
$N^{\text{\text{BG}}}_{i,j,k} = 0$ and $\sigma_{\text{sys}} =
\sigma_{\text{BG}} = 0$.  Only the statistical errors for signal
events are implemented.  Vanishing systematic errors is nothing but an
approximation for a wide range of situations in which they are much
smaller than statistical uncertainties.

\vspace{2mm}
\noindent
The symbol ``R'' is meant to be ``realistic'', while the ``O''
optimal.

We use the results of the analyses with two different recipes of the
experimental uncertainties to make our presentation most informative.
In Secs. \ref{sec:overview} and \ref{sec:degeneracy}, we rely on the
results of analysis based on Choice O of the $\chi^2$ parameters, that
is, no errors except for the statistical one.
In these sections we try to illuminate the structure of neutrino
oscillation with NSI in the golden channel.  Hence, we prefer to deal
with the allowed regions which are free from obstruction by the
experimental errors.

Whereas in Secs.~\ref{sec:disc} and \ref{sec:impact} where the
sensitivity contours of the NSI and the SI parameters are discussed,
we present the contours for both choices, Choice R and Choice O of the
$\chi^2$ parameters.
At the minimum, the comparison between these two choices reveals the
effects of the systematic uncertainties and the efficiencies on the
sensitivities, and therefore is useful.  If we are to prepare for
search for effects induced by dimension eight operators beyond
dimension six ones, as argued relevant in Sec.~\ref{sec:intro} (see
also \cite{Minakata:2009gh}), one may be interested in optimal
sensitivities that can be achieved by an ultimate detector.\footnote{
We give a caution here that the current estimation of experimental systematic 
uncertainties and backgrounds may not be sufficiently mature to address the 
question of sensitivities to such tiny values of NSI. 
}
While the concrete design of such a detector is not known, its first
approximation may be given by taking the limit of vanishing errors
under the current detector setting.

\section{Overview of the Sensitivities and Synergy between the Two  Detectors}
\label{sec:overview}

In this section we give an overview of the sensitivities to the NSI
elements, $\vert \varepsilon_{e \alpha} \vert$ and $\phi_{e \alpha}$
($\alpha=\mu, \tau$), and the SI parameters, $\delta$ and $\sin^2
2\theta_{13}$, to be achieved by the detectors at $L=3000$ km and
$L=7000$ km separately and in combination.  The principal purpose of
the discussions in this section is to understand the global features
of the sensitivities of the complex systems.
In fact, various viewpoints have to be involved to really understand 
what the complementarity between the near and far detectors means: 
\begin{itemize}

\item How prominent is the synergy between the near and the far
  detectors for determination of SI  and NSI
  parameters?  How it differs between the systems with $\varepsilon_{e
    \mu}$ and $\varepsilon_{e \tau}$? What about the dependence on
  values of the parameters, in particular on the size of the NSI
  elements?

\item How can the two-phase confusion be resolved?  Does the answer
  depend on which NSI element is turned on?

\item What is the nature of the parameter degeneracy in the system with
  NSI, and whether it can be resolved by the two-detector setting?  If
  so how can it be realized?

\end{itemize}

\noindent
The first two points will be discussed in this section and we will
devote the whole Sec.~\ref{sec:degeneracy} for the problem
of degeneracy.

\subsection{Data set} 
\label{subsec:data}

Toward the above stated goal, we systematically generated the contours
of allowed regions determined by the data taken in the near (3000 km)
and the far (7000 km) detectors and the one combined based on the
analysis method described in Sec.~\ref{sec:method}.  We call the set
of allowed contour figures the ``data set''.
To generate the data set we take the input parameters in the following
way: $\sin^2 2\theta_{13}= 10^{-4}$ and $10^{-3}$, $\delta=0, \pi/2,
\pi$ and $3\pi/2$, $\vert \varepsilon_{e \mu} \vert=5 \times 10^{-4},
10^{-3}$ and $5 \times 10^{-3}$, $\phi_{e \mu}=\pi/4, 3\pi/4, 5\pi/4$
and $7\pi/4$.  Altogether, there are $2 \times 4 \times 3 \times 4 =
96$ sheets.\footnote{
  A sheet, in principle, contains six two-dimensional plots because
  there are four parameters to be fit.  Practically, however, three of
  them appear to be enough; We have used the two-dimensional plots
  with axes $\vert \varepsilon_{e \mu} \vert - \phi_{e \mu}$, $\delta
  - \phi_{e \mu}$, and $\sin^2 2\theta_{13} - \delta$.
}
%
Then, the above parameters are varied to fit the data.  During the
fit, unless otherwise stated, the parameter space of both the normal
and the inverted neutrino mass hierarchies are swept.  The input is
always taken as the normal hierarchy except for the results shown in
Sec.~\ref{sec:impact-mh}.
The allowed regions at 2$\sigma$ and 3$\sigma$ CL are defined by 
the region corresponding to 
$\chi^2 - \chi^2_{\text{min}} < $ 6.18 and 11.8, respectively. 
We try to illuminate the dominant features of the complex system of
neutrino oscillation with NSI by analyzing the data set.

In the system with $\varepsilon_{e \tau}$, because of poorer
sensitivities compared to the system with $\varepsilon_{e \mu}$, we
use an order of magnitude larger values of the NSI parameters to
generate the data set:
$\vert \varepsilon_{e \tau} \vert=5 \times 10^{-3}, 10^{-2}$ and $2
\times 10^{-2}$, $\phi_{e \tau}=\pi/4, 3\pi/4, 5\pi/4$ and $7\pi/4$,
with the same values of $\theta_{13}$ and $\delta$.  The ranges of the
input values of $\varepsilon_{e \mu}$ and $\varepsilon_{e \tau}$ are
chosen such that they lead to the similar values of the probabilities
as can be seen by comparing the bi-probability plots of both systems
in Sec.~\ref{subsec:biprob}.  They roughly correspond to the
sensitivity regions obtained in \cite{NSI-nufact}.
Again there are 96 sheets as in the $\varepsilon_{e \mu}$ case.
Due to the lack of space, we only show a tiny fraction of such 
allowed contours in the following subsections but below we will 
give qualitative summary of our data set described above. 

\subsection{Classification scheme of the data set}
\label{subsec:class}

By reviewing $ 2 \times 96$ sheets we recognize that there exist 4
typical types:

\begin{itemize}

\item Type A: Measurement at the near detector (3000 km) cannot
  produce closed contours but the combination with the far detector
  (7000 km) makes the contours closed (at least for one of the 4
  variables), which determines the parameters within errors.
  This is the case where the synergy is most prominent.

\item Type B: Measurement at the near detector does produce closed
  contours, but multiple solutions are allowed, this is the phenomenon
  of parameter degeneracy.  In most cases in this category the
  combination with the far detector solves the degeneracy.

\item Type C: Measurement at the near detector itself produces closed
  contours without degeneracy.  The role of the far detector is to
  make the allowed region smaller.

\item Type D: Both the measurement at the near detector and the
  combination with the far detector fail to produce closed contours.

\end{itemize}

\noindent
The closing or not of contours, of course, depends upon the confidence
level.  We use 2 and 3 $\sigma$ CL in the tables and figures in this
section.
%

\begin{table}[h]
\vglue 0.3cm
\begin{tabular}{c|c|c|c|c|c}
\hline 
\hline
\ type and magnitude of NSI \ 
             & \ \ ~$\sin^2 2 \theta_{13}$~ \ \ 
             & \ \ Type A \ \ 
             & \ \ Type B \ \ 
             & \ \ Type C \ \ 
             & \ \ Type D \ \  \\
\hline
\ $\vert \varepsilon_{e\mu} \vert = 5 \times 10^{-4}$ \ 
                 & $10^{-4}$
                 & 5(2)
                 & 5(8)
                 & 1(5)
                 & 5(1)  \\
\hline
\ $\vert \varepsilon_{e\mu} \vert = 5 \times 10^{-4}$ \ 
                 & $10^{-3}$
                 & 3(2)
                 & 5(1)
                 & 3(8)
                 & 5(5)  \\
\hline
\ $\vert \varepsilon_{e\mu} \vert = 10^{-3}$ \ 
                 & $10^{-4}$
                 & 0(0)
                 & 15(8)
                 & 1(8)
                 & 0(0)  \\
\hline
\ $\vert \varepsilon_{e\mu} \vert = 10^{-3}$ \ 
                 & $10^{-3}$
                 & 2(4)
                 & 4(3)
                 & 6(8)
                 & 4(1)  \\
\hline
\ $\vert \varepsilon_{e\mu} \vert = 5 \times 10^{-3}$ \ 
                 & $10^{-4}$
                 & 4(0)
                 & 12(14)
                 & 0(2)
                 & 0(0)  \\
\hline
\ $\vert \varepsilon_{e\mu} \vert = 5 \times 10^{-3}$ \ 
                 & $10^{-3}$
                 & 0(0)
                 & 11(11)
                 & 5(5)
                 & 0(0)  \\
\hline
\hline
\end{tabular}
\caption[aaa]
{Presented is the number of sheets which fall into the categories of 
  Type A$-$D out of 96 sheets of the data set prepared for the system with NSI element 
  $\varepsilon_{e\mu}$. 
  The definition of Type A$-$D is described in the text. 
  The numbers in and out of parentheses are the ones at 2 and 3$\sigma$ CL, 
  respectively. The results shown here (or in this and next sections)
  correspond to the choice O (100\% detection efficiencies, 
  no background and no systematic uncertainties) defined in 
  Sec.~\ref{sec:method}. 
}
\label{table-emu}
\end{table}
%
\begin{table}[h]
\begin{tabular}{c|c|c|c|c|c}
\hline 
\hline
\ type and magnitude of NSI \ 
             & \ \ ~$\sin^2 2 \theta_{13}$~ \ \ 
             & \ \ Type A \ \ 
             & \ \ Type B \ \ 
             & \ \ Type C \ \ 
             & \ \ Type D \ \  \\
\hline
\ $\vert \varepsilon_{e\tau} \vert = 5 \times 10^{-3}$ \ 
                 & $10^{-4}$
                 & 16(16)
                 & 0(0)
                 & 0(0)
                 & 0(0)  \\
\hline
\ $\vert \varepsilon_{e\tau} \vert = 5 \times 10^{-3}$ \ 
                 & $10^{-3}$
                 & 16(16)
                 & 0(0)
                 & 0(0)
                 & 0(0)  \\
\hline
\ $\vert \varepsilon_{e\tau} \vert = 10^{-2}$ \ 
                 & $10^{-4}$
                 & 16(15)
                 & 0(1)
                 & 0(0)
                 & 0(0)  \\
\hline
\ $\vert \varepsilon_{e\tau} \vert = 10^{-2}$ \ 
                 & $10^{-3}$
                 & 14(11)
                 & 1(4)
                 & 1(1)
                 & 0(0)  \\
\hline
\ $\vert \varepsilon_{e\tau} \vert = 2 \times 10^{-2}$ \ 
                 & $10^{-4}$
                 & 16(15)
                 & 0(0)
                 & 0(1)
                 & 0(0)  \\
\hline
\ $\vert \varepsilon_{e\tau} \vert = 2 \times 10^{-2}$ \ 
                 & $10^{-3}$
                 & 8(4)
                 & 6(9)
                 & 2(3)
                 & 0(0)  \\
\hline
\hline
\end{tabular}
\caption[aaa]
{Same as in Table~\ref{table-emu} but for the system with NSI element 
  $\varepsilon_{e\tau}$. 
}
\label{table-etau}
\end{table}

In Table~\ref{table-emu} and \ref{table-etau} we present number of
sheets which fall into the categories of Type A$-$D of the data set
prepared for the system with NSI element $\varepsilon_{e\mu}$ and
$\varepsilon_{e\tau}$, respectively.  In the system with
$\varepsilon_{e \tau}$ the classification scheme works except that
there is no Type D as we will see below.
Here, we need some details of the criteria for the classification. 
Existence of a closed contour\footnote{
In doing computation, for technical reasons, the regions of parameter scan 
were limited to 
$10^{-4} \leq \vert \varepsilon_{e\mu} \vert \leq 6 \times 10^{-3} $ and 
$5 \times 10^{-4}  \leq \vert \varepsilon_{e\tau} \vert \leq 3 \times 10^{-2}$ 
in respective systems. 
Therefore, closed contours imply they are closed in the above regions.   
}
%
does not necessarily mean that the unique solution is contained in it;
It can occur that multiple solutions are involved in a single region
because they are too close to be resolved due to limited statistics,
for example.  Our classification scheme A$-$D above is, therefore,
purely ``phenomenological'' in nature.

It should also be noticed that classification of figures into A and D
has an ambiguity; It occurs in some cases the allowed contour is
closed for $\vert \varepsilon \vert$ but not for $\sin^2 2
\theta_{13}$, for example.
We define Type D such that the allowed contours of any one of the four
variables fail to be closed after combining the yields from the two
detectors.  Consistently, Type A is defined such that at least one of
the four variables has closed contours.  It makes classification
scheme a complete one; The number of pages add up to 16 when summed
over Type A$-$D for a given input values of $\vert \varepsilon \vert$
and $\sin^2 2 \theta_{13}$.

A glance over the tables illuminates the following features: 

\begin{itemize}

\item In the system with $\varepsilon_{e\tau}$ most of the sheets are
  classified into Type A except for the case of relatively large
  $\theta_{13}$ and $\vert \varepsilon_{e\tau} \vert$; Synergy between
  the two detectors are far more prominent compared to the system with
  $\varepsilon_{e\mu}$.

\item In the system with $\varepsilon_{e\mu}$ synergy effect exists
  but its effectiveness depends on $\theta_{13}$ and $\vert
  \varepsilon_{e\mu} \vert$ in an intricate way.  There is a general
  tendency that closed allowed region is more common in cases with
  large $\vert \varepsilon_{e\mu} \vert$.
  For a given value of $\vert \varepsilon_{e\mu} \vert$ synergy is
  stronger in case of smaller $\theta_{13}$, however with the notable
  exception of the case $\vert \varepsilon_{e\mu} \vert=10^{-3}$.
\end{itemize}

\noindent
One may ask why the synergy is so powerful in the
$\varepsilon_{e\tau}$ system.  The answer involves two reasonings;
Measurement at the near detector is not as powerful as in the
$\varepsilon_{e\mu}$ system because the two kinematic factors in the
coefficient of $\varepsilon_{e\tau}$ in (\ref{Pemu}), which have
similar magnitudes, tend to cancel, resulting the lower sensitivity to
$\varepsilon_{e\tau}$.  Then, powerfulness of the synergy is due to
the higher statistics of the far (7000 km) detector due to larger
$\vert \varepsilon_{e\tau} \vert $, which is enhanced by the former
reasoning.

\subsection{Features of sensitivities in system with $\varepsilon_{e \mu}$}
\label{subsec:sens}

In this and the next subsections we give further description of the
features of sensitivities to the NSI elements, $\vert \varepsilon_{e
  \alpha} \vert$ and $\phi_{e \alpha}$ ($\alpha=\mu, \tau$), and
$\delta$ and $\sin^2 2\theta_{13}$ by the near-far two-detector
setting.
In what follows we present only one figure for each category A$-$D.
The category B (the case with parameter degeneracy) will be discussed
in the Sec.~\ref{sec:degeneracy}.  We use 3$\sigma$ CL criterion for
the classification of types.

\begin{figure}[bhtp]
\begin{center}
\hglue -0.6cm
\includegraphics[width=0.82\textwidth]{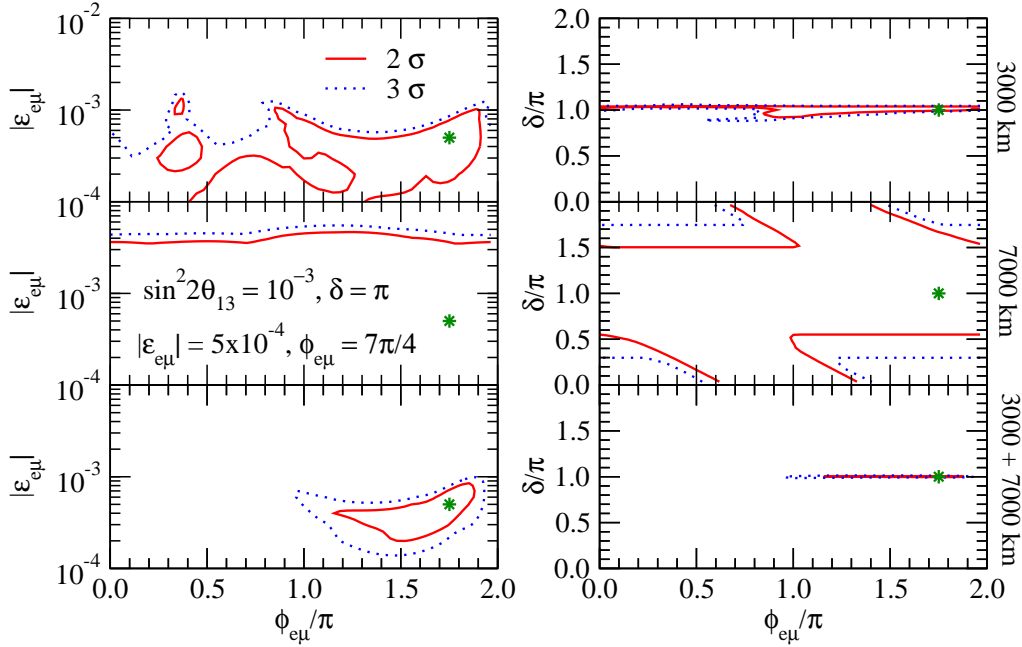}
\end{center}
\vglue -0.75cm
\caption{ An example of Type A. Allowed regions in the
  $\phi_{e\mu}-|\varepsilon_{e\mu}|$ plane (left panels) and
  $\phi_{e\mu}-\delta$ plane (right panels) corresponding to 2 and 3
  $\sigma$ CL obtained for the system with $\varepsilon_{e\mu}$.
  Upper (middle) panels correspond to the case where only a single
  detector at 3000 km (7000 km) is taken into account, whereas the
  lower panels correspond to the case where the results from the two
  detectors are combined.  The input parameters are taken as: $\sin^2
  2\theta_{13} = 10^{-3}$, $\delta=\pi$, $|\varepsilon_{e\mu}|= 5
  \times 10^{-4}$ and $\phi_{e \mu}=7\pi/4$ (indicated by the green
  asterisk), and the input mass hierarchy is normal.  No allowed
  regions exist in the inverted mass hierarchy regime.  All the
  results shown here (or in this and next sections) correspond to the
  choice O (100\% detection efficiencies, no background and no
  systematic uncertainties) defined in Sec.~\ref{sec:method}.
}
\label{synergy-em-4}
\end{figure}

Let us first consider the system with $\varepsilon_{e \mu}$.  As an
example of Type A, we present in Fig.~\ref{synergy-em-4} the regions
allowed by measurement at the near detector ($L=3000$ km, top panels),
the far detector ($L=7000$ km, middle panels), and the two detector
combined (bottom panels).  It is for the input values of the NSI
element $\vert \varepsilon_{e\mu} \vert =5 \times 10^{-4}$ and
$\phi_{e\mu} = 7\pi/4$ whereas the SI parameters are taken as $\delta
= \pi$ and $\sin^2 2\theta_{13} = 10^{-3}$ with the normal mass
hierarchy.
Left and right panels in this figures show allowed regions plotted in
$\vert \varepsilon_{e \mu} \vert - \phi_{e \mu} $ and $\delta -
\phi_{e \mu} $ space, respectively.  We note that for this case, no
allowed regions exist for the inverted mass hierarchy regime.

\begin{figure}[bhtp]
\vglue -0.2cm
\begin{center}
\hglue -0.6cm
\includegraphics[width=0.82\textwidth]{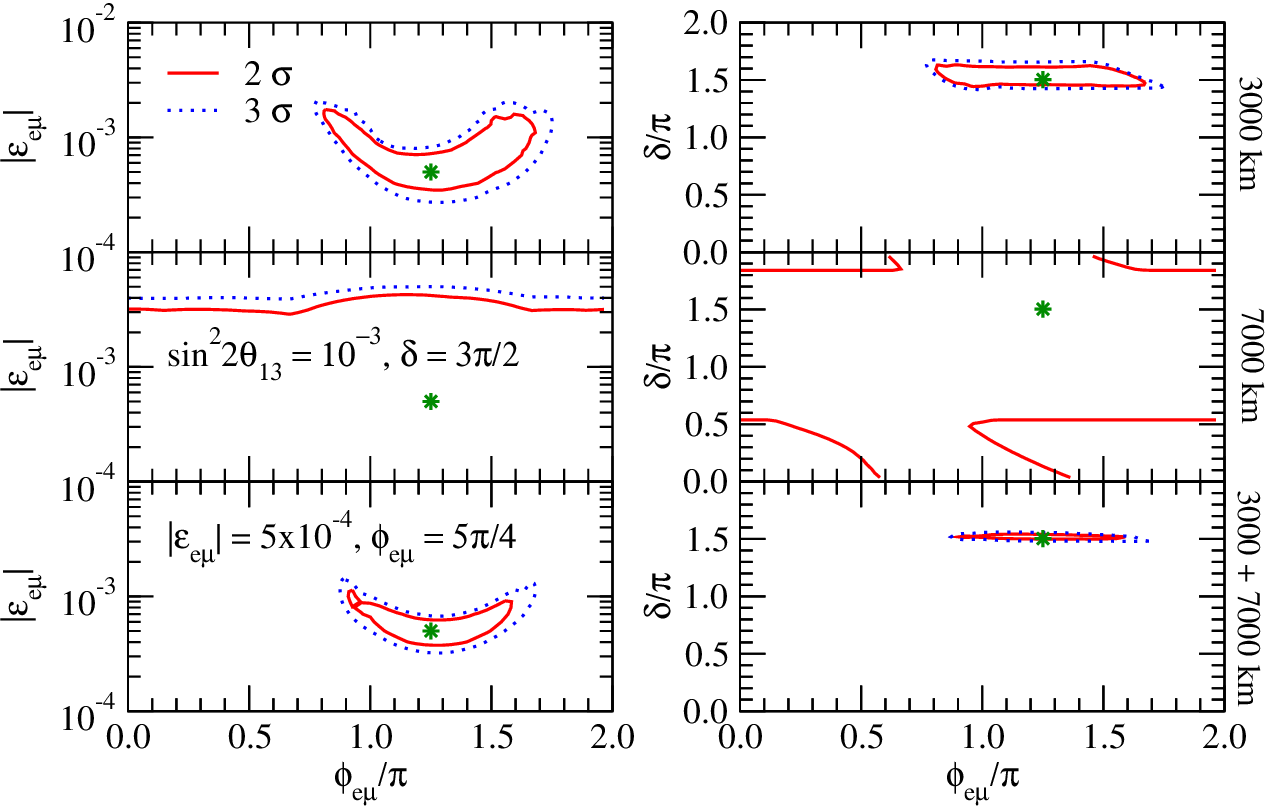}
\end{center}
\vglue -0.75cm
\caption{An example of Type C.  Same as for Fig.~\ref{synergy-em-4}
  but for the input parameters $\sin^2 2\theta_{13} = 10^{-3}$,
  $\delta=3\pi/2$, $|\varepsilon_{e\mu}|= 5 \times 10^{-4}$ and
  $\phi_{e \mu}=5\pi/4$, and the normal mass hierarchy.  No allowed
  regions exist in the inverted mass hierarchy regime.}
\label{modest-synergy-em-3}
\end{figure}
%

\begin{figure}[b]
\vglue -0.1cm
\begin{center}
\hglue -0.6cm
\includegraphics[width=0.82\textwidth]{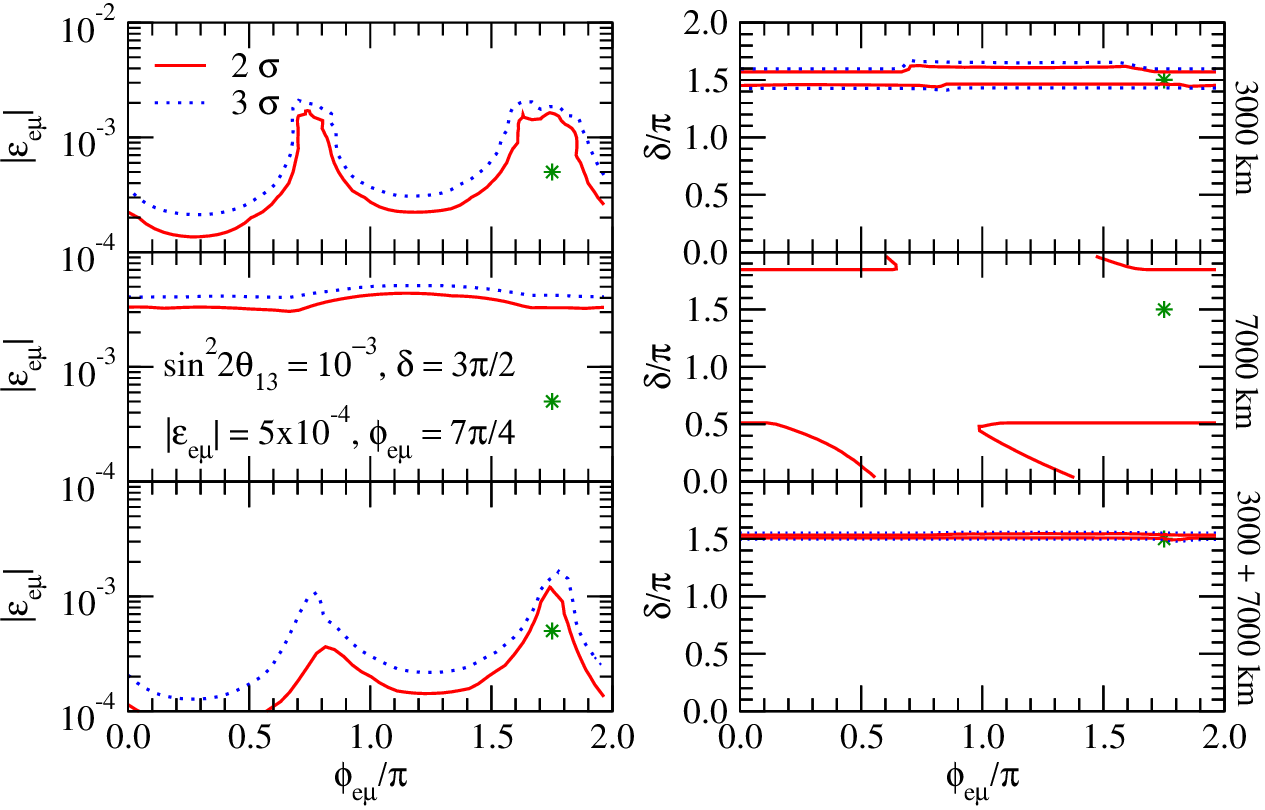}
\end{center}
\vglue -0.75cm
\caption{An example of Type D.  Same as for Fig.~\ref{synergy-em-4}
  but for the input parameters $\sin^2 2\theta_{13} = 10^{-3}$,
  $\delta=3\pi/2$, $|\varepsilon_{e\mu}|= 5 \times 10^{-4}$ and
  $\phi_{e \mu}=7\pi/4$, and the normal mass hierarchy.  No allowed
  regions exist in the inverted mass hierarchy regime.  }
\label{nosynergy-em-3}
\end{figure}

In this figure, it is indicated that measurement at the far detector,
though it itself does not appear to be so powerful, helps closing the
allowed regions, indicating prominent synergy between the two
detectors.  This feature is consistent with the one observed in
systems with NSI but with frozen phase degree of freedom discussed in
the previous paper \cite{NSI-nufact}.
However, Type A is relatively minor in the data set for 
large values of $\vert \varepsilon_{e\mu} \vert$ as seen in
Table~\ref{table-emu}, indicating less prominent synergy between the
two detectors in the $\varepsilon_{e \mu}$ system compared to the 
$\varepsilon_{e \tau}$'s. 
In Fig.~\ref{synergy-em-4} accuracy of determination of $\phi_{e\mu}$
is limited because of small $\vert \varepsilon_{e\mu} \vert$, though
$\delta - \phi_{e\mu}$ confusion is resolved.

An example of Type C is presented in Fig.~\ref{modest-synergy-em-3},
which was generated using the same values of $\vert \varepsilon_{e\mu}
\vert$ and $\theta_{13}$ as in Fig.~\ref{synergy-em-4} but different
phase values, $\delta = 3\pi/2$ and $\phi_{e\mu} = 5\pi/4$.  For
relatively large values of $\vert \varepsilon_{e\mu} \vert $, the case
of closed contours as the result of near detector measurement is most
popular, as can be seen in Table~\ref{table-emu}.
In fact, as long as NSI parameters are concerned, 
there is no example, within the cases examined, of unclosed 
contours (only upper bound) at 3000 km detector 
for the input value 
$\vert \varepsilon_{e\mu} \vert =5 \times 10^{-3}$. 
In the example in Fig.~\ref{modest-synergy-em-3}, we still see modest
synergy between the detectors. Interestingly, the sensitivity to
$\delta$ improves due to the far detector measurement at the magic
baseline.
The $\delta - \phi_{e\mu}$ confusion is clearly resolved though the 
accuracy of determination of $\phi_{e\mu}$ is modest. 

In Fig.~\ref{nosynergy-em-3} presented is an example of Type D
generated by using the same parameters as used in
Fig.~\ref{modest-synergy-em-3} except for $\phi_{e \mu} = 7\pi/4$,
showing that the synergy is least prominent among the categories
A$-$D.\footnote{ We note that for $\sin^2 2\theta_{13} = 10^{-4}$ (not
  shown), some cases classified as Type D at 3$\sigma$ CL become Type
  A or B even at 2$\sigma$ CL.  }
In this case the unclosed contours at 3000 km fail to be closed at
2$\sigma$ CL even after combination with 7000 km detector.
Again the far detector improves sensitivity to $\delta$, but not at
all to $\phi_{e\mu}$.

Finally, it may be worth to note that the characteristic feature at
the magic baseline that only the combination $\phi_{e \mu} + \delta$
is constrained \cite{NSI-nufact} is barely visible in most of the data
set.  The possible reasons are: The baseline is not exactly equal to
the magic baseline.  The statistics at the far detector is of course
much less and sensitivities to the parameters are close to the
sensitivity limit, lacking clear indication of $\delta - \phi$
correlation.
However, appearance of $\delta - \phi$ correlated oblique strip  
becomes frequent in case of relatively large 
$\vert \varepsilon_{e\mu} \vert = 5 \times 10^{-3}$. 
%

\subsection{Feature of sensitivities in system with $\varepsilon_{e \tau}$}
\label{subsec:sens-et}

 \begin{figure}[bhtp]
  \vglue -0.5cm
  \begin{center}
  \hglue -0.6cm
 \includegraphics[width=0.82\textwidth]{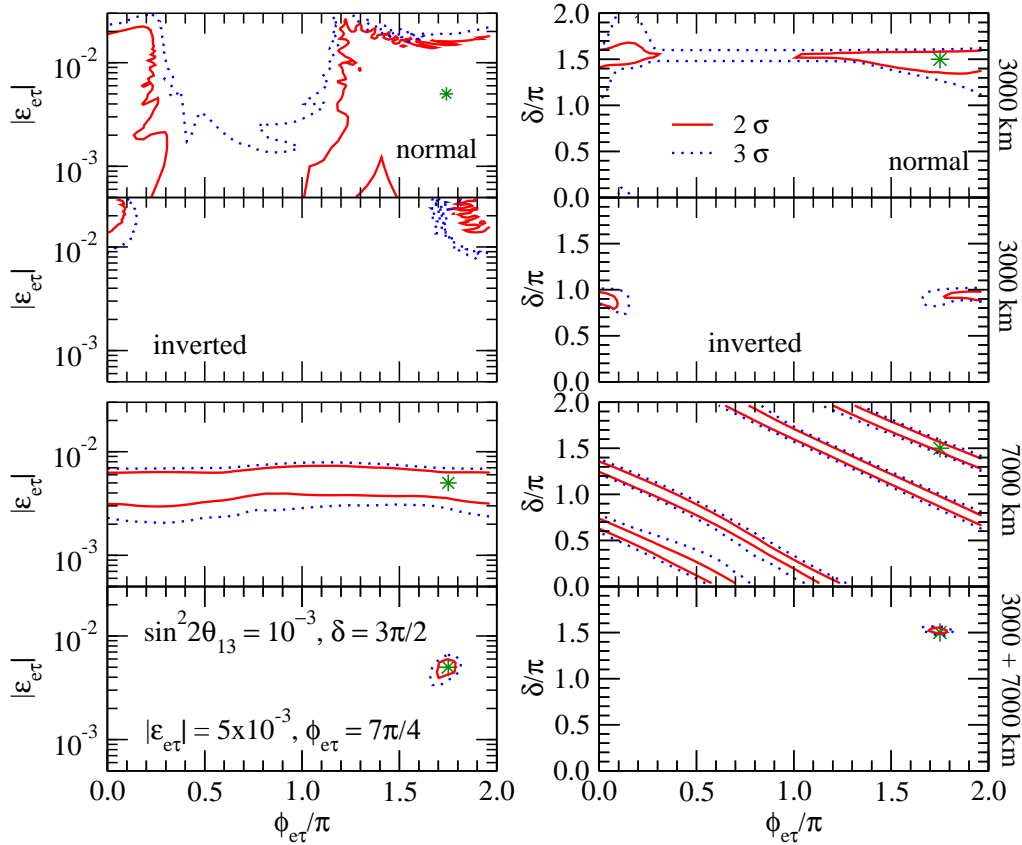}
  \end{center}
  \vglue -0.75cm
  \caption{An example of Type A.  Allowed regions obtained for the
    system with $\varepsilon_{e\tau}$.  The input parameters are taken
    as: $\sin^2 2\theta_{13} = 10^{-3}$, $|\varepsilon_{e\tau}|= 5
    \times 10^{-3}$, $\delta=3\pi/2$, and $\phi_{e \tau}=7\pi/4$ and
    normal mass hierarchy.  Unlike the plots shown in
    Figs~\ref{synergy-em-4}-\ref{nosynergy-em-3}, for the case where
    only the single detector at 3000 km is taken into account, allowed
    regions exist not only for the normal mass hierarchy regime but
    also for the inverted one, as shown in the panels in the second
    row.}
\label{mixed-sign-degeneracy-et-1}
\end{figure}

We now turn to the system with $\varepsilon_{e \tau}$.
Synergy between the two detectors is much more prominent 
in the system with $\varepsilon_{e \tau}$ than the 
$\varepsilon_{e \mu}$ system.
To demonstrate this point we present examples of Type A in 
Fig.~\ref{mixed-sign-degeneracy-et-1}
the allowed regions in 
$\vert \varepsilon_{e \tau} \vert - \phi_{e \tau} $ and 
$\delta - \phi_{e \tau} $ space
for the input parameters, 
$|\varepsilon_{e\tau}|= 5 \times 10^{-3}$, $\phi_{e \tau}=7\pi/4$, 
$\sin^2 2\theta_{13} = 10^{-3}$, $\delta=3\pi/2$ and 
the normal mass hierarchy. 
Within the currently used parameters most of the data set with 
$\varepsilon_{e\tau}$ fall into Type A except for the largest values of 
$\vert \varepsilon_{e \tau} \vert$ and $\theta_{13}$ in the set, 
$\vert \varepsilon_{e \tau} \vert = 2 \times 10^{-2}$ 
and $\sin^2 2\theta_{13} = 10^{-3}$, as indicated in Table~\ref{table-etau}. 
We note that in the system with $\varepsilon_{e \tau}$, 
for the case where only the near detector is used, the clone solution 
in the inverted mass hierarchy (the input is the normal one) exists 
very often,  
as shown in the second row of Fig.~\ref{mixed-sign-degeneracy-et-1}.
This will significantly lower the sensitivity to the mass hierarchy 
determination compared to the case where no NSI of 
the $\varepsilon_{e \tau}$ type is present, which 
will be demonstrated in Sec.~\ref{sec:impact-mh}.
%
\begin{figure}[bhtp]
\begin{center}
\hglue -0.6cm
\includegraphics[width=0.82\textwidth]{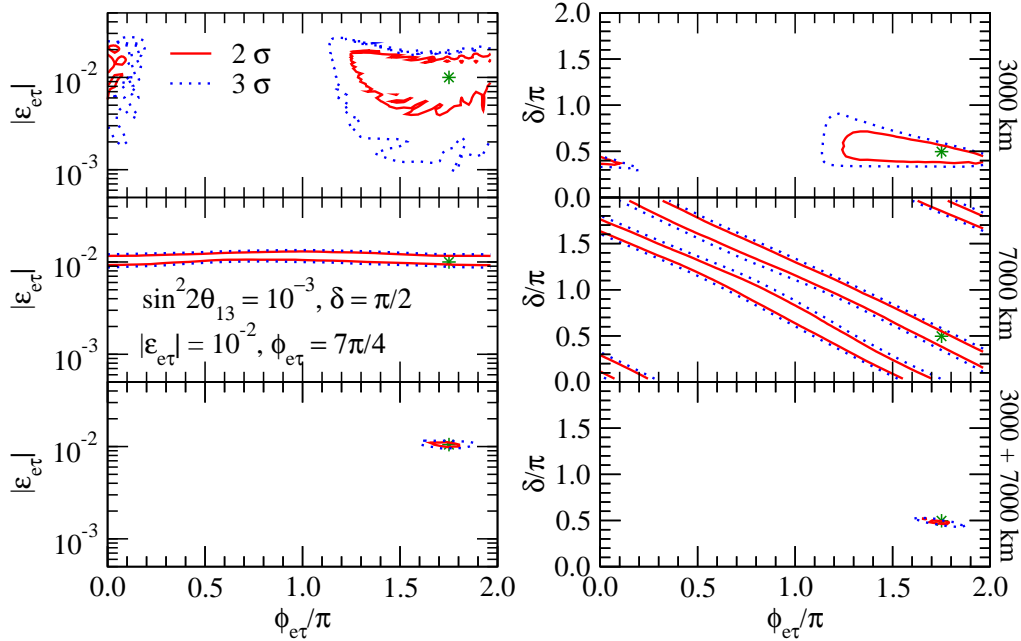}
\end{center}
\vglue -0.75cm
\caption{An example of Type C.  Allowed regions obtained for the
  system with $\varepsilon_{e\tau}$.  The input parameters are taken
  as: $\sin^2 2\theta_{13} = 10^{-3}$, $|\varepsilon_{e\tau}|= 2
  \times 10^{-2}$, $\delta=\pi/2$, and $\phi_{e \tau}=7\pi/4$ and
  normal mass hierarchy.  No allowed regions exist in the inverted
  mass hierarchy regime.  }
\label{modest-synergy-et-2}
\end{figure}
%
Type C is rare in the data set in the system with
$\varepsilon_{e\tau}$ for the currently used values of the parameters,
as indicated in Table~\ref{table-etau}.  An example is shown in
Fig.~\ref{modest-synergy-et-2} which was generated with the input
parameters $|\varepsilon_{e\tau}|= 2 \times 10^{-3}$, $\phi_{e
  \tau}=7\pi/4$, $\sin^2 2\theta_{13} = 10^{-3}$, $\delta=\pi/2$ and
the normal mass hierarchy.

Synergy is very remarkable in the system with $\varepsilon_{e\tau}$
apart from the case of the largest values of $\vert \varepsilon_{e
  \tau} \vert$ and $\theta_{13}$ mentioned above, as clearly seen in
Table~\ref{table-etau}.  Yet, effect of the far detector data is
visible even in relatively large $\vert \varepsilon_{e \tau} \vert$
case, making allowed regions tiny as a result of combination of both
detectors as shown in Fig.~\ref{modest-synergy-et-2}.  This feature is
generic for larger values of $\vert \varepsilon_{e \tau} \vert$.
The allowed contours in the $\phi_{e \tau} - \delta$ plane in 
Figs. \ref{mixed-sign-degeneracy-et-1} and \ref{modest-synergy-et-2}
due to the far detector measurement alone
clearly displays characteristic feature that only the combination 
$\phi_{e \tau} + \delta$ is constrained.

It is remarkable to find that there is no Type D in the data set,
representing an extremely powerful synergy between the two detectors
in the system with $\varepsilon_{e\tau}$.
It is also very significant that $\delta - \phi_{e \tau} $ confusion
is resolved in most of the figures for the system with
$\varepsilon_{e\tau}$.

Though we have achieved reasonably good understanding of the features
of the sensitivities, there is a curious feature in
Tables~\ref{table-emu} and \ref{table-etau}.
Let us compare Fig.~\ref{biP-3000km-em-20GeV} and
Fig.~\ref{biP-3000km-et-20GeV} in the right-bottom panel with $\sin^2
2\theta_{13} = 10^{-4}$, $\varepsilon_{e\mu}= 5 \times 10^{-3}$, and
$\varepsilon_{e\tau}= 2 \times 10^{-2}$.  The values of the NSI
parameters are chosen so that the size of the probabilities are
comparable.
Then, one would expect that the (relative) accuracy to which we can
determine (or constrain) NSI parameters is similar between the
$\varepsilon_{e\mu}$ and the $\varepsilon_{e\tau}$ systems; The
ellipses with different $\delta$ overlaps and the degree of
overlapping is very similar in both systems.
Nevertheless, the potential for parameter determination is in fact
very different between $\varepsilon_{e\mu}$ and $\varepsilon_{e\tau}$
systems as one observes by comparing between Table~\ref{table-emu} and
Table~\ref{table-etau}; Type B is dominant in the $\varepsilon_{e\mu}$
system, whereas Type A dominates leaving no Type B in the
$\varepsilon_{e\tau}$ system.  It clearly indicates that the
sensitivity is far better in the $\varepsilon_{e\mu}$ system compared
to the $\varepsilon_{e\tau}$ system.
On the other hand, the parameter degeneracy is much severer at the
near detector in systems with $\varepsilon_{e\mu}$ compared to the
ones with $\varepsilon_{e\tau}$ as is shown in the tables.

\begin{figure}[bhtp]
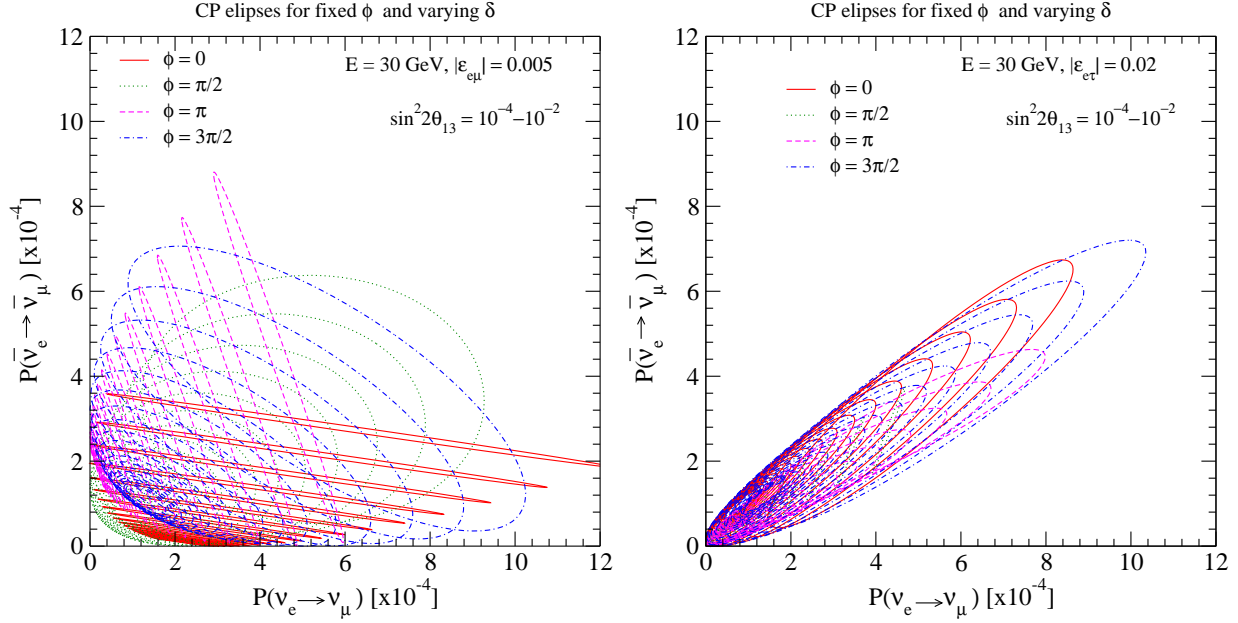

\vglue 0.2cm
\begin{center}
\hglue -0.6cm
\includegraphics[width=0.49\textwidth]{Prob_emu_vary_delta_theta13_0.001.eps}
\includegraphics[width=0.49\textwidth]{Prob_etau_vary_delta_theta13_0.001.eps}
\end{center}
\vglue -0.8cm
\caption{ Bi-probability plots drawn by continuously varying $\delta$,
  for four different values of $\phi$ for $L = 3000$ km.  For each
  value of $\phi$, the 20 different values of $\sin^2 2\theta_{13}$
  between $10^{-4}$ and $10^{-2}$ (logarithmically varied) are
  considered.  }
\label{Prob_vary_delta_theta13}
\end{figure}

A unified understanding of the puzzling features becomes possible once
one draws the bi-probability plots by varying $\delta$ for four
(twenty) different values of $\phi$ ($\theta_{13}$) as done in
Fig.~\ref{Prob_vary_delta_theta13}.  The degeneracy is severer in the
$\varepsilon_{e\mu}$ system because of the more dynamic behavior of
the bi-probability ellipses as shown in the left panel of
Fig.~\ref{Prob_vary_delta_theta13}.  Because the ellipses can locate
themselves essentially everywhere in the bi-probability space there
are chances that fake solutions can be produced at points far apart
from the true solution.
What about the difference in the sensitivities?  We observe in the
right panel of Fig.~\ref{Prob_vary_delta_theta13} that the ellipses in
the $\varepsilon_{e\tau}$ system remain in a much more compact region
when $\delta$ and $\theta_{13}$ are varied.  Because of the finite
resolution of the experimental data it appears that the dense
concentration of the ellipses with different parameters leads to
merging of many degenerate solutions, resulting in the lack of the
sensitivity in the $\varepsilon_{e\tau}$ system.

By comparing Figs.~\ref{synergy-em-4}-\ref{nosynergy-em-3}
for $\varepsilon_{e\mu}$ (which show no strong synergy by 
combining two detectors) and 
Figs.~\ref{mixed-sign-degeneracy-et-1} and \ref{modest-synergy-et-2}
for $\varepsilon_{e\tau}$ (which show strong synergy),
one might think that strong synergy for 
$\varepsilon_{e\tau}$ is simply because of the larger 
values we considered for this NSI element than that for $\varepsilon_{e\mu}$. 
We note, however, that if we consider larger value for  $\varepsilon_{e\mu}$,
the allowed NSI parameters obtained from the near detector alone 
are already restricted to small regions and we do not see any 
strong synergy when combined (see Fig.~\ref{mixed-sign-degeneracy-et})
with the far detector unlike the case of  $\varepsilon_{e\tau}$. 

\section{Parameter Degeneracy with NSI}
\label{sec:degeneracy}

In this section we try to understand the property of the degeneracy 
in systems with NSI, encountered in the previous section,
i.e., in the category Type B.
It was indicated that the structure of the conventional parameter degeneracy 
\cite{intrinsic,MNjhep01,octant}, 
(for reviews see e.g., \cite{MNP2,BMW02}), 
the intrinsic, the sign-$\Delta m^2_{31}$, 
and the $\theta_{23}$ octant degeneracies prevails but in a way involving 
NSI parameters, in particular, the phases of $\varepsilon$ parameters 
\cite{NSI-perturbation}. 
In fact, the degeneracy in systems with NSI turns out to be a highly 
complicated problem. 
In a general two $\varepsilon_{e \alpha}$ ($\alpha = \mu, \tau$) system 
there are total six parameters. 
Though some progress has been made in \cite{NSI-perturbation}, 
its complete treatment still eludes us.\footnote{
A completely new type of degeneracy, the solar-atmospheric variables 
exchange degeneracy, is found there. 
But, this degeneracy does not appear to enter into 
the discussion in this 
paper, and hence it is outside of the scope of this paper.
}
%
In our simplified system of a
single type of $\varepsilon$, however, we can achieve a good
understanding of the phenomenon.
After giving some examples to show characteristic features in this section, 
we give in Appendix~\ref{solutions} some semi-analytic treatment 
to obtain the degenerate solutions for a given true solution.

\subsection{$\phi$-degeneracy}
\label{subsec:phideg}

\begin{figure}[bhtp]
\begin{center}
\includegraphics[width=0.90\textwidth]{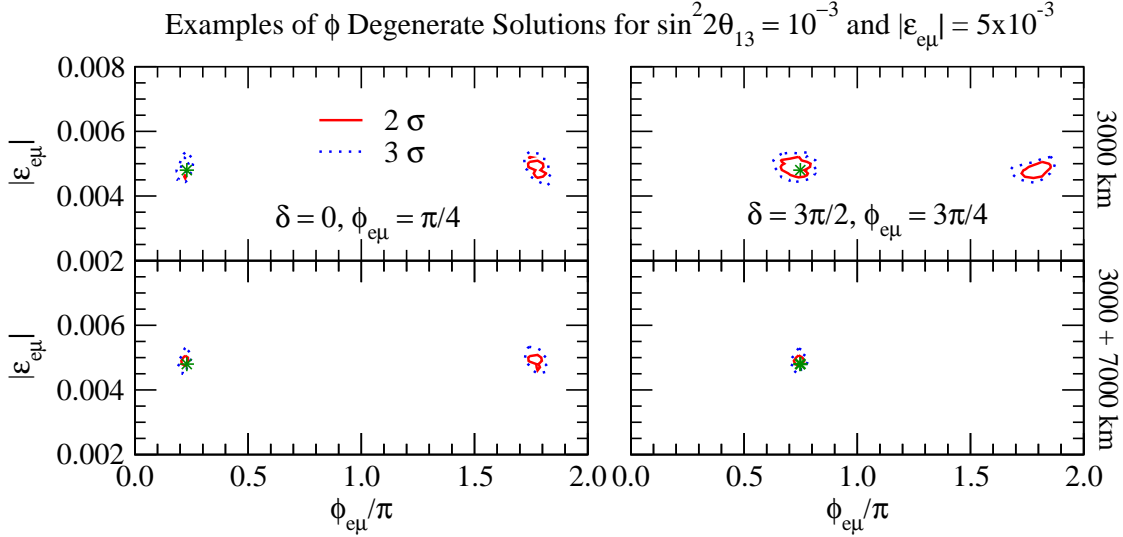}
\end{center}
\vglue -0.8cm
\caption{ Some examples of $\phi$-degeneracy.  The normal mass
  hierarchy is assumed.  The parameters are taken as $\sin^2 2
  \theta_{13} = 10^{-3}$, $\vert \varepsilon_{e \mu} \vert = 5 \times
  10^{-3}$ and the baseline is $L=3000$ km.  }
\label{phi-degeneracy}
\end{figure}

We start from the simplest case which may be called as the
$\phi$-degeneracy, in which the degenerate solutions are characterized
only by distinct values of $\phi$, and values of the remaining
variables are nearly equal.
In this subsection we restrict ourselves to the system with
$\varepsilon_{e\mu}$.  In Fig.~\ref{phi-degeneracy} some examples of
such degeneracy are presented.

Let us understand why and how such degeneracy arises.  We start from
the expressions of appearance oscillation probabilities in
(\ref{Pemu-biP-phi}).
Because the terms proportional to $\mathcal{C} $ and $\mathcal{I} $
(and corresponding coefficients for anti-neutrino, 
$\bar{ \mathcal{C} }$ and $\bar{ \mathcal{I} }$) are an order of
magnitude smaller than the other terms for typical values of the
parameters for relevant for neutrino factory, they can be neglected.
Then, it is interesting to examine a simplified model for the oscillation 
probabilities 
\begin{eqnarray}
P - \mathcal{A} &=& \pm \sqrt{ \mathcal{B}^2 + \mathcal{R}^2 + 2 \mathcal{B}\mathcal{R} \cos \delta } \cos (\phi + \alpha) 
\nonumber \\
\bar{P} -  \bar{ \mathcal{A} } &=& \pm 
\sqrt{ \bar{\mathcal{B}}^2 + \bar{\mathcal{R}}^2 + 2 \bar{\mathcal{B}}\bar{\mathcal{R}} \cos \delta } \cos (\phi + \bar{\alpha} ), 
\label{Pemu-model}
\end{eqnarray}
where the upper (lower) line corresponds to neutrino (anti-neutrino) channel
and the sign must be chosen correctly. 
$\alpha$ and $\bar{\alpha} $ are given by 
\begin{eqnarray}
\tan \alpha = \frac{\mathcal{B} \sin \delta } {\mathcal{B} \cos \delta + \mathcal{R} }, 
\hspace{1cm}
\tan \bar{\alpha} = \frac{ \bar{\mathcal{B}} \sin \delta } { \bar{\mathcal{B}} \cos \delta + \bar{\mathcal{R}} }.
\label{tan-alpha}
\end{eqnarray}

The features of the simplified model (\ref{Pemu-model}) can be best
illustrated by the bi-probability plot in $P - \bar{P}$ space.
See the top-right panel in the upper figure in Fig.~\ref{biP-3000km-em-20GeV} 
in particular, the case of $\delta = 0$, for one of the clearest example.
Namely, the ellipse shrinks into line due to the presence of only cosine 
dependence of the phase.
Then, it is evident that the degenerate solution exists; 
If we have a solution at $\phi=\phi_{1}$ then 
we have another solution at 
$\phi_{2} + \alpha = 2\pi - (\phi_{1} + \alpha)$, namely, 
$\phi_{2} = 2\pi - \phi_{1} - 2 \alpha$. 

The simplest situation arises at $\delta = 0$ where $\alpha =
\bar{\alpha} = 0$.  The second solution is given by $\phi_{2} = 2\pi -
\phi_{1}$ which nicely explains features of the left panel of
Fig.~\ref{phi-degeneracy}.
In the right panel of Fig.~\ref{phi-degeneracy}, 
the input value of $\delta$ is $\frac{3 \pi}{2}$, and $\alpha$ can be obtained 
by using (\ref{tan-alpha}) as $\alpha = - 0.88$, 
where a typical value of the energy $E=30$ GeV is used.  
Then, corresponding to solution $\phi_{1} = \frac{3 \pi }{4} = 2.36$ 
there is a second solution $\phi_{2} = 2\pi -
\phi_{1} - 2 \alpha = 5.7$, which explains the right panel of
Fig.~\ref{phi-degeneracy} 
in a reasonable accuracy.

Though this type of degeneracy can be solved in most cases with the
far detector measurement, it can fail in certain cases.  Since $\phi$
dependence at the magic baseline (\ref{Pemu-emu-magic}) takes a
particular form $\cos (\delta+\phi)$ the $\phi$-degeneracy cannot be
solved for $\delta \simeq 0$.  An example of this phenomenon is given
in the left panel of Fig.~\ref{phi-degeneracy}.

\subsection{Intrinsic and $\Delta m^2_{31}$-sign flipped degeneracies}
\label{sign-degeneracy}

\begin{figure}[h]
\vglue -0.4cm
\begin{center}
\hglue -0.6cm
\includegraphics[width=0.90\textwidth]{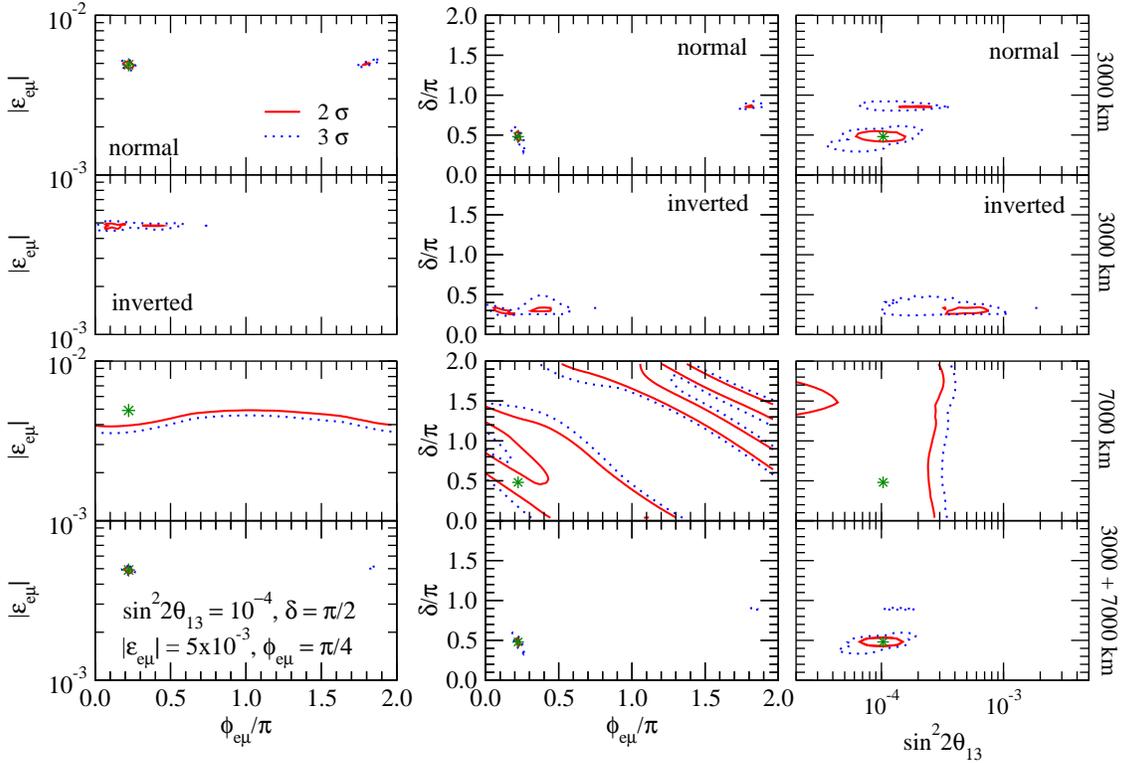}
\end{center}
\vglue -0.85cm
\caption{An example of Type B.  Allowed regions in the
  $\phi_{e\mu}-|\varepsilon_{e\mu}|$ plane (left column),
  $\phi_{e\mu}-\delta$ plane (middle column) and $\sin^2
  2\theta_{13}-\delta$ plane (right column) corresponding to 2 and 3
  $\sigma$ CL obtained for the system with $\varepsilon_{e\mu}$.
  Panels in the upper 2 rows (3rd row) correspond to the case where
  only a single detector at 3000 km (7000 km) is taken into account,
  whereas the ones in the 4th row correspond to the case where results
  from the two detectors are combined.  The input parameters are taken
  as: $\sin^2 2\theta_{13} = 10^{-4}$, $\delta=\pi/2$,
  $|\varepsilon_{e\mu}|= 5 \times 10^{-3}$ and $\phi_{e \mu}=\pi/4$
  (indicated by the green asterisk), and the mass hierarchy is normal.
  As in the case shown in Fig.~\ref{mixed-sign-degeneracy-et-1},
  allowed regions exist also in the inverted mass hierarchy regime
  when only a single detector at $L=3000$ km is considered, see the
  panels in the second row.}
\label{mixed-sign-degeneracy-em}
\end{figure}
%
\begin{figure}[h]
\begin{center}
\hglue -0.6cm
\includegraphics[width=0.90\textwidth]{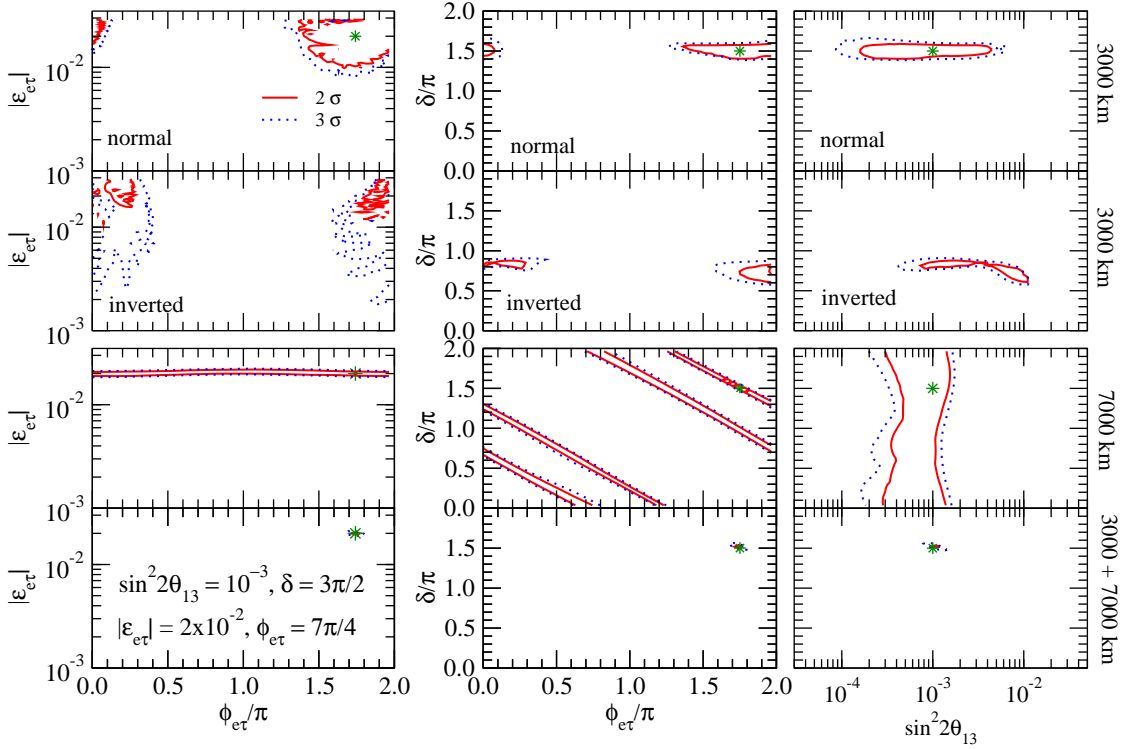}
\end{center}
\vglue -0.85cm
\caption{An example of Type B.  Allowed regions with
  $\varepsilon_{e\tau}$.  The input parameters are taken as: $\sin^2
  2\theta_{13} = 10^{-3}$, $|\varepsilon_{e\tau}|= 2 \times 10^{-2}$,
  $\delta=3\pi/2$, and $\phi_{e \tau}=7\pi/4$ and normal mass
  hierarchy.  }
\label{mixed-sign-degeneracy-et}
\end{figure}

Now, we discuss more generic form of the degeneracy, the intrinsic and
the $\Delta m^2_{31}$-sign flipped degeneracies which are similar to
those in the standard neutrino oscillation \cite{intrinsic,MNjhep01}.
The $\theta_{23}$ octant degeneracy \cite{octant} does not arise 
because we take the maximal $\theta_{23}$. 

In Fig.~\ref{mixed-sign-degeneracy-em}, we show, for the system with
$\varepsilon_{e\mu}$, the case (which is categorized as Type B) where
the measurement at the near detector at 3000 km leaves four regions at
2 $\sigma$, two of which correspond to the normal hierarchy and the
other two the inverted ones.
It is the manifestation of the sign-$\Delta m^2_{31}$ degeneracy in
systems with NSI where the phase of NSI parameter $\phi$ is indeed
actively involved.
Unlike in most cases this is one of the special cases in which the far
detector measurement fails to resolve the intrinsic degeneracy.  In
this particular case it occurs in the following way; Because the
relation $\cos (\delta_2 + \phi_2) = \cos (\delta_1+ \phi_1)$
approximately holds the far detector cannot solve the $\phi$
degeneracy.
Also, the energy spectra are very similar for the two solutions with
the same sign of $\Delta m_{31}^2$.

In Fig.~\ref{mixed-sign-degeneracy-et} an example of Type B for
the system with $\varepsilon_{e\tau}$ 
with degenerate solutions at the near detector is presented.  
As in the system with $\varepsilon_{e\mu}$ Type B takes place when $\vert
\varepsilon_{e\tau} \vert $ is relatively large.
Here, again the sign-$\Delta m^2_{31}$ degeneracy exists at the near 
detector (top panels), but it is completely resolved when combined with 
the far detector measurement. 
Most likely, the intrinsic degeneracy does not exist in this case, 
or is already resolved by the near detector measurement.

\subsection{Does spectrum information solve the degeneracies?}
\label{subsec:spectrum}

\begin{figure}[bhtp]
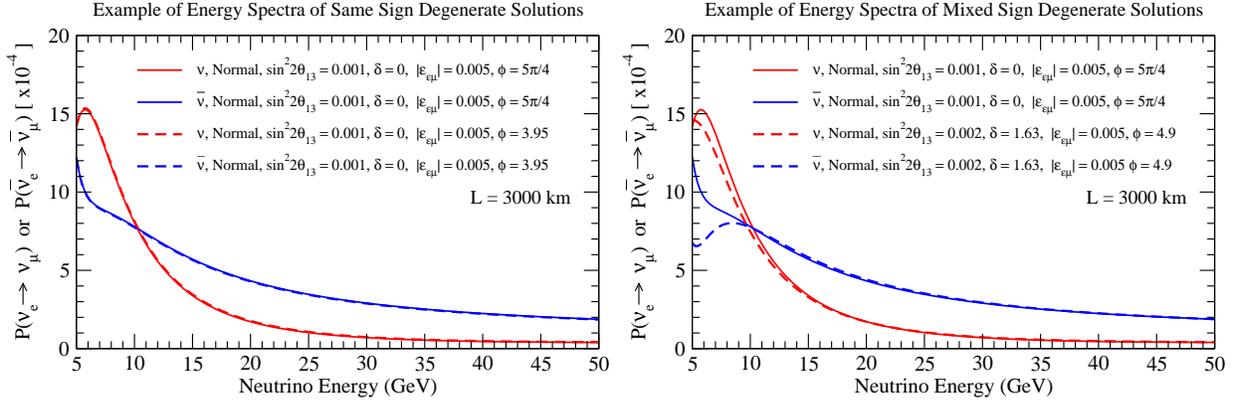

\vglue 0.3cm
\begin{center}
\includegraphics[width=0.49\textwidth]{P_em_same_dege_eneP70.eps}
\includegraphics[width=0.49\textwidth]{P_em_mixed_dege_eneP70.eps}
\end{center}
\vglue -0.6cm
\caption{ Energy spectra of the oscillation probability for a
  system with non-zero NSI element $\varepsilon_{e \mu}$ corresponding to the
  same $\Delta m^2_{31}$-sign intrinsic degeneracy (left panel) and
  the flipped $\Delta m^2_{31}$-sign degeneracy (right panel).  }
\label{spectrum-intrinsic-sign-degeneracy}
\end{figure}
%

One may ask the question, ``to what extent are the degeneracies we saw
in the previous subsection robust?''  It is well known that the
intrinsic degeneracy in the conventional neutrino oscillation with SI
is often fragile to the spectrum analysis.
To give an insight to this point we present in 
Fig.~\ref{spectrum-intrinsic-sign-degeneracy} 
the oscillation probabilities as a function of neutrino energy which 
correspond to the same $\Delta m^2_{31}$-sign intrinsic degeneracy 
(left panel) and the flipped $\Delta m^2_{31}$-sign degeneracy 
(right panel). 
From these figures it is obvious that the degeneracies we observe in 
the systems with $\varepsilon_{e \mu}$ or $\varepsilon_{e \tau}$ are quite 
robust against the spectrum analysis as long as we consider 
only a single detector at $L = 3000$ km. 
One can go through the similar analysis to understand the feature of the 
degenerate solutions by using the analytic framework and the results are 
presented in Appendix~\ref{solutions}.

\section{Discovery Potential of NSI Parameters}
\label{sec:disc}

In this and the next sections, we discuss the discovery potential of
the effects induced by NSI, and by the SI, respectively.  Throughout
these sections we present the sensitivity contours which are
calculated with two different choices of the $\chi^2$ assumptions,
Choice R and Choice O defined in Sec.~\ref{sec:method}.  We compare
the results of the two choices of the experimental uncertainties.
Typically, the sensitivities obtained with Choice R are worse by 
a factor of several to an order of magnitude than the ones with Choice O.

\subsection{Discovery potential of NSI}
\label{sec:disc-nsi}

Let us discuss first the sensitivity to discover the non-zero effect
of NSI.  We determine, for some fixed values of $\theta_{13}$ and
$\delta$, the 2 (3) $\sigma$ CL regions of the discovery potential of
NSI by the condition, $\chi^2_{\text{min}}(\varepsilon = 0) -
\chi^2_{\text{min}}(\text{true value of } \varepsilon \text{ and
}\phi) > 4 (9)$, for 1 DOF, by varying freely $\theta_{13}$ and
$\delta$ and the choice of mass hierarchy in fitting the input data.
If we cannot fit the input data generated with non-vanishing
$\varepsilon_{e \alpha}$ by hypotheses with arbitrary values of the SI
parameters without NSI, we conclude that we can disentangle NSI
effects from those of standard oscillations.
Note that the region determined by this condition includes the case
where $\phi$ is totally undetermined or unconstrained.

\begin{figure}[bhtp]
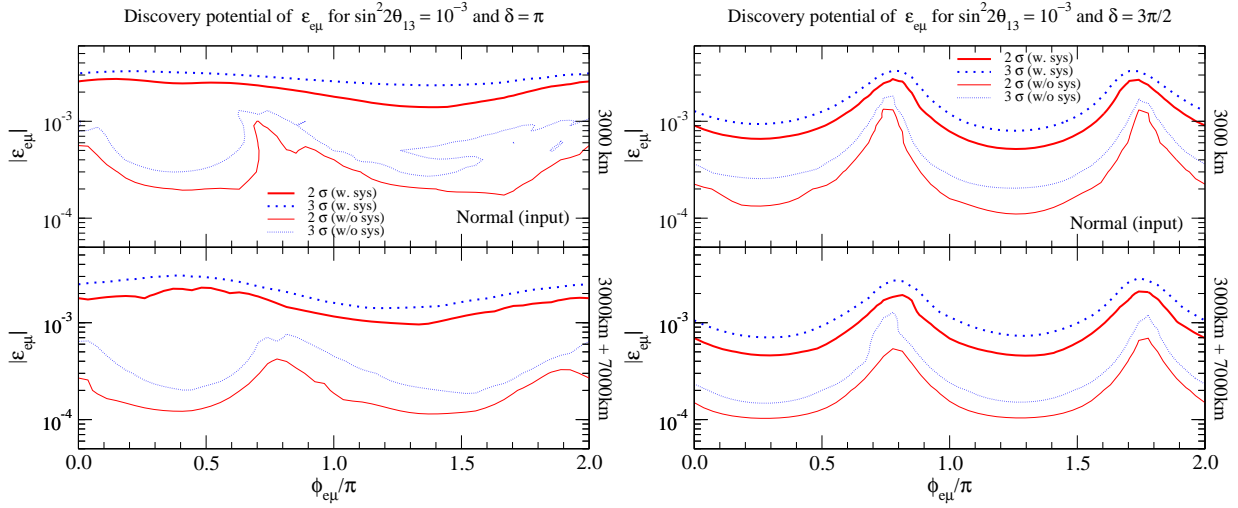

\begin{center}
\includegraphics[width=0.49\textwidth]{NSI_discovery_pi_em_v3.eps}
\includegraphics[width=0.49\textwidth]{NSI_discovery_3piby2_em_v3.eps}
\end{center}
\vglue -0.75cm
\caption{Regions where the non-zero NSI effect caused by $\varepsilon_{e\mu}$ 
can be identified for the case $\sin^2 2\theta_{13} = 10^{-3}$ and 
$\delta = \pi$ (left panel) and $\delta = 3\pi/2$ (right panel). 
The upper and the lower panels are for the cases with measurement at the 
near ($L=3000$ km) detector alone and the one combined with 
the far ($L=7000$ km) detector, respectively. 
The red solid and the blue dotted lines are for the sensitivities at
2$\sigma$ and 3$\sigma$ CL, respectively.  The thick and the thin
lines are for Choice R (with systematic uncertainties and
efficiencies) and Choice O of the $\chi^2$ analysis, respectively,
defined in Sec.~\ref{sec:method}.  }
\label{NSI-discovery-em}
\end{figure}
%
%
\begin{figure}[bhtp]
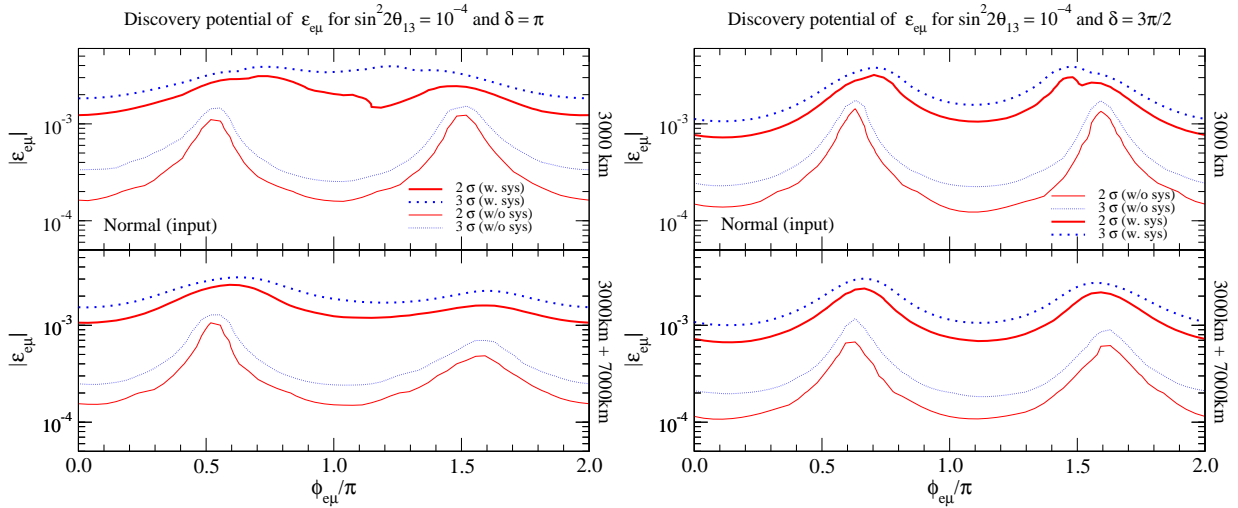

\begin{center}
\includegraphics[width=0.49\textwidth]{NSI_discovery_pi_em_0.0001_v3.eps}
\includegraphics[width=0.49\textwidth]{NSI_discovery_3piby2_em_0.0001_v3.eps}
\end{center}
\vglue -0.75cm
\caption{ The same as in Fig.~\ref{NSI-discovery-em} but with $\sin^2
  2\theta_{13} = 10^{-4}$.  }
\label{NSI-discovery-em-0.0001}
\end{figure}

In Figs.~\ref{NSI-discovery-em} and \ref{NSI-discovery-em-0.0001} we
show the regions where the non-zero NSI effect can be discovered in
the plane of true values of $\phi_{e\mu}$ and $| \varepsilon_{e\mu}|$
for the case where $\sin^2 2\theta_{13} = 10^{-3}$ and $\sin^2
2\theta_{13} = 10^{-4}$, respectively, for $\delta = \pi$ (left
panels) and $\delta = 3\pi/2$ (right panels).  The same values of
$\delta$ are consistently used throughout this section for
presentation of the allowed contours.  These two values are chosen
because one is CP conserving and the other CP violating. We think that
the choice is a rather conservative one, avoiding the point with the
best sensitivities.  In both figures we have taken the input mass
hierarchy to be the normal one.

First, we note that the sensitivity to $| \varepsilon_{e\mu}|$ by the
near detector at 3000 km is good enough so that adding the
contribution of the second detector at 7000 km does not improve much
the sensitivity to $\varepsilon_{e\mu}$.  The feature is indeed
expected as discussed in Secs.~\ref{sec:nsifeatures} and
\ref{sec:overview}.
Second, the sensitivities to $| \varepsilon_{e\mu}|$ depend very much
on $\phi_{e \mu}$, and varies from $\sim 10^{-4}$ to $\sim 10^{-3}$.
We have checked that the same general behavior is obtained for the
case of inverted mass hierarchy as input.
We note that in the case where detector efficiency, background and
systematic uncertainties are considered, indicated by the thick curves
in Fig.~\ref{NSI-discovery-em}, the sensitivities are worse by 
a factor of a few to one order of magnitude 
compared to the case where they are ignored.
%
\begin{figure}[bhtp]
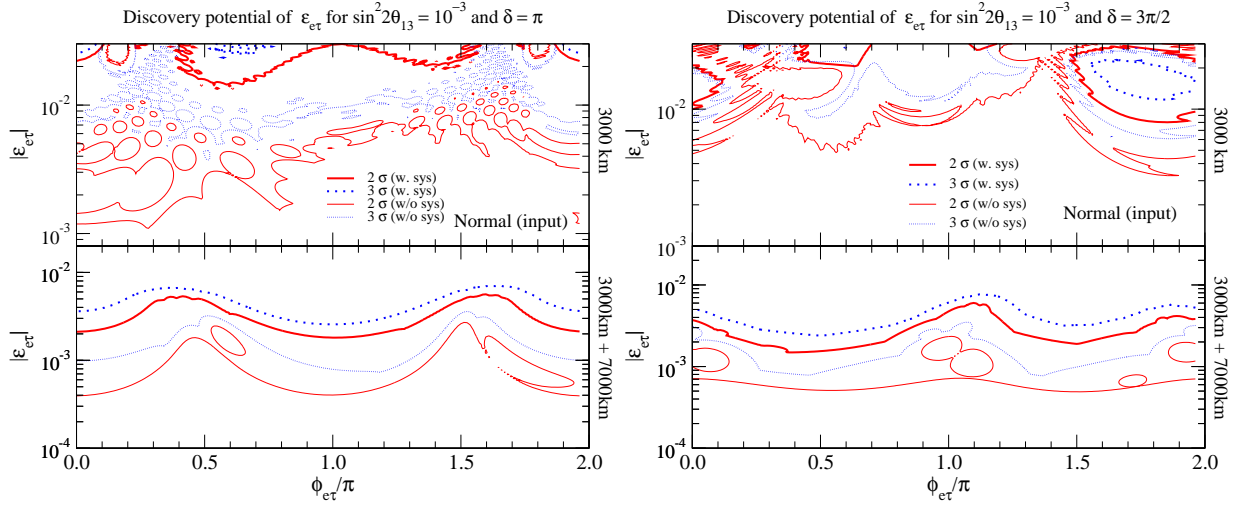

\begin{center}
\includegraphics[width=0.49\textwidth]{NSI_discovery_pi_et_v3_new.eps}
\includegraphics[width=0.49\textwidth]{NSI_discovery_3piby2_et_v3_new.eps}
\end{center}
\vglue -0.75cm
\caption{ Regions where the non-zero NSI effect caused by
  $\varepsilon_{e\tau}$ can be identified for the case $\sin^2
  2\theta_{13} = 10^{-3}$ and $\delta = \pi$ (left panel) and $\delta
  = 3\pi/2$ (right panel).  The upper and the lower panels are for the
  cases with measurement at the near ($L=3000$ km) and the far
  ($L=7000$ km) detectors, respectively.  The meaning of the
  color-type of the lines are the same as in
  Fig.~\ref{NSI-discovery-em}.  }
\label{NSI-discovery-et}
\end{figure}
%
\begin{figure}[bhtp]
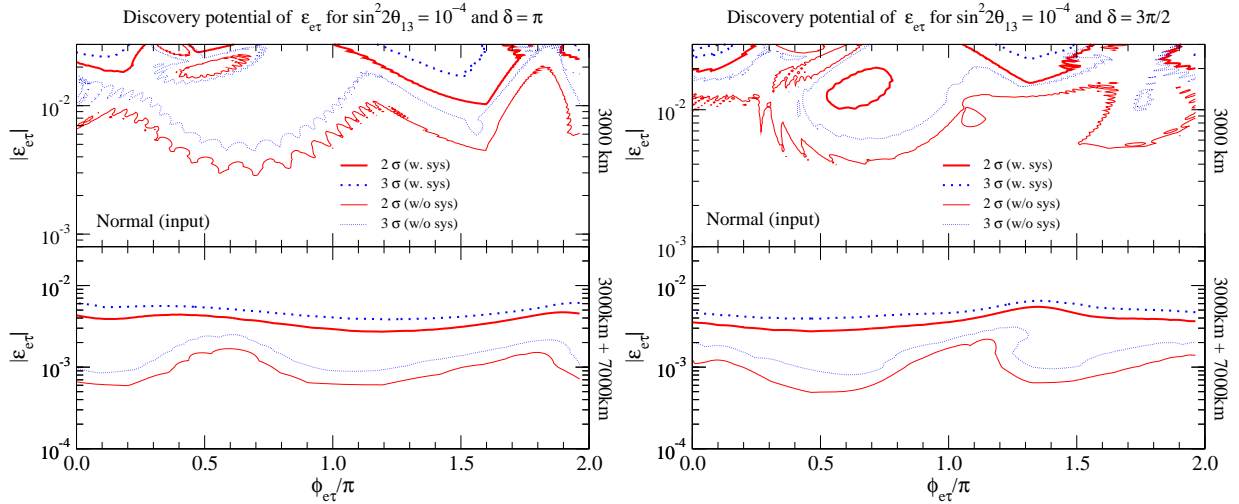

\begin{center}
\includegraphics[width=0.49\textwidth]{NSI_discovery_pi_et_0.0001_v3_new.eps}
\includegraphics[width=0.49\textwidth]{NSI_discovery_3piby2_et_0.0001_v3_new.eps}
\end{center}
\vglue -0.75cm
\caption{ The same as in Fig.~\ref{NSI-discovery-et} but with $\sin^2
  2\theta_{13} = 10^{-4}$.  }
\label{NSI-discovery-et-0.0001}
\end{figure}

In Figs.~\ref{NSI-discovery-et} and Fig.~\ref{NSI-discovery-et-0.0001}
we show similar plots but for the system with
$\varepsilon_{e\tau}$\footnote{ The readers may wonder if the jagged
  behavior in some of the panels in Figs.~\ref{NSI-discovery-et} and
  \ref{NSI-discovery-et-0.0001} might be due to the artifact of a too
  crude mesh of the parameter space used in the numerical analysis. We
  have checked that the small island structure does not go away,
  though with modified shape or merging, even if we increase the
  number of grids used in scanning the parameter space by a factor of
  10 to 15. The figures presented are the outcome of such an improved
  analysis. Therefore, we believe that some of the jagged behavior are
  real though the possibility of it being a numerical artifact cannot
  be completely ruled out.}.  In this system the sensitivity for the
detector at $L = 3000$ km has a rather complicated structure
%
which depends nontrivially on 
$\phi_{e\tau}$ and $\delta$.
It is clearly very difficult to discover 
$\vert \varepsilon_{e\tau}\vert \lsim$ few $10^{-2}$ 
with the near detector alone.  
However, as expected, the synergy of combining the two
detectors is so strong that the final sensitivity is comparable to
the case of $\varepsilon_{e \mu}$, {\em i.e.} 
$\vert \varepsilon_{e\tau}\vert \lsim$ a few $10^{-3}$.  
The sensitivity to
$\varepsilon_{e\tau}$ also depends on the true value of
$\phi_{e\tau}$, but the dependence is weaker than that for
$\varepsilon_{e\mu}$.  Again, we have checked that the same general
features are obtained for the discovery potential of
$\varepsilon_{e\tau}$ in the case of inverted mass hierarchy.  
Here again the role of detector efficiencies, background and systematic
uncertainties at the level exemplified in this paper, is to worse the
sensitivities several times compared to the case where 
they are ignored. 

We note that the results shown in this section can be compared with
the ones obtained in Refs.~\cite{kopp1,kopp3} where the discovery
potential of NSI effect by a future neutrino factory was studied.
The authors of Ref.~\cite{kopp1} considered both 
$\varepsilon_{e\mu}$ and $\varepsilon_{e\tau}$ (but only one of 
them was considered at a time) but for the case 
where only one detector at $L = 3000$ km is assumed, 
and it is found that for $\sin^2 2\theta_{13} = 10^{-3}$, 
the sensitivity to $\varepsilon_{e\mu}$ and $\varepsilon_{e\tau}$ 
are, respectively, a few $\times 10^{-3}$ and 
a few $\times 10^{-2}$ under a similar experimental setup to ours. 
By taking into account some differences of the analysis procedure and 
assumptions, results of our analysis for the detector at 3000 km 
alone seem to be consistent with the ones found in this reference. 

On the other hand, the authors of Ref.~\cite{kopp3} studied how one
can optimize the neutrino factory setting which gives better
sensitivities to constrain NSI parameters assuming two detectors with
different baselines.
In this reference, the sensitivity study was done by varying baselines
to the two detectors and it was found that the standard set up in
\cite{ISS-nufact} of two detectors at 4000 km and 7500 km, which is
similar to the set up considered in our previous paper
\cite{NSI-nufact} (and employed here), works quite well for
determining (or constraining) both standard and non-standard neutrino
parameters.  
The sensitivities to $|\varepsilon_{e\mu}|$ and $|\varepsilon_{e\tau}|$ 
found in Table 1 of \cite{kopp3} which are, respectively, 
$\sim 6 \times 10^{-3}$ and $\sim 2\times 10^{-2}$ seem to 
be consistent with ours if we take into account
some differences between their and our assumptions (higher
efficiencies, higher muon energy, larger total number of useful muon
decays, etc.) in this work.

\subsection{Discovery potential of the non-standard CP violation}
\label{sec:disc-nscpv}

From the results shown in the previous section, we have some ideas
about the parameter regions where the non-zero NSI effect can be
established.
Let us now ask another question; For which values of the NSI
parameters $| \varepsilon |$ and $\phi$, can we establish a new type
of CP violation due to NSI?  Hereafter, we denote such CP violation as
the non-standard CP violation.
To establish the discovery potential for such a new CP violation
effect, we determine the region of sensitivity with 2 (3) $\sigma$ CL
by the condition, $\chi^2_{\text{min}}(\phi = 0 \text{ or } \pi) -
\chi^2_{\text{min}} (\text{true value of } \varepsilon \text{ and
}\phi) > 4 (9)$, for 1 DOF, by varying freely all the parameters
except for $\phi$ in fitting the input data.

\begin{figure}[h]
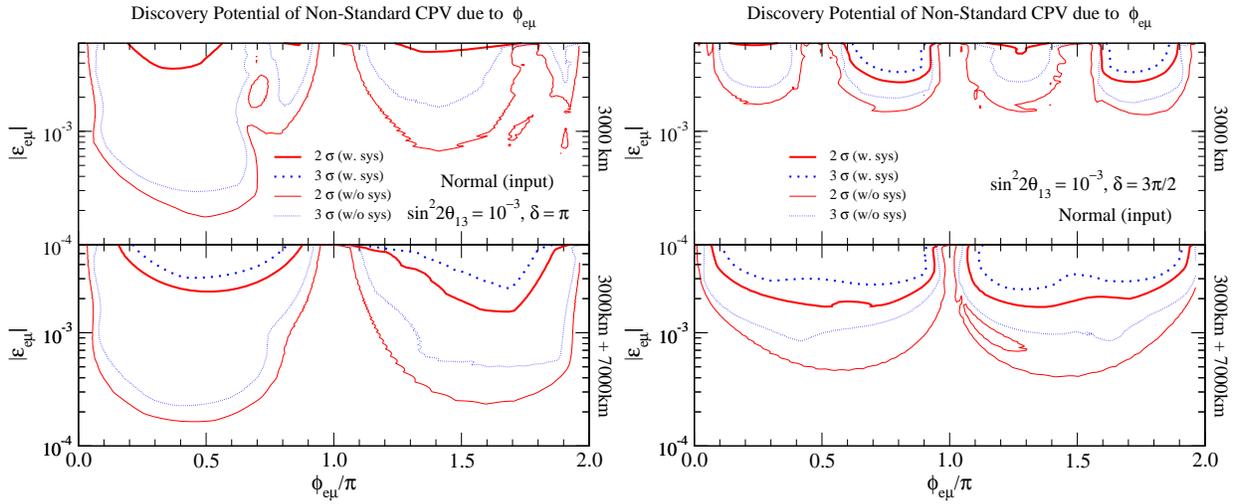

\begin{center}
\includegraphics[width=0.49\textwidth]{NSI_NonStandard-CPV_pi_em_v3_new.eps}
\includegraphics[width=0.49\textwidth]{NSI_NonStandard-CPV_3piby2_em_v3_new.eps}
\end{center}
\vglue -0.75cm
\caption{Regions where the non-standard CP violation caused by
  $\phi_{e \mu} \ne 0$ or $\phi_{e \mu} \ne \pi$ can be established
  for the case $\sin^2 2\theta_{13} = 10^{-3}$, $\delta = \pi$ (left
  panel) and $\delta = 3\pi/2$ (right panel).  Mass hierarchy was
  taken to be normal in the input.  The meaning of the color-type of
  the lines are the same as in Fig.~\ref{NSI-discovery-em}. }
\label{NSI-CPV-sensitivity-em}
\end{figure}

\begin{figure}[h]
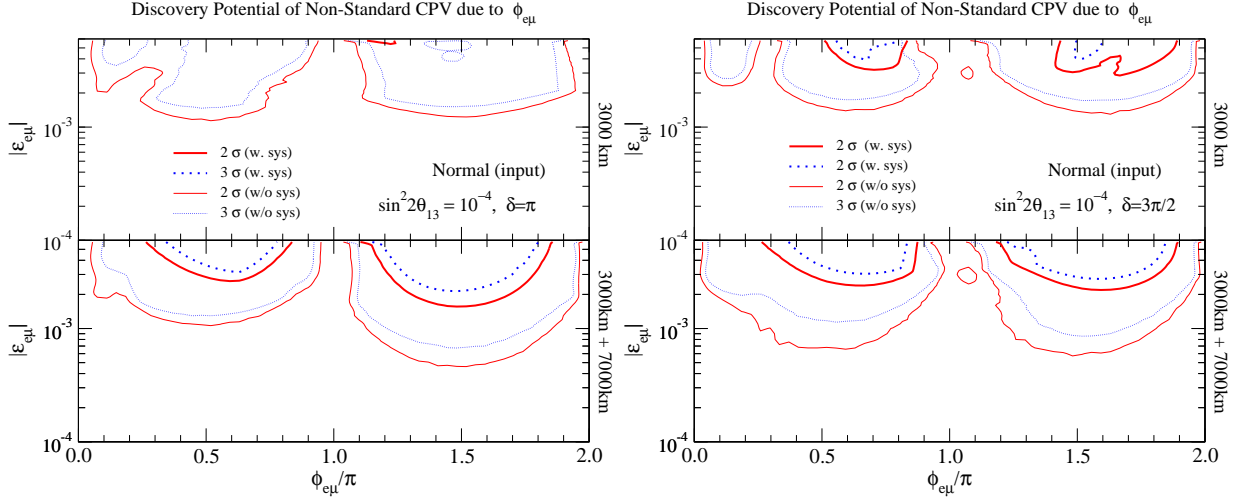

\begin{center}
\includegraphics[width=0.49\textwidth]{NSI_NonStandard-CPV_pi_em_0.0001_v3.eps}
\includegraphics[width=0.49\textwidth]{NSI_NonStandard-CPV_3piby2_em_0.0001_v3.eps}
\end{center}
\vglue -0.75cm
\caption{The same as in Fig.~\ref{NSI-CPV-sensitivity-em} but for
  $\sin^2 2\theta_{13} = 10^{-4}$.}
\label{NSI-CPV-sensitivity-em-0.0001}
\end{figure}

In Fig.~\ref{NSI-CPV-sensitivity-em}
(Fig.~\ref{NSI-CPV-sensitivity-em-0.0001}) we show the region where
non-standard CP violation associated with $\phi_{e\mu}$ can be
established for $\sin^2 2\theta_{13}= 10^{-3}$ ($\sin^2 2\theta_{13} =
10^{-4}$) and $\delta = \pi$ (shown in the left panel), or $\delta =
3\pi/2$ (in the right panel).
We observe that in addition to the significant improvement of the sensitivity 
in the case where two detectors are combined (see lower panels), 
there are qualitative differences in the behavior of the
sensitivity contours for $L = 3000$ km and the combined case. 

Let us see the right panel of Fig.~\ref{NSI-CPV-sensitivity-em}, 
 the case without systematic uncertainties.  
At $L=3000$ km, in spite of the maximal non-standard CP violation
($\phi_{e \mu}= \pi/2, 3 \pi/2$) we cannot establish the non-standard
CP violation even though $|\varepsilon_{e \mu}|$ is large.
Though it might look curious it is easy to understand why this feature
arises; It is due to the $\phi$-degeneracy (see
Sec.~\ref{subsec:phideg}).  Using (\ref{Pemu-model}) with input
$\delta=3\pi/2, \sin^2 2\theta_{13}=10^{-3}$, and $E=20$ GeV, one can
show that the clone solution of $\phi_{1}=\pi/2~(3\pi/2)$ is located
very close to $\phi_{2}=0~(\pi)$. So we are unable to distinguish the
maximal CP violation and CP conservation owing to the
$\phi$-degeneracy.  Fortunately, the degeneracy is resolved when
informations from the far detector is combined.

%
%
\begin{figure}[bhtp]
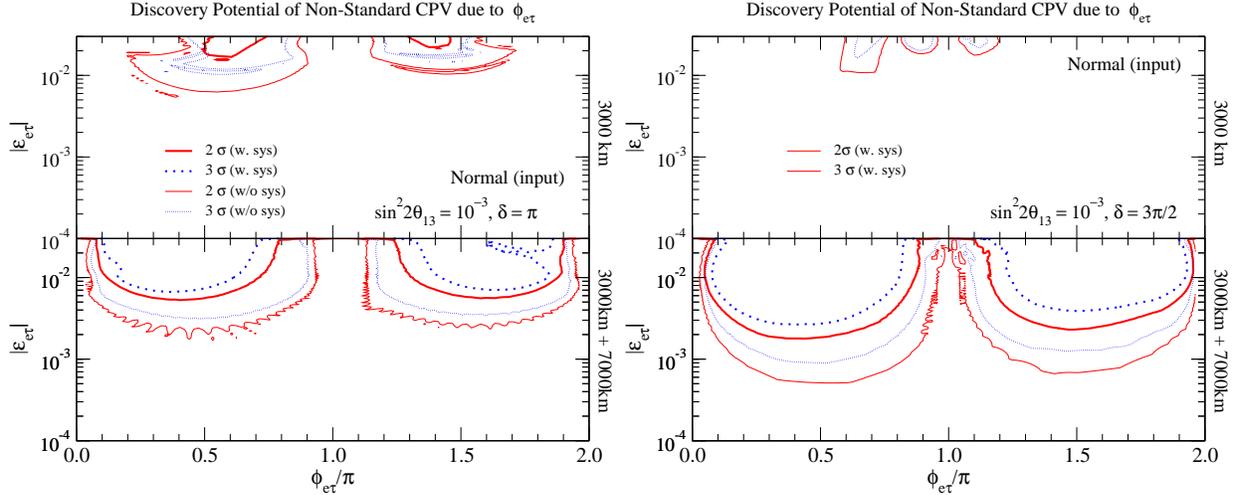

\begin{center}
\includegraphics[width=0.49\textwidth]{NSI_NonStandard-CPV_pi_et_v3_new.eps}
\includegraphics[width=0.49\textwidth]{NSI_NonStandard-CPV_3piby2_et_v3_new.eps}
\end{center}
\vglue -0.75cm
\caption{The same plots as in Fig.~\ref{NSI-CPV-sensitivity-em} but
  for the non-standard CP violation caused by $\phi_{e \tau} \ne 0$ or
  $\phi_{e \tau} \ne \pi$.  }
\label{NSI-CPV-sensitivity-et}
\end{figure}
%
\begin{figure}[bhtp]
\vglue -0.6cm
\begin{center}
\includegraphics[width=0.49\textwidth]{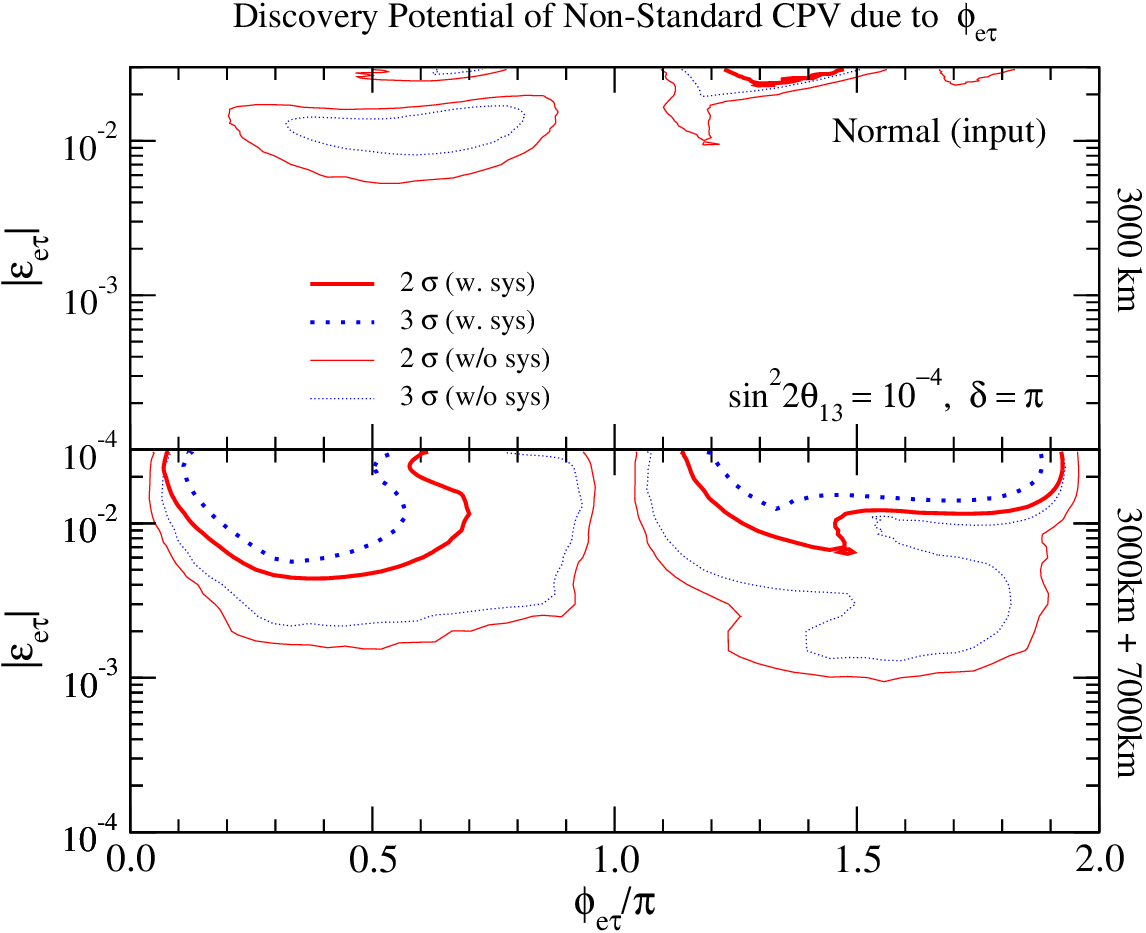}
\includegraphics[width=0.49\textwidth]{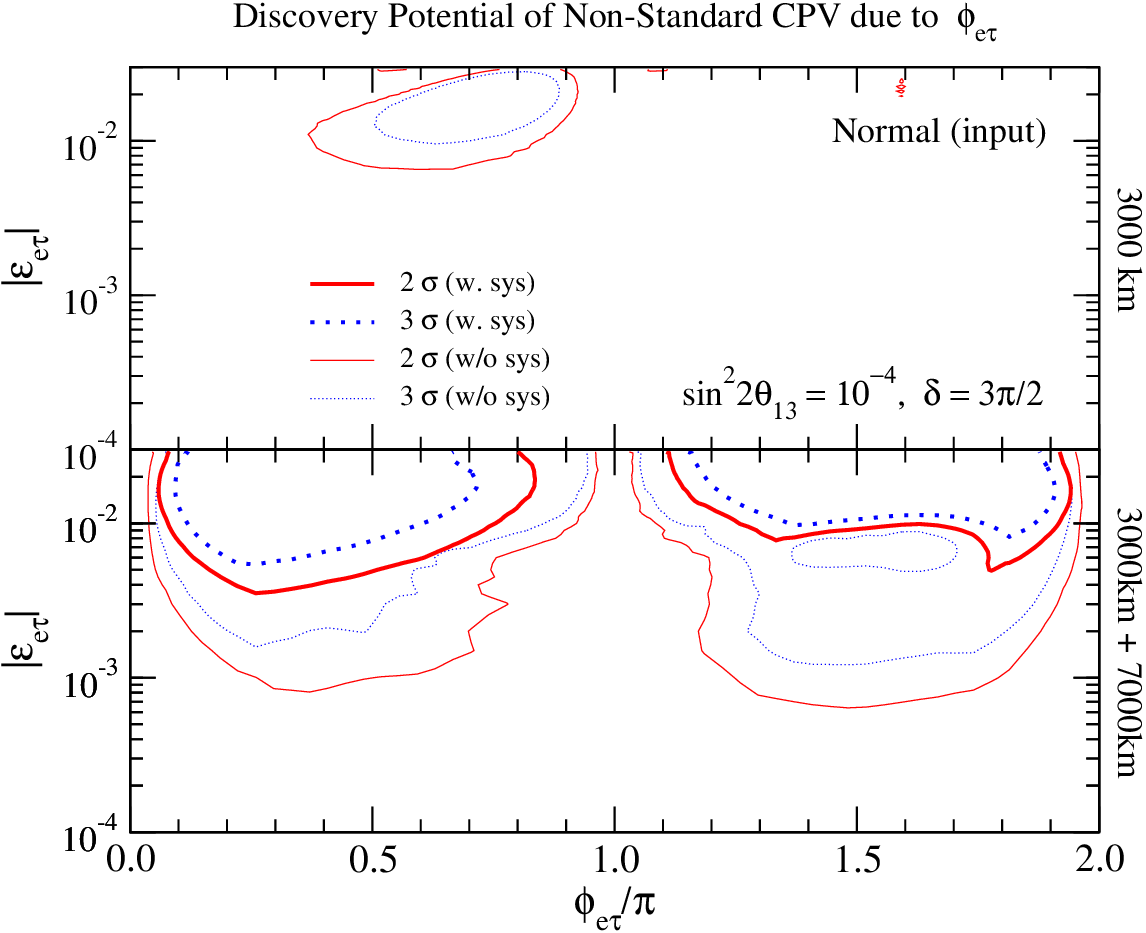}
\end{center}
\vglue -0.75cm
\caption{ The same as in Fig.~\ref{NSI-CPV-sensitivity-et} but for
  $\sin^2 2\theta_{13} = 10^{-4}$.  }
\label{NSI-CPV-sensitivity-et-0.0001}
\end{figure}

After combining the results from two detectors, we conclude that if
$0.2 \lsim \phi_{e\mu}/\pi \lsim 0.8$ or 
$1.2 \lsim \phi_{e\mu}/\pi \lsim 1.9 $ 
we can identify the effect of non-standard CP violation
down to $\vert \varepsilon_{e\mu}\vert \sim$ a few to several $\times 10^{-3}$
at 3 $\sigma$ CL for Choice R of the $\chi^2$ analysis, depending on
the values of $\delta$ and $\sin^2 2\theta_{13}$.  
%
For Choice O the sensitivity is better by a factor up to several
depending upon the SI parameters.
We observe that if $\phi_{e\mu}/\pi \lsim 0.2$ or $0.8 \lsim
\phi_{e\mu}/\pi \lsim 1.2$ or $\phi_{e\mu}/\pi \gsim 1.9$ it 
seems practically impossible to establish non-standard CP violation.
Here the input choice for the mass hierarchy is normal, but the 
results are qualitatively the same in case we take the input mass hierarchy 
to be inverted and simultaneously change the input 
$\delta \to \pi -\delta$.

In Figs.~\ref{NSI-CPV-sensitivity-et} and
\ref{NSI-CPV-sensitivity-et-0.0001} we show similar plots but for the
discovery of non-standard CP violation associated with $\phi_{e\tau}$.
We note that for these cases the effect of the synergy of combining
two detectors is even larger, as expected.
We observe that, compared to the $\varepsilon_{e\mu}$ system, despite
the discovery potential of non-standard CP violation with the detector
at 3000 km alone is rather poor especially for $\phi_{e\tau}/\pi \gsim
1$, after combining two detectors, the final sensitivities are not
very much different from that for the $\varepsilon_{e\mu}$ system.
For $0.1 \lsim \phi_{e\tau}/\pi \lsim 0.8$ or $1.2 \lsim
\phi_{e\tau}/\pi \lsim 1.9$, with Choice R, we can identify the effect
of non-standard CP violation down to $\vert \varepsilon_{e\tau}\vert
\sim 10^{-2}$ or smaller at 3 $\sigma$ CL, depending on the values of
$\delta$ and $\sin^2 2\theta_{13}$.
Similar to the case for the system with $\varepsilon_{e\mu}$,
for $\phi_{e\tau}/\pi \lsim 0.1$, $0.8 \lsim \phi_{e\tau}/\pi \lsim 1.2$, or 
$\phi_{e\tau}/\pi \gsim 1.9$, it seems practically impossible to 
establish non-standard CP violation. 

The results for the sensitivity to non-standard CP violation for the 
inverted mass hierarchy is similar to that for the case of 
the normal mass hierarchy shown in this paper.

We observe that including the detector efficiencies, event backgrounds
and the systematic uncertainties reduce the sensitivities to NSI and
the non-standard CP violation by a factor of a few to several (even an
order of magnitude in a rare case), as expected.  But, their effects
do not appear to change the qualitative features of the sensitivity
contours.

Our results for $\varepsilon_{e\tau}$ can be compared with the ones
found in \cite{winter-nonstandardCP}.  From Fig. 1 of this reference
the non-standard CP violation (due to $\phi_{e\tau} \ne 0, \pi $) can
be established at 3 $\sigma$ for $|\varepsilon_{e\tau}|$ larger than
$\sim (7-10) \times 10^{-3}$ when the true value of $\phi_{e\tau}$ is
not so close to $0$ or $\pi$ for the case where the true value of
$\theta_{13} =0$.  The sensitivity obtained in
\cite{winter-nonstandardCP} is in good agreement with ours with Choice R.

\section{Impact of NSI on Standard CP violation and Mass Hierarchy}
\label{sec:impact}


At this point it is important to examine whether NSI can obscure the 
discovery of the standard CP violation due to $\delta$ and the neutrino 
mass hierarchy, and if yes to what extent.

\subsection{Impact of NSI on the establishment of the standard 
CP violation}
\label{sec:impact-cpv}

\begin{figure}[bhtp]
\begin{center}
\includegraphics[width=0.58\textwidth]{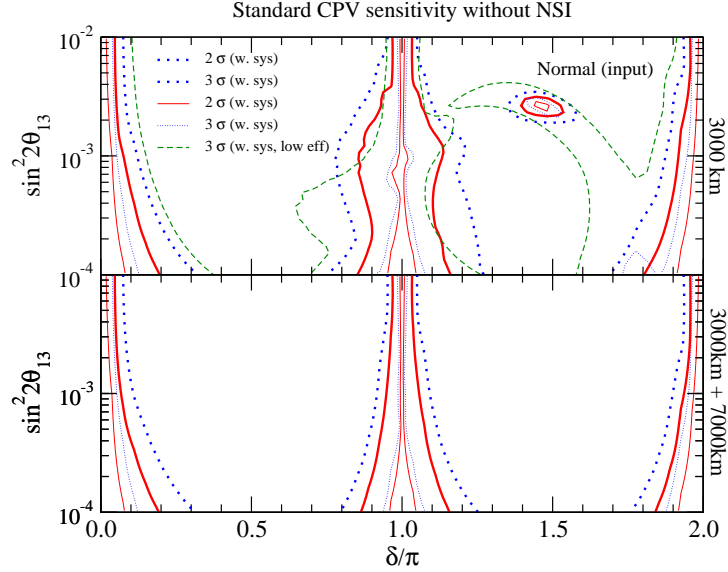}
\end{center}
\vglue -0.6cm
\caption{ Sensitivity to discovery of standard CP violation. Here no
  effect of NSI is assumed both in the input data and in the fit.  The
  upper panel shows the case where only the detector at 3000 km is
  considered, whereas the lower panel is the case corresponding to the
  combination of detectors at two different baselines.  For the sake
  of comparison, only for the case of 3000 km, we also show, by the
  green dashed curves, the case with lower detection efficiencies as
  for Fig. 81 of ~\cite{ISS-nufact}.  }
\label{cpv-sensitivity-noNSI}
\end{figure}

First let us consider the sensitivity of our setup to $\delta$ without
NSI.  We show, in Fig.~\ref{cpv-sensitivity-noNSI}, in the plane of
true values (input) of $\delta$ and $\sin^2 2\theta_{13}$, the regions
where CP violation can be established at 2 and 3 $\sigma$ CL for 1
DOF.  For this plot we assumed that there is no effect of the NSI both
in the input data and in the fit.  We determine the 2 (3) $\sigma$ CL
regions by freely varying $\theta_{13}$, $\delta$ and the mass
hierarchy in fitting the input data with the condition,
$\chi^2_{\text{min}}(\delta = 0 \text{ or } \pi) -
\chi^2_{\text{min}}(\text{true value of } \delta) > 4 (9)$, for 1 DOF.
The upper panel in Fig.~\ref{cpv-sensitivity-noNSI} shows the case where 
only the detector at 3000 km is considered whereas the lower panel is the 
case corresponding to the combination of detectors at two baselines. 
Note that, if $\delta$ is very close to 0 or $\pi$, it is impossible
to establish CP violation, a well known fact. 
In the case where only one detector at 3000 km is assumed, there is a
small region at around $\delta \sim 3\pi/2$ and $\sin^2 2\theta_{13}
\sim 3 \times 10^{-3}$ where the sensitivity is significantly reduced.
The loss of CP sensitivity at these particular values of $\delta$ and
$\theta_{13}$ occurs when the mass hierarchy is unknown. This fact has
been noted before, see Fig.~85 in Ref.~\cite{ISS-nufact} and comments
in this reference.

%
\begin{figure}[bhtp]
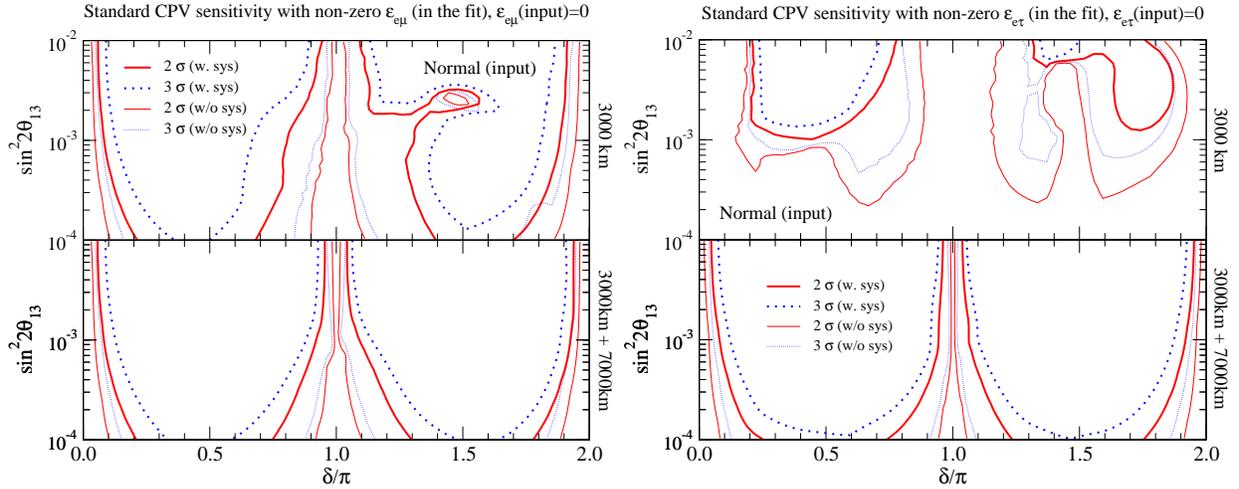

\begin{center}
\includegraphics[width=0.49\textwidth]{CPV_NSI_em_v3.eps}
\includegraphics[width=0.49\textwidth]{CPV_NSI_et_v3.eps}
\end{center}
\vglue -0.75cm
\caption{ Similar plots as in Fig.~\ref{cpv-sensitivity-noNSI} but
  with non-zero NSI allowed in the fit; The input data was generated
  without NSI but non-zero values of $\varepsilon_{e\mu}$ (left panel)
  and $\varepsilon_{e\tau}$ (right panel) were allowed in the fit.  }
\label{cpv-sensitivity-NSI}
\end{figure}

In passing we remark that there exist significant differences between
the sensitivity regions given in upper panels of
Fig.~\ref{cpv-sensitivity-noNSI} (i.e., blue dotted $3\sigma$ curves)
and Fig.~85 in Ref.~\cite{ISS-nufact}.  We note, however, that if we
assume the same detection efficiency as used in
Ref.~\cite{ISS-nufact}, we obtained roughly the same regions as
shown by the green dashed curves in
Fig.~\ref{cpv-sensitivity-noNSI}.\footnote{
As can be seen in Fig. 81 of \cite{ISS-nufact}, the efficiency used in 
this reference is roughly a factor of 2 lower than the one we 
take in this paper. 
Our efficiency is an energy independent approximation to the one given in 
Ref.~\cite{Abe:2007bi}.  
}
%
In comparing our results to the one in \cite{ISS-nufact} we
should also take into account the difference in baseline; Ours is
3000 km whereas the one in Fig.~81 of Ref.~\cite{ISS-nufact} is 4000
km.

Let us now discuss how this result can be affected by the presence of
NSI effects below the discovery sensitivity.  We show in
Fig.~\ref{cpv-sensitivity-NSI}, similar plots as seen in
Fig.~\ref{cpv-sensitivity-noNSI} but assuming the presence of non-zero
$\varepsilon_{e\mu}$ (left panel) and $\varepsilon_{e\tau}$ (right
panel) in the fit.
At each point of $\delta$ and $\sin^2 2\theta_{13}$ in this plot, as
in the previous case, we have generated the input data without NSI but
in fitting the data, we vary freely not only $\delta$, $\theta_{13}$
and the mass hierarchy choice, but also the values of
$\varepsilon_{e\mu}$ (or of $\varepsilon_{e\tau}$). This is done in
order to see to what extent the presence of NSI in the fit aggravates
the discovery potential for (standard) CP violation.

For the case where non-zero $\varepsilon_{e\mu}$ is assumed in the
fit, by comparing the results shown in
Fig.~\ref{cpv-sensitivity-noNSI} with the ones shown in the left panel
of Fig.~\ref{cpv-sensitivity-NSI}, we can conclude that the
sensitivity regions for discovery of standard CP violation are not
strongly affected by the presence of small NSI effects (below their
discovery level). Neither if we have only one detector at 3000 km but
particularly so if an additional one at 7000 km is combined.
However, this conclusion is somewhat different when a non-zero
$\varepsilon_{e\tau}$ is assumed in the fit, as we can see in the
right panel of Fig.~\ref{cpv-sensitivity-NSI}.  In this case, we
conclude that the region of sensitivity to standard CP violation is
significantly diminished if only the near detector at 3000 km is
considered.  It is even more so for Choice R of the $\chi^2$
analysis.  Fortunately, a similarly good sensitivity to CP
violation as in the case without NSI is recovered after the
combination with the far detector, as can be seen in the right-lower
panel in Fig.~\ref{cpv-sensitivity-NSI}.

We have repeated the same exercise for the case where the input mass
hierarchy is inverted. In this case, even in the standard scenario (no
NSI at all) the sensitivity region is reduced if only the near
detector is considered, but if one includes the far detector the
sensitivity becomes comparable to the one shown in
Fig.~\ref{cpv-sensitivity-noNSI} for the normal hierarchy. In fact it
is even better for $\delta \sim 0$ and slightly worse for $\delta \sim
\pi$.  Here, again the effect of non-zero $\varepsilon_{e\mu}$ and
$\varepsilon_{e\tau}$ at the single or the combined baselines is similar 
to the one for the normal mass hierarchy.

Let us also consider the case where the input values of NSI parameters
take non-zero values which are within the discovery reach.  Here, as
in the previous case, we vary freely the SI and NSI parameters in the
fit.  We have obtained qualitatively similar results to those without
input value of NSI (since they are similar, we do not show the plots).
We observe in this exercise that the difference from the previous
results without input NSI at 3000 km is quite small for
$\varepsilon_{e\mu}$, but it is not so small for the case with
$\varepsilon_{e\tau}$ for the values of the input NSI parameters
considered.
Fortunately, when two detectors are combined, the impact of non-zero
NSI on CP violation sensitivity does not appear to be so large within
values in the discovery reach of our setup.
We have verified that for the inverted mass hierarchy in the case 
on non-zero input NSI we also reach  very similar conclusions.

\subsection{Impact of NSI on the resolution of neutrino mass hierarchy}
\label{sec:impact-mh}

It is generally believed that the neutrino mass hierarchy can be
determined ``relatively easily'' in neutrino factory measurements
because of sufficiently strong earth matter effect due to the long
baseline.  This is in fact true when only SI are at play.  However,
once a new type of degeneracy, which involves both the normal and the
inverted mass hierarchies, is uncovered in systems subject to NSI, it
is legitimate to ask whether it can affect resolution of the mass
hierarchy, and if yes, how seriously.
Because of the feature discussed in Sec.~\ref{subsec:sens-et}, we anticipate that the presence of $\varepsilon_{e\tau}$ can be harmful.  
Even if we assume $\vert \varepsilon_{e\tau}\vert$ to be below its discovery
potential value, the solution of the mass hierarchy with a single
detector at 3000 km is unsure and the need for a second detector at
7000 km seems to be imperative.

We study the region in the input parameter plane $\delta - \sin^2
2\theta_{13}$ where the neutrino mass hierarchy can be determined by
generating a data set assuming a certain input mass hierarchy and
trying to fit the data with opposite hierarchy.  We say that the
mass hierarchy can be determined at 2 (3) $\sigma$ CL for 1 DOF if
$\chi^2_{\text{min}}(\text{opposite hierarchy})-
\chi^2_{\text{min}}(\text{input hierarchy})> 4 (9)$.

\begin{figure}[bhtp]
\begin{center}
\includegraphics[width=0.95\textwidth]{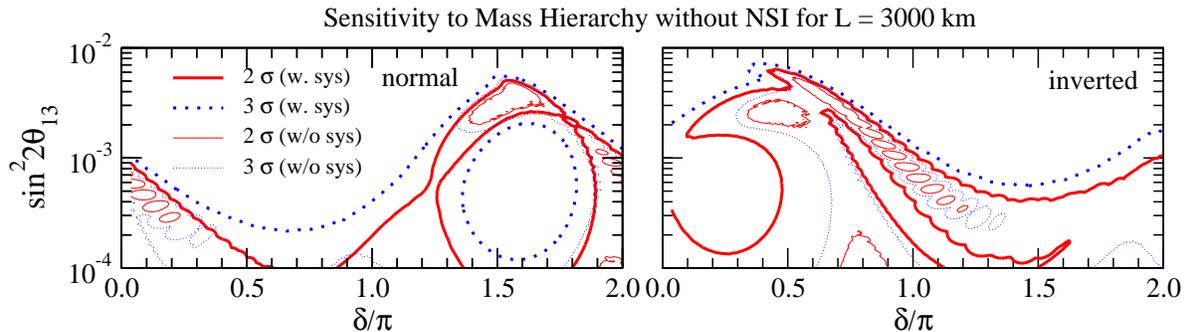}
\end{center}
\vglue -0.75cm
\caption{Regions in the $\delta$ versus $\sin^2 2\theta_{13}$ plane
  where the neutrino mass hierarchy can be established at 2 and 3
  $\sigma$ CL for 1 DOF with a single detector at 3000 km, for the
  case where no NSI effect is present.  In the left (right) panel the
  input mass hierarchy is taken to be normal (inverted).  It can be
  shown that the mass hierarchy is resolved in the whole region if the
  far detector is combined.  }
\label{Mass-sensitivity-noNSI}
\end{figure}

First let us investigate the sensitivity without considering NSI,
which means that it is absent not only in the input but also in the
fit.  In Fig.~\ref{Mass-sensitivity-noNSI} we show the sensitivity
region in the $\delta - \sin^2 2\theta_{13}$ plane where the mass
hierarchy can be determined at 2 and 3 $\sigma$ CL for 1 DOF with a
single detector at 3000 km.  In the left and the right panels the
input mass hierarchy was taken to be normal and inverted,
respectively.  We observe that in both cases there are small regions
in this plane in which the mass hierarchy cannot be determined.
However, we have checked that if an additional detector at 7000 km is
included in the analysis the mass hierarchy can be solved in the
entire plane covered in the analysis.
For the mass hierarchy determination, the impact of the non-perfect
detection efficiencies, non-zero backgrounds and systematic
uncertainties is larger compared to the standard CP violation
sensitivities (see Fig.~\ref{cpv-sensitivity-noNSI}).
%

%
\begin{figure}[bhtp]
\begin{center}
\vglue -0.5cm
 \includegraphics[width=0.95\textwidth]{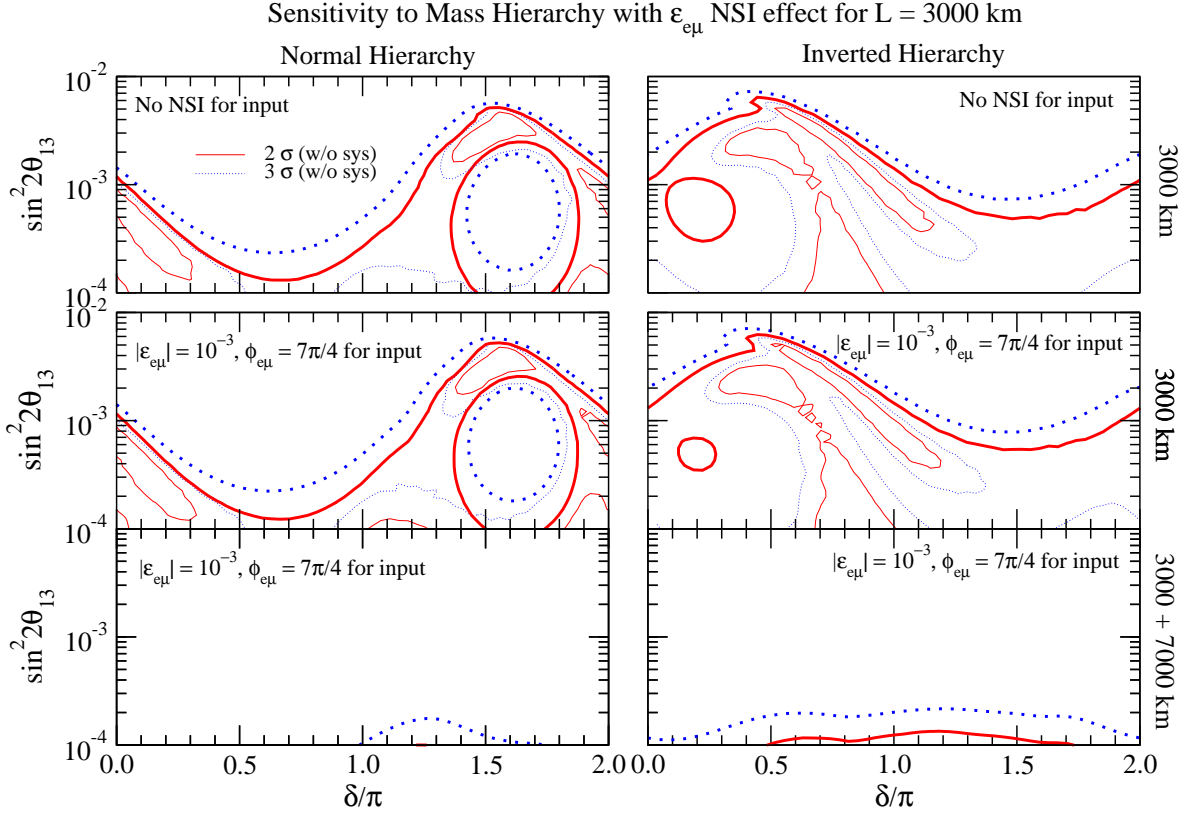}
\end{center}
\vglue -0.75cm
\caption{Same as in Fig.~\ref{Mass-sensitivity-noNSI} but for non-zero
  $\varepsilon_{e\mu}$. In the upper panels the input
  $\varepsilon_{e\mu}=0$, but we allow it to be non-zero in the fit.
  In the lower panels the input is $\vert \varepsilon_{e\mu} \vert=
  10^{-3}$ and $\phi_{e\mu}=7\pi/4$.  }
\label{Mass-sensitivity-NSI}
\end{figure}
%
\begin{figure}[bhtp]
\begin{center}
\vglue -0.5cm
\includegraphics[width=0.95\textwidth]{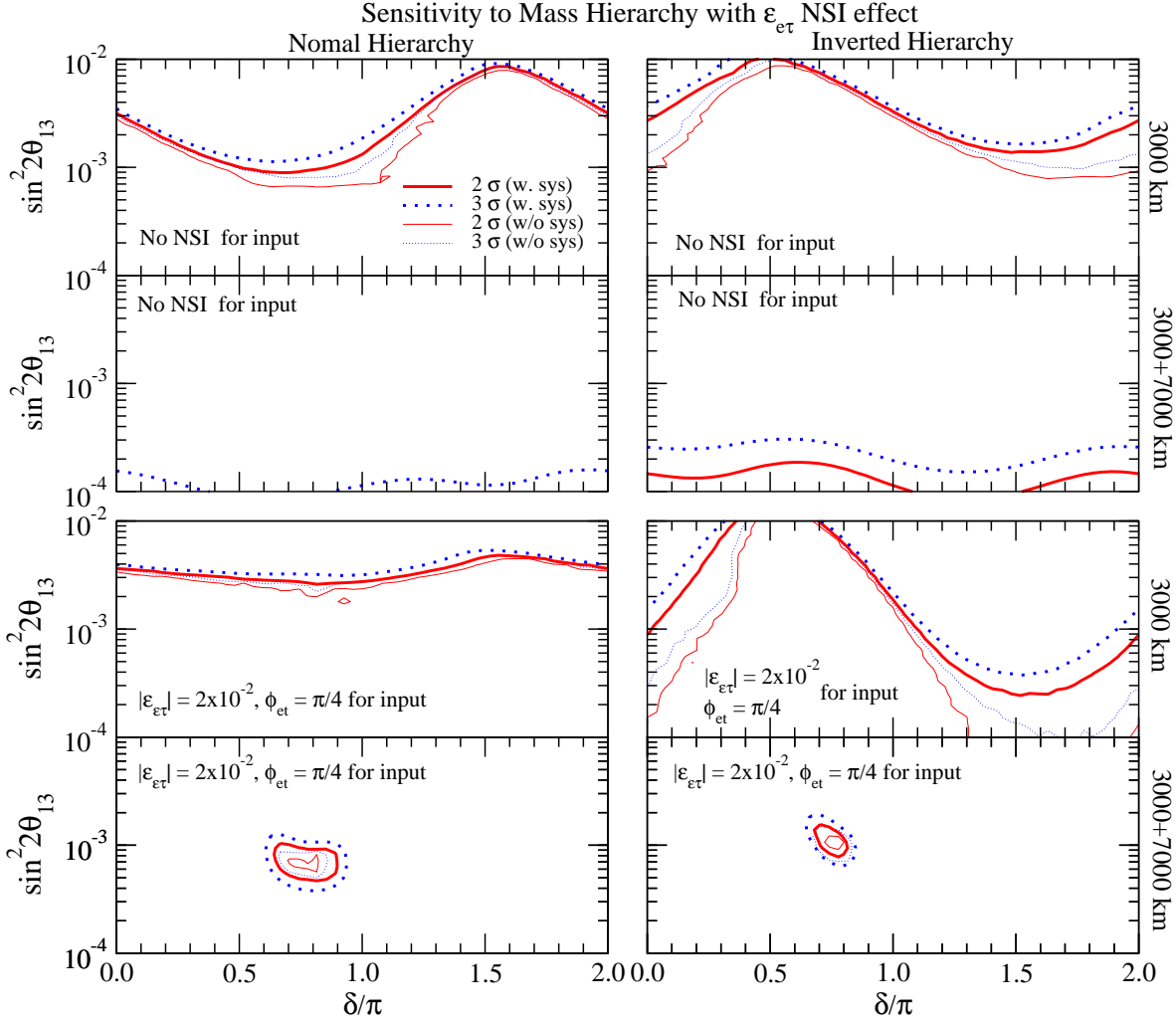}
\end{center}
\vglue -0.7cm
\caption{Similar plots as in Fig.~\ref{Mass-sensitivity-NSI} but for
  non-zero $\varepsilon_{e\tau}$. In the upper panels the input
  $\varepsilon_{e\tau}=0$, but we allow it to be non-zero in the fit.
  In the middle panels the input is $\vert \varepsilon_{e\tau} \vert=
  2\times 10^{-2}$ and $\phi_{e\tau}=\pi/4$. In the lower panels we
  present the result of the combination of the detectors at 3000 and
  7000 km for the input ($\vert \varepsilon_{e\tau} \vert= 2\times
  10^{-2}$, $\phi_{e\tau}=\pi/4$). }
\label{Mass-sensitivity-NSI2}
\end{figure}
%
Let us now switch on NSI effects and see their impact on the mass
hierarchy determination.  In Fig.~\ref{Mass-sensitivity-NSI}, we show
similar plots as in Fig.~\ref{Mass-sensitivity-noNSI} for non-zero
$\varepsilon_{e\mu}$.  In the upper panels of
Fig.~\ref{Mass-sensitivity-NSI} we have considered the input
$\varepsilon_{e\mu}$ to be below its discovery limit (null for all
practical purposes) but allowed non-vanishing $\varepsilon_{e\mu}$ in
the fit.  In the lower panels of Fig.~\ref{Mass-sensitivity-NSI} we
take the input $\varepsilon_{e\mu}$ to be in the discovery region of
our setup, $\vert \varepsilon_{e\mu} \vert= 10^{-3}$ and
$\phi_{e\mu}=7\pi/4$, and vary freely the SI and NSI parameters in the
fit.
In both cases, in the region of $\varepsilon_{e\mu}$ below or within
discovery reach, we observe a slight but not very significant decrease
of the sensitivity for both the normal (left panels) and the inverted
(right panels) mass hierarchies. The regions where sensitivity loss is
observed are limited to the area around no sensitivity region without
NSI in Fig.~\ref{Mass-sensitivity-noNSI}.
We note that the impact of turning on the systematic uncertainties in
the manner of Choice R produces a sizable effect.
We note that if $\sin^2 2\theta_{13} \lsim 10^{-4}$, for some range 
of $\delta$ the mass hierarchy can not be determined even if 
we include the far detector at 7000 km
(see the bottom panels of Fig.~\ref{Mass-sensitivity-NSI}).

Now let us examine the case where $\varepsilon_{e\tau}$ is turned on,
whose results are shown in Fig.~\ref{Mass-sensitivity-NSI2}.  
The organization of Fig.~\ref{Mass-sensitivity-NSI2} is as follows; 
The 1st (2nd) row panels are for case where no NSI, or the one 
below discovery limit, is assumed in the input, 
for the near detector alone (near and far ones combined).
The 3rd (4th) row panels are for non-vanishing
input value of $\varepsilon_{e\tau}$ within the discovery region both
at the near detector alone (combined ones).  

Let us first look at the case where no NSI is considered in the input. 
If we consider only the near detector at $L=3000$ km, the region where the 
mass hierarchy can be disentangled severely shrunk when NSI is switched on. 
It is the case for both input hierarchies. 
We note that, compared to the case of $\varepsilon_{e\mu}$,  
the effects of the systematic uncertainties are 
least prominent for the near detector alone (see 
the 1st row of Fig.~\ref{Mass-sensitivity-NSI2}). 
It becomes impossible to resolve it if $\sin^2 2\theta_{13} \lsim 10^{-3}$, 
and the sensitivity is lost for some values of $\delta$ at larger values of 
$\sin^2 2\theta_{13}$. 
However, by looking the 2nd row of 
Fig.~\ref{Mass-sensitivity-NSI2}
we observe that the power of the synergy of the two
detectors is quite strong so that the combination of the two detector 
allows us to determine the mass hierarchy for most of the parameter regions
we considered in this paper 
even in the presence of very small $\varepsilon_{e\tau}$.

Then, what happens if (input vale of) 
$\varepsilon_{e\tau}$ is not so small?  
As seen in the 3rd row panels of Fig.~\ref{Mass-sensitivity-NSI2}, 
it strongly aggravates the mass hierarchy determination.
For instance, if the input hierarchy is normal (left panel) the
hierarchy is undetermined if $\sin^2 2\theta_{13} \lsim $ a few
$\times 10^{-3}$.  If the input hierarchy is inverted (right panel)
sensitivity to the hierarchy very much depends on $\delta$ and it can
be determined down to $\sin^2 2\theta_{13} \lsim 10^{-3}$ only in a
limited region of $\delta$.
In the bottom panels of Fig.~\ref{Mass-sensitivity-NSI2} we show the result 
of combining the near and the far detectors. Even with this combination 
there is a small region where the mass hierarchy is not determined.

We note that compared to the $\varepsilon_{e\mu}$ element, the impact
of the detection efficiencies, background and systematic uncertainties
is relatively small.  It appears that this is partly because the
impact of non-zero $\varepsilon_{e\tau}$ itself is already large
before inclusion of the uncertainties.

\section{Conclusions}
\label{sec:conclusions}

In this paper, we have studied the question of how to distinguish
between physics effects due to SI and NSI in neutrino
oscillations.  They include, most notably, discriminating effects of
possible non-standard CP violation due to phases associated with the
NSI elements from the standard CP violation caused by the lepton KM
phase $\delta$, and vice versa.
They also include a related question of how NSI affects the  
determination of the neutrino mass hierarchy. 
Our study was done in the context of a future neutrino factory 
endowed with an intense muon storage ring of 50 GeV,
delivering $10^{21}$ useful $\mu$-decays a year and 
operating for 4 years in neutrino and 4 years in antineutrino modes. 
We assumed two magnetized iron detectors, one at 3000 km and the other 
at 7000 km from the neutrino source, each with 50 kton fiducial mass. 
We considered only data from the golden channels $\nu_e \to \nu_\mu$
and $\bar \nu_e \to \bar\nu_\mu$.

In this work we have considered NSI effects only in the neutrino
propagation, and turned on only one of the relevant NSI elements for
the golden channel, $\varepsilon_{e\mu}=\vert \varepsilon_{e\mu}\vert
e^{i\phi_{e\mu}}$ or $\varepsilon_{e\tau}=\vert
\varepsilon_{e\tau}\vert e^{i\phi_{e\tau}}$.
Since neutrino oscillation in such a system is very complicated we
have studied the effect of a single $\varepsilon$ (two real
parameters) at a time.  We have fixed the values of the standard
oscillation parameters at their current best fit values, expect for
$\sin^2 2\theta_{13}$, $\delta$ and sign of $\Delta m^2_{31}$, which
we also vary freely in our analysis.

Prior to the presentation of the full discovery potential of NSI and
SI parameters we have analyzed the structure of neutrino oscillation
with NSI.  We have utilized the bi-probability plot in $P(\nu_e
\rightarrow \nu_\mu) - P(\bar{\nu}_e \rightarrow \bar{\nu}_\mu) $
space, which are drawn by varying either NSI phase $\phi$ or the KM
phase $\delta$, to illuminate the characteristic properties of the
neutrino oscillation in systems with NSI. It revealed to be a powerful
tool to understand the synergy between the two detectors, and the
difference between the $\varepsilon_{e\mu}$ and $\varepsilon_{e\tau}$
systems.
Furthermore, we have given an overview of the sensitivities to NSI and the 
SI parameters by classifying the data set of allowed region contours into 
four different types depending upon the degree of synergy between the near and 
far detector measurements. 

Most notably, we have observed in the near detector measurement the
phenomenon of parameter degeneracy which is similar to the one in
standard neutrino oscillation but with active participation of NSI
effects.  Though it is a highly complicated problem, we were able to
control it analytically in the restricted setting in the presence of
only a single type of $\varepsilon$.  The NSI enriched parameter
degeneracy discussed in this paper contains the $\Delta m^2_{31}$-sign
flipped degeneracy, the intrinsic one, and the one called the
$\phi-$degeneracy. The last one is the special case of the more
generic cases, but it is best characterized as the $\phi-$degeneracy
because the solutions are different essentially only by the values of
$\phi$.  We have shown that these degeneracies are very robust against
spectrum analysis and so can potentially disrupt the resolution of CP
violation and the mass hierarchy.
However, it is also observed that in many cases the parameter
degeneracies can be resolved by adding the far detector.  This is why,
especially for NSI due to $\varepsilon_{e\tau}$,  the inclusion of the
7000 km detector is imperative.

We have analyzed the discovery potential of NSI with our setup, by
performing two typical cases of $\chi ^2$ analyses, Choice R and
Choice O as defined in Sec.~\ref{sec:method}, with and without
detection efficiency function, background and systematic
uncertainties, respectively.  In the range of SI parameters covered in
this paper and independent of the neutrino mass hierarchy, we have
established that measurements performed by a single detector at 3000
km can discover NSI due to $\varepsilon_{e\mu}$ down to
$\vert\varepsilon_{e\mu}\vert \sim 10^{-3}-10^{-4}$ (Choice O) and a
few to several times larger than these for Choice R, the exact value
depending on $\phi_{e\mu}$.  In the $\varepsilon_{e\mu}$ system,
inclusion of the far detector does not improve this much.
However, in the case of the discovery of NSI due to
$\varepsilon_{e\tau}$ the synergy between the two detectors is very
strong.  Although one can only hope to reach down to $\vert
\varepsilon_{e\tau}\vert \sim$ a few $\times 10^{-2}$ (Choice R) with
the near detector at 3000 km, the inclusion of the far detector makes
the sensitivity to $\varepsilon_{e\tau}$ similar to that for
$\varepsilon_{e\mu}$.  We have identified through the discussions in
Secs.~\ref{sec:nsifeatures} and \ref{sec:overview} why such disparity
between sensitivities in the $\varepsilon_{e\mu}$ and the
$\varepsilon_{e\tau}$ systems arises.  We note that taking into
account the non-perfect detection efficiencies, non-zero backgrounds
and systematic uncertainties change sensitivities a few to one order 
of magnitude.

We have also investigated the potential to unravel CP violation
associated with NSI using such an experiment. We have concluded that
to establish CP violation associated with $\phi_{e\mu}$ the addition
of the second detector indeed helps.  For both mass hierarchies, if
$0.2 \lsim \phi_{e\mu}/\pi \lsim 0.8$ or $1.2 \lsim \phi_{e\mu}/\pi
\lsim 1.9 $ we can discover non-standard CP violation promoted by 
this phase down to $\vert \varepsilon_{e\mu}\vert \sim 
\text{(a few -  several)}\times 10^{-3} $ (for Choice R, depending on 
$\sin^2 2\theta_{13}$ and $\delta$) at 3 $\sigma$ CL.
For CP violation associated with $\phi_{e\tau}$ the synergy between
the two detectors is more efficient than the one in the $\phi_{e\mu}$
system.  The near detector alone gives poor sensitivities but after
combining two detectors, if $0.1 \lsim \phi_{e\tau}/\pi \lsim 0.8$ or
$1.2 \lsim \phi_{e\tau}/\pi \lsim 1.9$, we can identify the effect of
non-standard CP violation down to $\vert \varepsilon_{e\tau}\vert \sim
\text{ a few } \times 10^{-3} - 10^{-2}$ (for Choice R, depending on
$\sin^2 2\theta_{13}$ and $\delta$) at 3 $\sigma$ CL, independently of
the neutrino mass hierarchy.  Improvement of the systematic uncertainties 
does affect the sensitivity to non-standard CP violation by a factor
of a few but not so dramatically.

We have also checked to what extent the existence of NSI can aggravate
the discovery potential of standard CP violation in the $\delta -
\sin^2 2\theta_{13}$ plane.  We have observed that NSI effects induced
by $\varepsilon_{e\mu}$ (either with a magnitude below or above the
discovery reach of our setup) will not spoil much the discovery
potential of standard CP violation with respect to the standard case,
even if we have a single detector at 3000 km.  Nevertheless, NSI
effects induced by $\varepsilon_{e\tau}$ are potentially harmful if
only the near detector is considered, even for values of $\vert
\varepsilon_{e\tau}\vert$ below the sensitivity reach.  Fortunately,
the discovery potential possessed by the setting without NSI is
essentially recovered by the inclusion of an additional detector at
7000 km.  We have checked that these conclusions do not depend on the
input neutrino mass hierarchy.

Finally, we have studied the impact of NSI on the resolution of the
neutrino mass hierarchy. For non-zero $\varepsilon_{e\mu}$, with the
near detector at 3000 km, we observe a slight but not very significant
decrease of the region in the $\delta - \sin^2 2\theta_{13}$ plane
where the neutrino mass hierarchy can be established.  By adding the
far detector at 7000 km, one can distinguish the mass hierarchy for
almost all values of $\delta$ if $\sin^2 2\theta_{13} \gsim 10^{-4}$. 
For non-zero $\varepsilon_{e\tau}$, the region where the mass hierarchy
can be determined with the near detector alone severely shrinks.  Here
again, the power of the combination of two detectors is sufficiently
strong, allowing the mass hierarchy to be determined in almost the whole
parameter space of $\delta$ and $\theta_{13}$ considered in this work
(except for a small region if $\vert\varepsilon_{e\tau}\vert$ is
rather large).
We observe that the impact of non-perfect detection, background and
systematic uncertainties on the determination of the mass hierarchy is
somewhat larger for the case where $\varepsilon_{e\mu}$ is present
than for the case where $\varepsilon_{e\tau}$ is, though the effect
of the latter is already large without these factors.

\vspace{-0.3cm}
\begin{acknowledgments} 
  \vspace{-0.3cm} We would like to thank N. Cipriano Ribeiro for his
  contribution in the early stages of this work.
Three of us (H.M., H.N. and  R.Z.F.) are grateful for 
the hospitality of the Theory Group of the
Fermi National Accelerator Laboratory during the summer of 2008,
where part of this research was completed. R.Z.F. is also grateful
for the hospitality of the Department of Physics of Tokyo
Metropolitan University where some part of this project was developed 
in November of 2008.  
A. M. G wants to thank the hospitality of the Instituto de 
F\'{\i}sica da Universidade de S\~ao Paulo, during his visit in 2007 
where the initial part of this project was developed.
This work was supported in part by KAKENHI, Grant-in-Aid for
Scientific Research No. 19340062, Grant-in-Aid for JSPS Fellows 
No. 209677, Japan Society for the Promotion of Science, 
Funda\c{c}\~ao de Amparo \`a Pesquisa do Estado de S\~ao Paulo (FAPESP), 
Funda\c{c}\~ao de Amparo \`a Pesquisa do Estado de Rio de Janeiro (FAPERJ),  
Conselho  Nacional de Ci\^encia e Tecnologia (CNPq),
HELEN Project, and Direccion Academica de Investigacion 
(DAI)-Pontificia Universidad Catolica del Peru.

\end{acknowledgments}

\appendix 

\section{Expression of the Appearance oscillation 
probability for the Bi-Probability plot}
\label{Pemu-biPform}

The appearance oscillation probability in the neutrino channel 
in system with $\varepsilon_{e\alpha}$ ($\alpha=\mu, \tau$) are given by 
\begin{eqnarray}
P( \nu_{e} \rightarrow \nu_{\mu} ) &=& 
\mathcal{A}_{e\alpha} + \mathcal{B}_{e\alpha} \cos (\delta + \phi)+ \mathcal{C}_{e\alpha} \sin (\delta + \phi) + 
\mathcal{R}_{e\alpha} \cos \phi+ \mathcal{I}_{e\alpha} \sin \phi,  
\label{Pemu-compact}
\end{eqnarray}
with the coefficients 
\begin{eqnarray} 
\mathcal{A}_{e\mu}  &=& 
4 s^2_{23} s^2_{13} 
\left( \frac{ \Delta m^2_{31} }{ a - \Delta m^2_{31} } \right)^2 
\sin^2 \left( \frac{ a - \Delta m^2_{31} }{ 4E } L \right) 
\nonumber \\
&+& 8 J_r \frac { \Delta m^2_{31}  \Delta m^2_{21}  }{  a ( a - \Delta m^2_{31} )} 
\sin \left(\frac{  a L}{4E}  \right) 
\sin \left( \frac{ a - \Delta m^2_{31} }{ 4E } L \right)
\cos \left(  \delta -  \frac{  \Delta m^2_{31} L}{4E}   \right) 
\nonumber \\
&+& 
4 c^2_{12} s^2_{12} c^2_{23} 
\left( \frac{ \Delta m^2_{21} } { a } \right)^2 
\sin^2 \left(\frac{ a L}{4E}  \right) 
\nonumber \\
&-& 4 s_{23}^4  |\varepsilon_{e\mu}|^2 
\left(  \frac{ a }{ a - \Delta m_{31}^2 } \right)^2 
\sin\left( \frac{\Delta m_{31}^2 L}{4E} \right) 
\cos\left( \frac{a L}{4E} \right) 
\sin  \left( \frac{ a - \Delta m^2_{31} }{ 4E } L \right) 
  \nonumber \\ 
&+& 4 |\varepsilon_{e\mu}|^2 
\left( \frac{ a - c_{23}^2 \Delta m_{31}^2 }{ a - \Delta m_{31}^2 } \right)^2  
\cos\left( \frac{\Delta m_{31}^2 L}{4E} \right) 
 \sin\left( \frac{a L}{4E} \right) 
 \sin \left( \frac{ a - \Delta m^2_{31} }{ 4E } L \right) 
  \nonumber \\ 
&+& 4 c_{23}^4 |\varepsilon_{e\mu}|^2 
 \sin \left( \frac{\Delta m_{31}^2 L}{4E} \right) 
 \sin\left( \frac{a L}{4E} \right) 
 \cos \left( \frac{ a - \Delta m^2_{31} }{ 4E } L \right), 
\end{eqnarray}
\begin{eqnarray}
\mathcal{B}_{e\mu} &=& 
8  s_{23} s_{13} \vert \varepsilon_{e\mu} \vert 
\frac{ \Delta m_{31}^2  }{ (a - \Delta m_{31}^2)^2}
 \sin \left( \frac{ a - \Delta m^2_{31} }{ 4E } L \right) 
\nonumber \\ 
&\times & 
\left[ 
(a - c_{23}^2 \Delta m_{31}^2) 
\cos\left( \frac{\Delta m_{31}^2 L}{4E} \right) 
 \sin\left( \frac{a L}{4E} \right) 
 - a s_{23}^2 
\sin\left( \frac{\Delta m_{31}^2 L}{4E} \right) \cos\left( \frac{a L}{4E} \right) 
\right], 
\end{eqnarray}
\begin{eqnarray}
\mathcal{C}_{e\mu} &=& 
8 c^2_{23} s_{23}  s_{13}  \vert \varepsilon_{e\mu} \vert 
\frac{ \Delta m_{31}^2  } {(a - \Delta m_{31}^2)} 
 \sin\left( \frac{\Delta m_{31}^2 L}{4E} \right) 
  \sin\left( \frac{a L}{4E} \right)
 \sin \left( \frac{ a - \Delta m^2_{31} }{ 4E } L \right), 
\end{eqnarray}
\begin{eqnarray}
\mathcal{R}_{e\mu} &=& 
8    c_{12} s_{12} c_{23}  \vert \varepsilon_{e\mu} \vert  
 \left( \frac{ \Delta m_{21}^2 } {a } \right) 
 \sin\left( \frac{a L}{4E} \right) 
  \nonumber \\ 
&& 
\hspace{-14mm}
\times \left[ 
\left( \frac{ a - c_{23}^2 \Delta m_{31}^2 }{ a - \Delta m_{31}^2 } \right)  
\cos\left( \frac{\Delta m_{31}^2 L}{4E} \right) 
 \sin \left( \frac{ a - \Delta m^2_{31} }{ 4E } L \right) 
+ c_{23}^2
\sin  \left( \frac{\Delta m_{31}^2 L}{4E} \right) 
 \cos \left( \frac{ a - \Delta m^2_{31} }{ 4E } L \right) 
\right], 
\nonumber \\ 
\end{eqnarray}
\begin{eqnarray}
\mathcal{I}_{e\mu} &=& 
 - 8 c_{12} s_{12} c_{23} s^2_{23}  
 \vert \varepsilon_{e\mu} \vert 
\left( \frac{ \Delta m_{21}^2 }{ a - \Delta m_{31}^2 } \right) 
 \sin \left( \frac{\Delta m_{31}^2 L}{4E} \right) 
  \sin \left( \frac{a L}{4E} \right)
 \sin \left( \frac{ a - \Delta m^2_{31} }{ 4E } L \right). 
\label{coefficient-emu}
\end{eqnarray}

The analogous expression of $P( \nu_{e} \rightarrow \nu_{\mu} )$ with
$ \varepsilon_{e\tau} $ can be obtained from (\ref{Pemu}), or by
noting that they originally appear in the particular combination of
the generalized atmospheric and the solar variables
\cite{NSI-perturbation} though it is obscured a little in
(\ref{Pemu}). (See Eq.~(6.5) in \cite{NSI-perturbation}.)
The probability for anti-neutrino can be obtained 
from the one for neutrinos by doing transformations 
$a \to -a$, $\delta \to - \delta$, and $\phi \to - \phi$.

\section{Understanding rotating ellipses in the bi-probability plot} 
\label{rotating}

It can be easily shown that the bi-probability trajectory drawn by
varying $\phi$ holding $\delta$ fixed (or, vice versa) takes the form
of ellipse.  We obtain the equation which determines the major (minor)
axis from (\ref{Pemu-biP-phi})
\begin{eqnarray}
\frac{\partial}{\partial \phi} \bigg[ (\mathcal{S} \cos \phi + \mathcal{T} \sin \phi)^2
 + (\bar{\mathcal{S}} \cos \phi - \bar{\mathcal{T}} \sin \phi)^2 \bigg]
=
0,
\end{eqnarray}
where
\begin{eqnarray}
\mathcal{S} 
&\equiv&
\mathcal{R} + \mathcal{B} \cos \delta + \mathcal{C} \sin \delta \nonumber \\
\mathcal{T}
&\equiv&
\mathcal{I} - \mathcal{B} \sin \delta + \mathcal{C} \cos \delta \nonumber\\
\bar{\mathcal{S}} 
&\equiv&
\bar{\mathcal{R}} + \bar{\mathcal{B}} \cos \delta - \bar{\mathcal{C}} \sin \delta \nonumber \\
\bar{\mathcal{T}}
&\equiv&
\bar{\mathcal{I}} + \bar{\mathcal{B}} \sin \delta + \bar{\mathcal{C}} \cos \delta.
\end{eqnarray}
Hence, the slope of the major axis, $\alpha_{\text{vary} \phi}$, is given by 
\begin{eqnarray}
\alpha_{\text{vary}\phi}
&=&
\frac{\bar{\mathcal{S}} \cos \phi_{\text{max}} - \bar{\mathcal{T}} \sin \phi_{\text{max}}}
 {\mathcal{S} \cos \phi_{\text{max}} + \mathcal{T} \sin \phi_{\text{max}}},
\label{slope-phi}
\end{eqnarray}
where
\begin{eqnarray}
\tan 2 \phi_{\text{max}} 
= 
2 \frac{\mathcal{S} \mathcal{T} - \bar{\mathcal{S}} \bar{\mathcal{T}}}
 {\mathcal{S}^2 - \mathcal{T}^2 + \bar{\mathcal{S}}^2 - \bar{\mathcal{T}}^2} ,
\end{eqnarray}
so the slope of the axis is independent of the value of
 $|\varepsilon_{e \alpha}|$.

\begin{figure}[bhtp]
\begin{minipage}{0.49\textwidth}
\begin{center}
\includegraphics[width=1.0\textwidth]{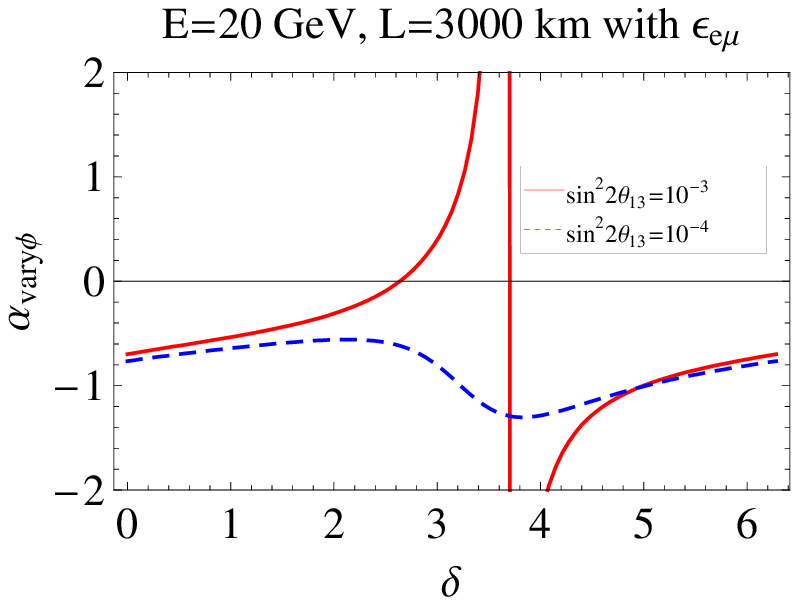}
\end{center}
\vglue 0.4cm
\end{minipage}
\begin{minipage}{0.49\textwidth}
\begin{center}
\includegraphics[width=1.0\textwidth]{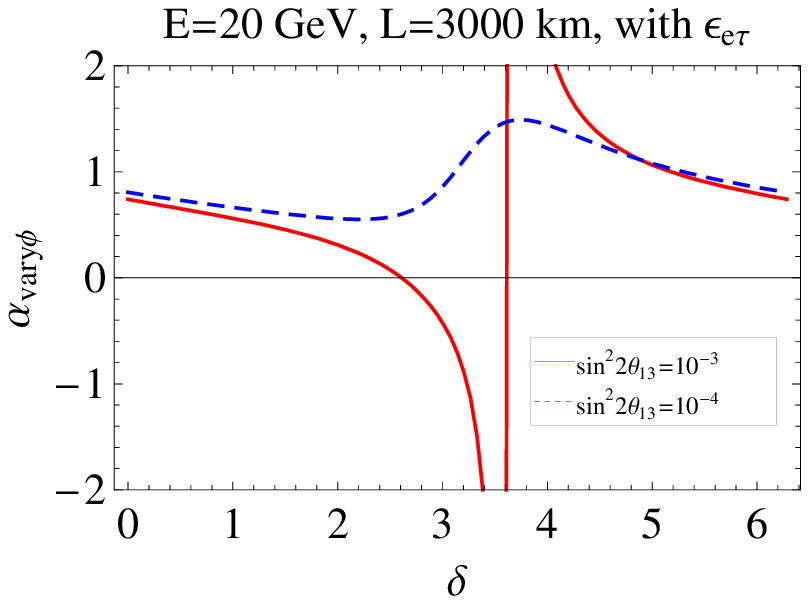}
\end{center}
\vglue 0.4cm
\end{minipage}
\vglue -0.8cm
\caption{
The slope of the major axis of the ellipse which is made by varying 
$\phi$ in the $\varepsilon_{e \mu}(\varepsilon_{e \tau})$ system. 
$\sin^2 2\theta_{13}=10^{-3}(10^{-4})$ for the red solid (blue dashed) line. 
$\alpha_{\text{vary} \phi}$ is independent of $|\varepsilon_{e \alpha}|$.
}
\label{alpha-phi}
\end{figure}

The $\delta$ dependence of $\alpha_{\text{vary} \phi}$ are plotted in 
Fig.~\ref{alpha-phi}. 
The slope of the major axis is mostly negative and positive in the 
$\varepsilon_{e\mu}$ and $\varepsilon_{e\tau}$ systems, respectively.
When $\theta_{13}$ is small, the coefficients of $\delta$ independent 
term ($\mathcal{R}$ and $\mathcal{I}$) relatively play the dominant role in 
(\ref{slope-phi}), the rotating behavior of the ellipse is soft. 
On the other hand, the slope of the major axis changes dramatically
in the region around $\delta = \pi$ with large $\theta_{13}$ as we 
saw in the Figs.~\ref{biP-3000km-em-20GeV} and \ref{biP-3000km-et-20GeV}. 

%
\begin{figure}[bhtp]
\vglue -0.2cm
\begin{minipage}{0.49\textwidth}
\begin{center}
\includegraphics[width=1.0\textwidth]{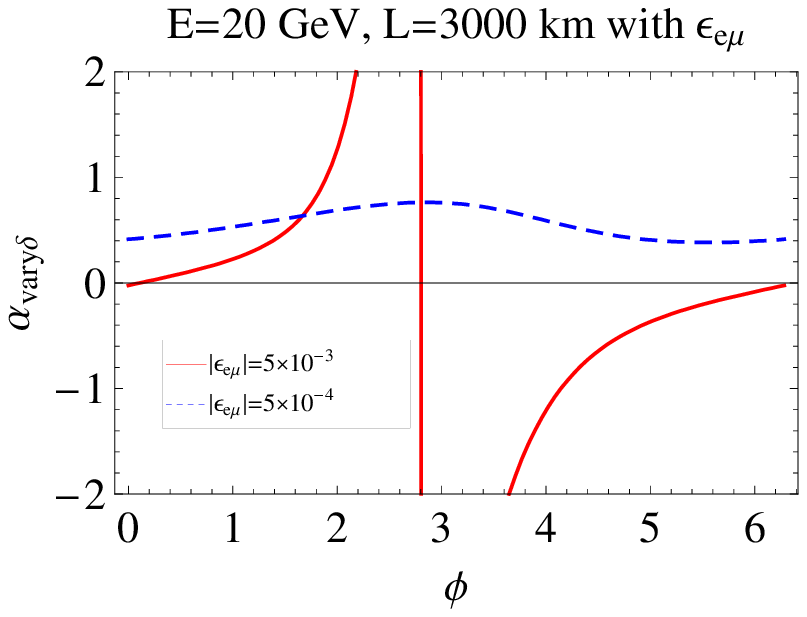}
\end{center}
\vglue 0.4cm
\end{minipage}
\begin{minipage}{0.49\textwidth}
\begin{center}
\includegraphics[width=1.0\textwidth]{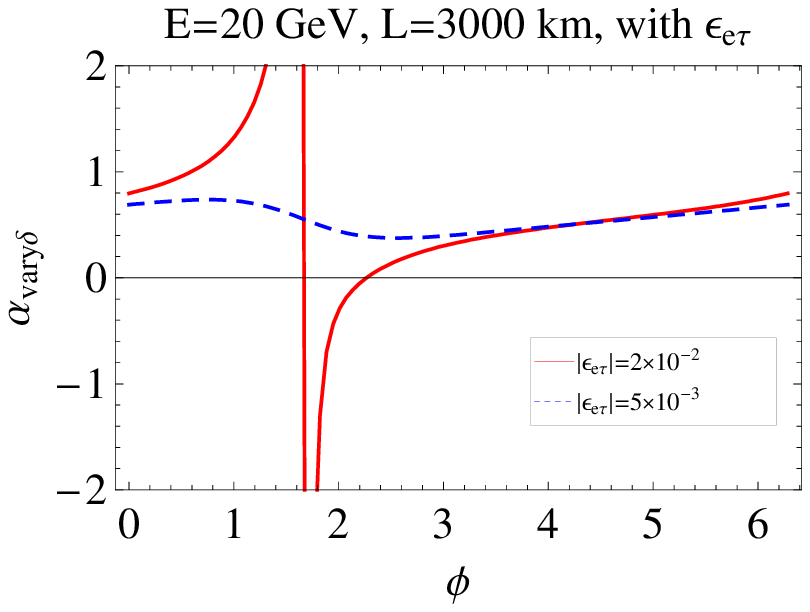}
\end{center}
\vglue 0.4cm
\end{minipage}
\vglue -0.8cm
\caption{
The slope of the major axis of the ellipse which is made by varying 
$\delta$ in the $\varepsilon_{e \mu}(\varepsilon_{e \tau})$ system. 
$|\varepsilon_{e \mu}|=5 \times 10^{-3}(5 \times 10^{-4})$ and 
$|\varepsilon_{e \tau}|=2 \times 10^{-2}(5 \times 10^{-3})$ for the 
red solid (blue dashed) line. 
$\alpha_{\text{vary} \delta}$ is independent of $\theta_{13}$.
}
\label{alpha-del}
\end{figure}

Similarly, we can work out the behavior of slopes as a function of $\phi$ 
in the case of the bi-probability trajectory drawn by varying $\delta$ 
by holding $\phi$. 
Skipping details, we present the results of the slope $\alpha_{\text{vary} \delta} $ 
of $\delta$-varied bi-probability plots  in Fig.~\ref{alpha-del}.
We observe the similar rotating behavior of the bi-probability diagram 
as in the $\delta$-varied ellipses.

\section{Various degenerate solutions and analytic solutions}
\label{solutions}

We note that for a given true solution, it is possible to obtain
degenerate solutions using analytic expressions as follows.  To do
this let us take the analytic expression of the appearance oscillation
probabilities $P(\nu _e \rightarrow \nu _\mu )$ in (\ref{Pemu}) (and
its anti-neutrino counterpart) with the notation $P(\nu _e \rightarrow
\nu _\mu ) \equiv P_{e \mu}(\delta, \theta_{13}, |\epsilon_{e
  \alpha}|, \phi_{e \alpha}; E )$ ($\alpha = \mu, \tau$).
Since the number of unknown parameters are four, 
$\theta_{13}$, $\delta$, $\vert \varepsilon \vert$, and $\phi$, 
we need four observable quantities. 
We take the oscillation probabilities $P(\nu _e \rightarrow \nu _\mu )$ and 
$P(\bar{\nu} _e \rightarrow \bar{\nu} _\mu )$ at two different energies 
$E_{1}$ and $E_{2}$ for these four inputs. 
Taking the assumed input values of the four parameters given in the caption of 
Fig.~\ref{mixed-sign-degeneracy-em} and 
Fig.~\ref{mixed-sign-degeneracy-et}, 
we solve the equations
\begin{eqnarray}
 P_{e \mu}(\delta^{true}, \theta_{13}^{true}, |\epsilon_{e \alpha}^{true}|, \phi_{e \alpha}^{true}, E_{i})
 &=&
 P_{e \mu}(\delta^{D}, \theta_{13}^{D}, |\epsilon_{e \alpha}^{D}|, \phi_{e \alpha}^{D}, E_{i})\nonumber \\
 \bar{P}_{e \mu}(\delta^{true}, \theta_{13}^{true}, |\epsilon_{e \alpha}^{true}|, \phi_{e \alpha}^{true}, E_{i})
 &=&
 \bar{P}_{e \mu}(\delta^{D}, \theta_{13}^{D}, |\epsilon_{e \alpha}^{D}|, \phi_{e \alpha}^{D}, E_{i}) 
\hspace{0.4cm}
 (\alpha = \mu, \tau;  i = 1, 2) 
\label{Dequation}
\end{eqnarray}
numerically to obtain the degenerate solutions to where the
superscript ``D'' is attached.
To solve (\ref{Dequation}) we arbitrarily take two reference energies
as $E_{1}=10$ GeV and $E_{2}=20$ GeV.

\begin{table}[h]
\vglue 0.5cm
\begin{tabular}{c|c|c|c|c|c}
\hline 
\hline
\ nature of the solution \ 
             & \ hierarchy \ 
             & \ \ ~$\sin^2 2 \theta_{13}$~ \ \ 
             & \ \ ~~$\delta$~~ \ \ 
             & \ \ ~~$\vert \varepsilon_{e \mu} \vert$~~ \ \
             & \ \ ~~$\phi_{e \mu}$~~ \ \  \\
\hline
\ input  \ 
                 & normal
                 & $0.001$
                 & $0$ 
                 & $0.005$ 
                 & $\frac{3}{4} \pi = 2.4$  \\
\hline
\ solution of (\ref{Dequation})  \ 
                 & normal
                 & $0.001$
                 & $0$ 
                 & $0.005$ 
                 & $3.9$  \\
\hline
\ approximate solution of (\ref{Dequation}) \ 
                 & inverted
                 & $0.0035$
                 & $1.5$ 
                 & $0.0046$ 
                 & $3.4$  \\
\hline
\ approximate solution of (\ref{Dequation}) \ 
                 & inverted
                 & $0.0025$
                 & $1.4$ 
                 & $0.0048$ 
                 & $5.1$  \\
\hline
\hline
\end{tabular}
\caption[aaa]
{
  Presented are solutions of the degeneracy equation (\ref{Dequation}) 
  for input parameters similar to the ones used in Fig.~\ref{mixed-sign-degeneracy-em} 
  but with $\phi_{e\mu} = 3\pi/4$ as given in the first row.  
  See the text for explanation of what  ``approximate solution of (\ref{Dequation})'' 
  means in the first column of the Table. 
}
\label{page68}
\end{table}
%
%
%
\begin{table}[h]
\begin{tabular}{c|c|c|c|c|c}
\hline 
\hline
\ nature of the solution \ 
             & \ hierarchy \ 
             & \ \ ~$\sin^2 2 \theta_{13}$~ \ \ 
             & \ \ ~~$\delta$~~ \ \ 
             & \ \ ~~$\vert \varepsilon_{e \tau} \vert$~~ \ \
             & \ \ ~~$\phi_{e \tau}$~~ \ \  \\
\hline
\ input (Fig.~\ref{mixed-sign-degeneracy-et}) \ 
                 & normal
                 & $0.001$
                 & $\frac{3}{2} \pi$ 
                 & $0.02$ 
                 & $\frac{7}{4} \pi$  \\
\hline
\ solution of (\ref{Dequation})  \ 
                 & inverted
                 & $0.0016$
                 & $2.77$ 
                 & $0.022$ 
                 & $0.51$  \\
\hline
\hline
\end{tabular}
\caption[aaa]
{
Presented are solutions of the degeneracy equation (\ref{Dequation}) for input parameters corresponding to Fig.~\ref{mixed-sign-degeneracy-et} 
given in the first row.  
}
\label{page96}
\end{table}
%
%
%
\begin{table}[h]
\begin{tabular}{c|c|c|c|c|c}
\hline 
\hline
\ nature of the solution \ 
             & \ hierarchy \ 
             & \ \ ~$\sin^2 2 \theta_{13}$~ \ \ 
             & \ \ ~~$\delta$~~ \ \ 
             & \ \ ~~$\vert \varepsilon_{e \mu} \vert$~~ \ \
             & \ \ ~~$\phi_{e \mu}$~~ \ \  \\
\hline
\ input  \ 
                 & normal
                 & $0.001$
                 & $0$ 
                 & $0.005$ 
                 & $\frac{5}{4} \pi = 3.9$  \\
\hline
\ solution of (\ref{Dequation})  \ 
                 & normal
                 & $0.001$
                 & $0$ 
                 & $0.005$ 
                 & $2.4$  \\
\hline
\ approximate solution of (\ref{Dequation}) \ 
                 & inverted
                 & $0.0034$
                 & $1.5$ 
                 & $0.0047$ 
                 & $3.4$  \\
\hline
\ approximate solution of (\ref{Dequation}) \ 
                 & inverted
                 & $0.0025$
                 & $1.4$ 
                 & $0.0049$ 
                 & $5.1$  \\
\hline
\hline
\ input  \ 
                 & normal
                 & $0.0001$
                 & $\frac{1}{2} \pi = 1.6$ 
                 & $0.005$ 
                 & $\frac{5}{4} \pi = 3.9$  \\
\hline
\ solution of (\ref{Dequation})  \ 
                 & normal
                 & $0.0002$
                 & $0.45$ 
                 & $0.004$ 
                 & $1.8$  \\
\hline
\ approximate solution of (\ref{Dequation}) \ 
                 & inverted
                 & $0.0015$
                 & $0.6$ 
                 & $0.0052$ 
                 & $6.0$  \\
\hline
\hline
\end{tabular}
\caption[aaa]
{
Similar table as TABLE \ref{page68}.
}
\label{page77em}
\end{table}
%
\begin{table}[h]
\begin{tabular}{c|c|c|c|c|c}
\hline 
\hline
\ nature of the solution \ 
             & \ hierarchy \ 
             & \ \ ~$\sin^2 2 \theta_{13}$~ \ \ 
             & \ \ ~~$\delta$~~ \ \ 
             & \ \ ~~$\vert \varepsilon_{e \tau} \vert$~~ \ \
             & \ \ ~~$\phi_{e \tau}$~~ \ \  \\
\hline
\ input\ 
                 & normal
                 & $0.001$
                 & $\pi = 3.1$ 
                 & $0.02$ 
                 & $\frac{1}{4} \pi = 0.79$  \\
\hline
\ solution of (\ref{Dequation})  \ 
                 & normal
                 & $0.001$
                 & $3.1$ 
                 & $0.022$ 
                 & $5.4$  \\
\hline
\ solution of (\ref{Dequation})  \ 
                 & inverted
                 & $0.001$
                 & $4.5$ 
                 & $0.016$ 
                 & $3.7$  \\
\hline
\hline
\ input \ 
                 & normal
                 & $0.001$
                 & $\pi = 3.1$ 
                 & $0.02$ 
                 & $\frac{7}{4} \pi = 5.5$  \\
\hline
\ solution of (\ref{Dequation})  \ 
                 & normal
                 & $0.001$
                 & $3.1$ 
                 & $0.018$ 
                 & $0.63$  \\
\hline
\ solution of (\ref{Dequation})  \ 
                 & inverted
                 & $0.001$
                 & $4.5$ 
                 & $0.016$ 
                 & $3.9$  \\
\hline
\hline
\end{tabular}
\caption[aaa]
{
Similar table as TABLE \ref{page96}.
}
\label{page88et}
\end{table}

In Tables~\ref{page68} and ~\ref{page96}
we present examples of such degenerate solutions for the system with 
$\varepsilon_{e\mu}$ and $\varepsilon_{e\tau}$, respectively.  
The first column of Tables~\ref{page68}  and ~\ref{page96}
is to specify the nature of the solutions. 
The label ``approximate solution of (\ref{Dequation})'' implies the following 
situation: 
By solving (\ref{Dequation}) with the input parameters in the second 
column we obtain a complex solution which cannot be regarded as the 
physical one. 
The solutions given in Table~\ref{page68} are real numbers which are 
close enough to the complex solutions. 
It should also be noticed that the degeneracy equations 
(\ref{Dequation}) sometimes have solutions which do not survive 
in a form of allowed contours as a results of analysis of neutrino factory 
measurement at $L=3000$ km. 
For example, there is a solution 
$\sin^2 2\theta_{13} = 0.0028$, $\delta = 3.5$, 
$\vert \varepsilon_{e \mu} \vert = 0.008$, and $\phi_{e \mu}=5.1$ 
for the same input as given in Table~\ref{page68}. 
We have confirmed that this solution indeed solves the equation 
(\ref{Dequation}) at $E=10$ GeV and $E=20$ GeV, but the oscillation 
probabilities deviate from the input ones at elsewhere in the energy spectra. 
Therefore, the degenerate solution was lifted by the spectral informations 
used by our numerical analysis.

In Tables \ref{page77em}-\ref{page88et} we also show 
some other examples. 
Table \ref{page77em} is similar table as Table \ref{page68}. Input values 
of parameters are close to the first solution of Table \ref{page68}. 
Therefor the solutions of this set are very similar to Table \ref{page68}'s.
These relationship exists in the system with $\epsilon_{e \tau}$ too. 
In Table \ref{page88et}, input values of parameters of upper set are close 
to the first solution of lower set with similar another solution.

\end{document}